\documentclass[]{aastex631}
\usepackage{CJKutf8}

\newcommand{\Caltech}{\affil{California Institute of Technology, 1200 E. California Blvd., MC 249-17, Pasadena, CA 91125, USA}}

\newcommand{\GEMINI}{\affil{Gemini Observatory/NSF’s NOIRLab, 950 N. Cherry Avenue, Tucson, AZ, 85719, USA
}}
\newcommand{\ASCL}{\affil{Astrophysics Source Code Librar
y, Michigan Technological University, 1400 Townsend Drive, Houghton, MI 49931}}

\newcommand{\OSU}{\affil{Department of Astronomy, The Ohio State University, 140 West 18th Avenue, Columbus, Ohio 43210, USA}}

\newcommand{\Alberta}{\affil{Department of Physics, University of Alberta, Edmonton, AB T6G 2E1, Canada}}

\newcommand{\ANU}{\affil{Research School of Astronomy and Astrophysics, Australian National University, Canberra, ACT 2611, Australia}}

\newcommand{\IPARCOS}{\affil{Instituto de F\'{\i}sica de Part\'{\i}culas y del Cosmos, Universidad Complutense de Madrid, E-28040 Madrid, Spain}}

\newcommand{\Carnegie}{\affil{Observatories of the Carnegie Institution for Science, 813 Santa Barbara Street, Pasadena, CA 91101, USA}}

\newcommand{\CCAPP}{\affil{Center for Cosmology and AstroParticle Physics, 191 West Woodruff Avenue, Columbus, OH 43210, USA}}

\newcommand{\CfA}{\affil{Center for Astrophysics $\mid$ Harvard \& Smithsonian , 60 Garden Street, Cambridge, MA 02138, USA}}

\newcommand{\CITEVA}{\affil{Centro de Astronomía (CITEVA), Universidad de Antofagasta, Avenida Angamos 601, Antofagasta, Chile}}

\newcommand{\ESO}{\affil{European Southern Observatory, Karl-Schwarzschild Stra{\ss}e 2, D-85748 Garching bei M\"{u}nchen, Germany}}

\newcommand{\Heidelberg}{\affil{Astronomisches Rechen-Institut, Zentrum f\"{u}r Astronomie der Universit\"{a}t Heidelberg, M\"{o}nchhofstra\ss e 12-14, D-69120 Heidelberg, Germany}}

\newcommand{\COOL}{\affil{Cosmic Origins Of Life (COOL) Research DAO, coolresearch.io}}

\newcommand{\ICRAR}{\affil{International Centre for Radio Astronomy Research, University of Western Australia, 35 Stirling Highway, Crawley, WA 6009, Australia}}

\newcommand{\IRAM}{\affil{Institut de Radioastronomie Millim\'{e}trique (IRAM), 300 Rue de la Piscine, F-38406 Saint Martin d'H\`{e}res, France}}

\newcommand{\IRAP}{\affil{CNRS, IRAP, 9 Av. du Colonel Roche, BP 44346, F-31028 Toulouse cedex 4, France}}

\newcommand{\UPS}{\affil{Universit\'{e} de Toulouse, UPS-OMP, IRAP, F-31028 Toulouse cedex 4, France}}

\newcommand{\ITA}{\affil{Universit\"{a}t Heidelberg, Zentrum f\"{u}r Astronomie, Institut f\"{u}r Theoretische Astrophysik, Albert-Ueberle-Str 2, D-69120 Heidelberg, Germany}}

\newcommand{\IWR}{\affil{Universit\"{a}t Heidelberg, Interdisziplin\"{a}res Zentrum f\"{u}r Wissenschaftliches Rechnen, Im Neuenheimer Feld 205, D-69120 Heidelberg, Germany}}

\newcommand{\JHU}{\affil{Department of Physics and Astronomy, The Johns Hopkins University, Baltimore, MD 21218, USA}}

\newcommand{\MPE}{\affil{Max-Planck-Institut f\"{u}r extraterrestrische Physik, Giessenbachstra{\ss}e 1, D-85748 Garching, Germany}}

\newcommand{\MPIA}{\affil{Max-Planck-Institut f\"{u}r Astronomie, K\"{o}nigstuhl 17, D-69117, Heidelberg, Germany}}

\newcommand{\OAN}{\affil{Observatorio Astron\'{o}mico Nacional (IGN), C/Alfonso XII, 3, E-28014 Madrid, Spain}}

\newcommand{\UToledo}{\affil{University of Toledo, 2801 W. Bancroft St., Mail Stop 111, Toledo, OH, 43606}}

\newcommand{\UBonn}{\affil{Argelander-Institut f\"ur Astronomie, Universit\"at Bonn, Auf dem H\"ugel 71, 53121 Bonn, Germany}}

\newcommand{\UChile}{\affil{Departamento de Astronom\'{i}a, Universidad de Chile, Camino del Observatorio 1515, Las Condes, Santiago, Chile}}

\newcommand{\UCM}{\affil{Departamento de F\'{\i}sica de la Tierra y Astrof\'{\i}sica, Universidad Complutense de Madrid, E-28040 Madrid, Spain}}

\newcommand{\UCSD}{\affil{Center for Astrophysics and Space Sciences, Department of Physics,  University of California,\\ San Diego, 9500 Gilman Drive, La Jolla, CA 92093, USA}}

\newcommand{\ULyon}{\affil{Univ Lyon, Univ Lyon 1, ENS de Lyon, CNRS, Centre de Recherche Astrophysique de Lyon UMR5574,\\ F-69230 Saint-Genis-Laval, France}}

\newcommand{\UWyoming}{\affil{Department of Physics and Astronomy, University of Wyoming, Laramie, WY 82071, USA}}

\newcommand{\UGent}{\affil{Sterrenkundig Observatorium, Universiteit Gent, Krijgslaan 281 S9, B-9000 Gent, Belgium}}

\newcommand{\STScI}{\affil{Space Telescope Science Institute, 3700 San Martin Drive, Baltimore, MD 21218, USA}}

\newcommand{\STScIESA}{\affiliation{AURA for the European Space Agency (ESA), Space Telescope Science Institute, 3700 San Martin Drive, Baltimore, MD 21218, USA}}

\newcommand{\INAF}{\affil{INAF -- Osservatorio Astrofisico di Arcetri, Largo E. Fermi 5, I-50157, Firenze, Italy}}

\newcommand{\LERMA}{\affil{Observatoire de Paris, PSL Research University, CNRS, Sorbonne Universit\'es, 75014 Paris}}

\newcommand{\SAIMSU}{\affil{Sternberg Astronomical Institute, Lomonosov Moscow State University, Universitetsky pr. 13, 119234 Moscow, Russia}}

\newcommand{\StockholmOKC}{\affil{The Oskar Klein Centre for Cosmoparticle Physics, Department of Physics, Stockholm University, AlbaNova, Stockholm, SE-106 91, Sweden}}

\newcommand{\obstime}{112.6}

\received{03-Dec-2022}
\accepted{XXX}

\submitjournal{ApJL}

\shorttitle{The PHANGS-JWST Survey}
\shortauthors{PHANGS Collaboration}

\graphicspath{{./}{Figures/}}

\begin{document}

\title{The PHANGS-JWST Treasury Survey: \\Star Formation, Feedback, and Dust \underline{P}hysics at \underline{H}igh \underline{A}ngular resolution in \underline{N}earby \underline{G}alaxie\underline{S}}

\correspondingauthor{Janice C. Lee}
\email{janice.lee@noirlab.edu}
 
\author[0000-0002-2278-9407]{Janice C. Lee}
\GEMINI
\affiliation{Steward Observatory, University of Arizona, 933 N Cherry Ave,Tucson, AZ 85721, USA}

\author[0000-0002-4378-8534]{Karin M. Sandstrom}
\UCSD

\author[0000-0002-2545-1700]{Adam K. Leroy}
\OSU
 
\author[0000-0002-8528-7340]{David A. Thilker}
\JHU
 
\author[0000-0002-3933-7677]{Eva Schinnerer}
\MPIA

\author[0000-0002-5204-2259]{Erik~Rosolowsky}
\Alberta

\author[0000-0003-3917-6460]{Kirsten~L.~Larson}
\STScIESA
 
\author[0000-0002-4755-118X]{Oleg~V.~Egorov}
\Heidelberg
\SAIMSU
 
\author[0000-0002-0012-2142]{Thomas G. Williams}
\MPIA
 
\author[0000-0002-2617-5517]{Judy Schmidt}
\ASCL

\author[0000-0002-6155-7166]{Eric Emsellem}
\ESO
\ULyon
 
\author[0000-0002-5259-2314]{Gagandeep S. Anand}
\STScI
 
\author[0000-0003-0410-4504]{Ashley~T.~Barnes}
\UBonn
 
\author[0000-0002-2545-5752]{Francesco Belfiore}
\INAF
 
\author[0000-0003-0583-7363]{Ivana Be\v{s}li\'c}
\affiliation{Argelander-Institut f\"{u}r Astronomie, Universit\"{a}t Bonn, Auf dem H\"{u}gel 71, 53121 Bonn, Germany}
 
\author[0000-0003-0166-9745]{Frank Bigiel}
\UBonn
 
\author[0000-0003-4218-3944]{Guillermo A. Blanc}
\Carnegie
\UChile
 
\author[0000-0002-5480-5686]{Alberto D. Bolatto}
\affiliation{Department of Astronomy and Joint Space-Science Institute, University of Maryland, College Park, MD 20742, USA}
 
\author[0000-0003-0946-6176]{M\'ed\'eric~Boquien}
\CITEVA
 
\author[0000-0002-8760-6157]{Jakob den Brok}
\affiliation{Argelander-Institut f\"{u}r Astronomie, Universit\"{a}t Bonn, Auf dem H\"{u}gel 71, 53121, Bonn, Germany}
 
\author[0000-0001-5301-1326]{Yixian Cao}
\affiliation{Max-Planck-Institut f\"ur Extraterrestrische Physik (MPE), Giessenbachstr. 1, D-85748 Garching, Germany}
 
\author[0000-0003-0085-4623]{Rupali Chandar}
\UToledo
 
\author[0000-0002-5235-5589]{J\'er\'emy~Chastenet}
\UGent
 
\author[0000-0002-5635-5180]{M\'elanie Chevance}
\ITA
\COOL
 
\author[0000-0003-2551-7148]{I-Da Chiang \begin{CJK*}{UTF8}{bkai}(江宜達)\end{CJK*}}%
\affiliation{Institute of Astronomy and Astrophysics, Academia Sinica, No. 1, Sec. 4, Roosevelt Road, Taipei 10617, Taiwan}

\author[0000-0002-8549-4083]{Enrico Congiu}
\UChile
 
\author[0000-0002-5782-9093]{Daniel A. Dale}
\UWyoming
 
\author[0000-0003-1943-723X]{Sinan Deger}
\Caltech
\StockholmOKC
 
\author[0000-0002-1185-2810]{Cosima Eibensteiner}
\affiliation{Argelander-Institut für Astronomie, Universität Bonn, Auf dem Hügel 71, 53121 Bonn, Germany}
 
\author[0000-0001-5310-467X]{Christopher M. Faesi}
\affiliation{University of Connecticut, Department of Physics, 196A  Auditorium Road, Unit 3046, Storrs, CT, 06269}
 
\author[0000-0001-6708-1317]{Simon C.~O. Glover}
\ITA
 
\author[0000-0002-3247-5321]{Kathryn Grasha}
\ANU
 
\author[0000-0002-9768-0246]{Brent Groves}
\ICRAR
\ANU
 
\author[0000-0002-8806-6308]{Hamid Hassani}
\Alberta
 
\author[0000-0001-7448-1749]{Kiana~F.~Henny}
\UWyoming
 
\author[0000-0001-9656-7682]{Jonathan~D.~Henshaw}
\affiliation{Astrophysics Research Institute, Liverpool John Moores University, 146 Brownlow Hill, Liverpool L3 5RF, UK}
\MPIA 

\author[0000-0001-8040-4088]{Nils~Hoyer}
\affiliation{Donostia International Physics Center, Paseo Manuel de Lardizabal 4, E-20118 Donostia-San Sebasti{\'{a}}n, Spain}
\MPIA
\affiliation{Universit{\"{a}}t Heidelberg, Seminarstrasse 2, D-69117 Heidelberg, Germany}

\author[0000-0002-9181-1161]{Annie~Hughes}
\IRAP
\UPS
 
\author[0000-0002-4232-0200]{Sarah Jeffreson}
\CfA
 
\author[0000-0002-9165-8080]{Mar\'ia J. Jim\'enez-Donaire}
\OAN
\affiliation{Centro de Desarrollos Tecnológicos, Observatorio de Yebes (IGN), 19141 Yebes, Guadalajara, Spain}
 
\author[0000-0002-0432-6847]{Jaeyeon Kim}
\ITA
 
\author[0000-0003-4770-688X]{Hwihyun~Kim}
\affiliation{Gemini Observatory/NSF's NOIRLab, 950 N. Cherry Avenue, Tucson, AZ, USA}
 
\author[0000-0002-0560-3172]{Ralf S.\ Klessen}
\ITA
\IWR
 
\author[0000-0001-9605-780X]{Eric W. Koch}
\CfA
 
\author[0000-0001-6551-3091]{Kathryn Kreckel}
\Heidelberg
 
\author[0000-0002-8804-0212]{J.~M.~Diederik Kruijssen}
\COOL
 
\author[0000-0002-4825-9367]{Jing~Li}
\affiliation{Astronomisches Rechen-Institut, Zentrum f\"{u}r Astronomie der Universit\"{a}t Heidelberg, M\"{o}nchhofstra\ss e 12-14, 69120 Heidelberg, Germany}
 
\author[0000-0001-9773-7479]{Daizhong Liu}
\MPE
 
\author[0000-0002-1790-3148]{Laura A. Lopez}
\affiliation{Department of Astronomy, The Ohio State University, 140 West 18th Avenue, Columbus, Ohio 43210, USA}
\affiliation{Center for Cosmology and Astroparticle Physics, 191 West Woodruff Avenue, Columbus, OH 43210, USA}
\affiliation{Flatiron Institute, Center for Computational Astrophysics, NY 10010, USA}
 
\author[0000-0001-6038-9511]{Daniel Maschmann}
\affiliation{Steward Observatory, University of Arizona, 933 N Cherry Ave,Tucson, AZ 85721, USA}
\affiliation{Sorbonne {Universit\'e},
LERMA, Observatoire de Paris, PSL university, CNRS, F-75014, Paris, France}
 
\author[0000-0002-5993-6685]{Ness Mayker Chen}
\affiliation{Department of Astronomy, The Ohio State University, 140 West 18th Avenue, Columbus, Ohio 43210, USA}
\affiliation{Center for Cosmology and Astroparticle Physics, 191 West Woodruff Avenue, Columbus, OH 43210, USA}
 
\author[0000-0002-6118-4048]{Sharon E. Meidt}
\UGent
 
\author[0000-0001-7089-7325]{Eric J. Murphy}
\affiliation{National Radio Astronomy Observatory, 520 Edgemont Road, Charlottesville, VA 22903, USA}
 
\author[0000-0002-3289-8914]{Justus Neumann}
\MPIA 
 
\author[0000-0002-6922-2598]{Nadine~Neumayer}
\MPIA 
 
\author[0000-0002-1370-6964]{Hsi-An Pan}
\affiliation{Department of Physics, Tamkang University, No.151, Yingzhuan Road, Tamsui District, New Taipei City 251301, Taiwan}
 
\author[0000-0002-0873-5744]{Ismael Pessa}
\MPIA 
\affiliation{Leibniz-Institut f\"{u}r Astrophysik Potsdam (AIP), An der Sternwarte 16, 14482 Potsdam, Germany}
 
\author[0000-0003-3061-6546]{J\'er\^ome Pety}
\IRAM
\LERMA
 
\author[0000-0002-0472-1011]{Miguel Querejeta}
\OAN
 
\author[0000-0001-5965-3530]{Francesca~Pinna}
\MPIA 
 
\author[0000-0002-0579-6613]{M. Jimena Rodr\'{\i}guez}
\affiliation{Steward Observatory, University of Arizona, 933 N Cherry Ave,Tucson, AZ 85721, USA}
\affiliation{Instituto de Astrof\'{\i}sica de La Plata, CONICET--UNLP, Paseo del Bosque S/N, B1900FWA La Plata, Argentina }
 
\author[0000-0002-2501-9328]{Toshiki Saito}
\affiliation{National Astronomical Observatory of Japan, 2-21-1 Osawa, Mitaka, Tokyo, 181-8588, Japan}
 
\author[0000-0003-0651-0098]{Patricia S\'anchez-Bl\'azquez}
\UCM
\IPARCOS
 
\author[0000-0002-6363-9851]{Francesco Santoro}
\MPIA
 
\author[0000-0002-5783-145X]{Amy Sardone}
\OSU \CCAPP
 
\author[0000-0002-0820-1814]{Rowan J. Smith}
\affiliation{Jodrell Bank center for Astrophysics, Department of Physics and Astronomy, University of Manchester, Oxford Road, Manchester M13 9PL, UK}
 
\author[0000-0001-6113-6241]{Mattia C. Sormani}
\ITA
 
\author[0000-0003-2707-4678]{Fabian Scheuermann}
\Heidelberg
 
 
\author[0000-0002-9333-387X]{Sophia K. Stuber}
\MPIA 
 
\author[0000-0002-9183-8102]{Jessica Sutter}
\affiliation{Center for Astrophysics \& Space Sciences, Department of Physics, University of California, San Diego, 9500 Gilman Drive, San Diego, CA 92093, USA}
 zz
\author[0000-0003-0378-4667]{Jiayi~Sun \begin{CJK*}{UTF8}{gbsn}(孙嘉懿)\end{CJK*}}
\affiliation{Department of Physics and Astronomy, McMaster University, 1280 Main Street West, Hamilton, ON L8S 4M1, Canada}
\affiliation{Canadian Institute for Theoretical Astrophysics (CITA), University of Toronto, 60 St George Street, Toronto, ON M5S 3H8, Canada}
 
\author[0000-0003-4209-1599]{Yu-Hsuan Teng}
\affiliation{Center for Astrophysics and Space Sciences, Department of Physics, University of California, San Diego, 9500 Gilman Drive, La Jolla, CA 92093, USA}
 
\author[0000-0002-9483-7164]{Robin G. Tre{\ss}}
\affiliation{Institute of Physics, Laboratory for galaxy evolution and spectral modelling, EPFL, Observatoire de Sauverny, Chemin Pegais 51, 1290 Versoix, Switzerland.}
 
\author[0000-0003-1242-505X]{Antonio Usero}
\OAN

\author[0000-0002-7365-5791]{Elizabeth~J.~Watkins}
\Heidelberg

\author[0000-0002-3784-7032]{Bradley~C.~Whitmore}
\STScI

\author[0000-0001-7876-1713]{Alessandro Razza}
\UChile
 

\begin{abstract}
The PHANGS collaboration has been building a reference dataset for the multi-scale, multi-phase study of star formation and the interstellar medium in nearby galaxies.  With the successful launch and commissioning of JWST, we can now obtain high-resolution infrared imaging to probe the youngest stellar populations and dust emission on the scales of star clusters and molecular clouds ($\sim$5-50 pc).  In Cycle 1, PHANGS is conducting an 8-band imaging survey from 2-21$\mu$m of 19 nearby spiral galaxies. 
CO(2--1) mapping, optical integral field spectroscopy, and UV-optical imaging for all 19 galaxies have been obtained through large programs with ALMA, VLT/MUSE, and Hubble. PHANGS-JWST enables a full inventory of star formation, accurate measurement of the mass and age of star clusters, identification of the youngest embedded stellar populations, and characterization of the physical state of small dust grains.  When combined with Hubble catalogs of $\sim$10,000 star clusters, MUSE spectroscopic mapping of $\sim$20,000 HII regions, and $\sim$12,000 ALMA-identified molecular clouds, it becomes possible to measure the timescales and efficiencies of the earliest phases of star formation and feedback, build an empirical model of the dependence of small dust grain properties on local ISM conditions, and test our understanding of how dust-reprocessed starlight traces star formation activity, all across a diversity of galactic environments.  Here we describe the PHANGS-JWST Treasury survey, present the remarkable imaging obtained in the first few months of science operations, and provide context for the initial results presented in the first series of PHANGS-JWST publications.

\end{abstract}


\section{Introduction}
\label{sec:intro}

The discovery of infrared emission in the Orion Nebula revealed that the earliest phases of star formation occur in the densest, dust-enshrouded cores of molecular clouds which are not observable in the optical \citep[][]{becklin67, kleinmann67}.  
Beginning with IRAS in the 1980s \citep{neugebaueretal84}, a series of cryogenic infrared space missions including ISO, Spitzer, and Herschel \citep{iso,werneretal04,pilbrattetal10} mapped emission from star formation and dust with increasing resolution and coverage across the mid- and far-infrared \citep[e.g. ][and references therein]{soifer87,soifer08,kennicutt12}. These missions demonstrated that infrared observations are a requisite component for understanding the process of star formation in galaxies and the physics of the interstellar medium.  With the transformative infrared capabilities of JWST, studies of star formation and dust in galaxies are again entering a new era.  New breakthroughs in this field will rely on JWST together with 
the collective effort of the astronomy community to conduct observations across the electromagnetic spectrum which capture all major stages of the star formation cycle, from gas to stars.

For nearby galaxies, JWST is finally extending the high spatial resolution infrared studies of star formation and dust previously only possible in the Milky Way and Local Group to a more representative sample of galaxies within a few tens of Mpc.  Here, we introduce the PHANGS-JWST Cycle 1 Treasury survey, which was designed to use JWST's order-of-magnitude improvement in sensitivity and angular resolution to map the infrared emission from the youngest stellar populations and dusty interstellar medium across the disks of galaxies on scales of tens of parsecs.  These are the physical scales where gas fragmentation is expected (e.g.\ due to the Jeans instability), and where the key physical processes that drive, regulate, and extinguish star formation can be investigated for individual star clusters ($\sim$5 pc), HII regions ($\sim$10 pc), and molecular clouds ($\sim$50 pc).  

Imaging in 8-bands from 2-21$\mu$m is being obtained of 19 nearby (d$<$20 Mpc) spiral galaxies. All have UV-optical imaging from HST \citep[][]{phangs-hst}, optical integral field spectroscopy from VLT/MUSE \citep[][]{phangs-muse}, and CO(2--1) mapping from ALMA \citep[][]{phangs-alma}, through survey programs led by the PHANGS collaboration \citep{schinnerer19}. All data are publicly available from the respective observatory's archives, and HST catalogs of $\gtrsim12{,}000$ star clusters and associations \citep[][]{thilker22,deger22,whitmore21,larson22}, MUSE catalogs of $\gtrsim20{,}000$ \ion{H}{2} regions \citep[][Congiu et al., subm.]{santoro22, GROVES_HIICAT}, and ALMA catalogs of $\gtrsim10{,}000$ molecular clouds \citep{sun18,rosolowsky21,sun22} have been produced.\footnote{\url{www.phangs.org}} 

The PHANGS-JWST sample covers the range of star formation, gas, dust, structural, and dynamical properties found in present-day massive galaxies on the star-forming main sequence, enabling studies that connect small-scale gas and star formation physics to the broader context of galactic structure and galaxy evolution. The combined PHANGS multiwavelength observational programs yield the first dataset that accesses all key stages of the star formation cycle at the resolution of individual star-forming regions across a sample that reflects the galactic environments where the majority of star formation occurs at $z=0$ \citep[e.g.][]{brinchmann04,salim07,Saintonge17}.

PHANGS targets were among the first to be observed soon after after the successful completion of JWST commissioning in June 2022. Within the first two months of science operations, data for four galaxies (NGC~7496, IC~5332, NGC~628, NGC~1365) were obtained.  As a Treasury program, the PHANGS-JWST imaging data have no exclusive access period. The release of the data, shortly after observation, revealed the architecture of the dusty ISM in exquisite detail, capturing the attention of not only the science community, but also of the general public and press.

Analysis by the PHANGS collaboration is underway, and focuses on the following science goals:
\begin{enumerate}
    \item characterization of the youngest embedded stellar populations that are inaccessible at optical-UV wavelengths, completion of the inventory of newborn stars and definitive measurement of star cluster mass functions and formation efficiencies, to probe the early evolution of stars and star cluster populations;
    \item characterization of ISM bubble and shell features, and measurement of the duration of dust embedded star formation and star formation efficiencies to identify the sources and timescales of star formation feedback;
    \item determination of how local interstellar conditions influence the properties, evolution, and processing of the smallest dust grains, in particular the polycyclic aromatic hydrocarbons (PAHs), to enable a physically robust interpretation of dust emission in nearby galaxies;
    \item development of new dust-based high resolution tracers of the neutral gas that are complementary to HI, CO, CI, and C$^+$ mapping, to resolve the multi-scale filamentary structure of the ISM;
    \item establishment and calibration of robust mid-IR diagnostics of star formation activity that account for the youngest stars, to compute star formation rates from cloud- to galaxy scales.
\end{enumerate}

This paper is intended to provide a general reference for the parameters of the survey, as well as an introduction for the papers in this Issue, which showcases some of the science possible based on PHANGS-JWST data for the first four galaxies observed. The remaining sections are organized as follows.  
In Section \ref{sec:sample}, we describe the galaxy sample and provide a brief summary of the supporting multi-wavelength PHANGS datasets.  Our observing strategy with NIRCam and MIRI is presented in Section \ref{sec:observations}.  Next in Section \ref{sec:redux}, we describe our early data reduction efforts to enable first science, focusing on modifications made to the standard STScI JWST Science Calibration Pipeline,  
and note some limitations for the use of these early data. In Section \ref{sec:results}, we present the imaging for NGC~7496, IC~5332, NGC~628, and NGC~1365. In Section \ref{sec:goals}, we describe the scientific goals that motivate the PHANGS-JWST survey, and highlight results presented in this Issue.  In Section \ref{sec:products}, we describe the image products that will be released by the PHANGS team, and provide examples of additional JWST high level science products which would be valuable to be developed from the ensemble of panchromatic PHANGS datasets.  We conclude with a short summary in Section \ref{sec:products}. 
\section{Galaxy Sample} \label{sec:sample}

The PHANGS-JWST Treasury Survey targets  the 19 PHANGS galaxies that have the full complement of UV-optical imaging from PHANGS--HST \citep{phangs-hst}, 
optical spectral mapping from PHANGS--MUSE \citep{phangs-muse}, 
and millimeter-wave spectral mapping from PHANGS--ALMA \citep{phangs-alma}.  
In addition to these major survey datasets from JWST, HST, VLT-MUSE, and ALMA, 
a wealth of additional supporting data from the PHANGS collaboration has been and continues to be obtained, e.g., Astrosat far-ultraviolet/near-ultraviolet imaging (PI: E.~Rosolowsky), HST H$\alpha$ narrowband imaging (PI: R.~Chandar), ground-based wide-field H$\alpha$ narrowband imaging (PI: G.~Blanc, I-T.~Ho), and \ion{H}{1} 21-cm observations from team programs with the VLA and MeerKAT.  A complete listing of PHANGS observations and data products is provided at \url{www.phangs.org/data}. 

\begin{figure*}
    \plotone{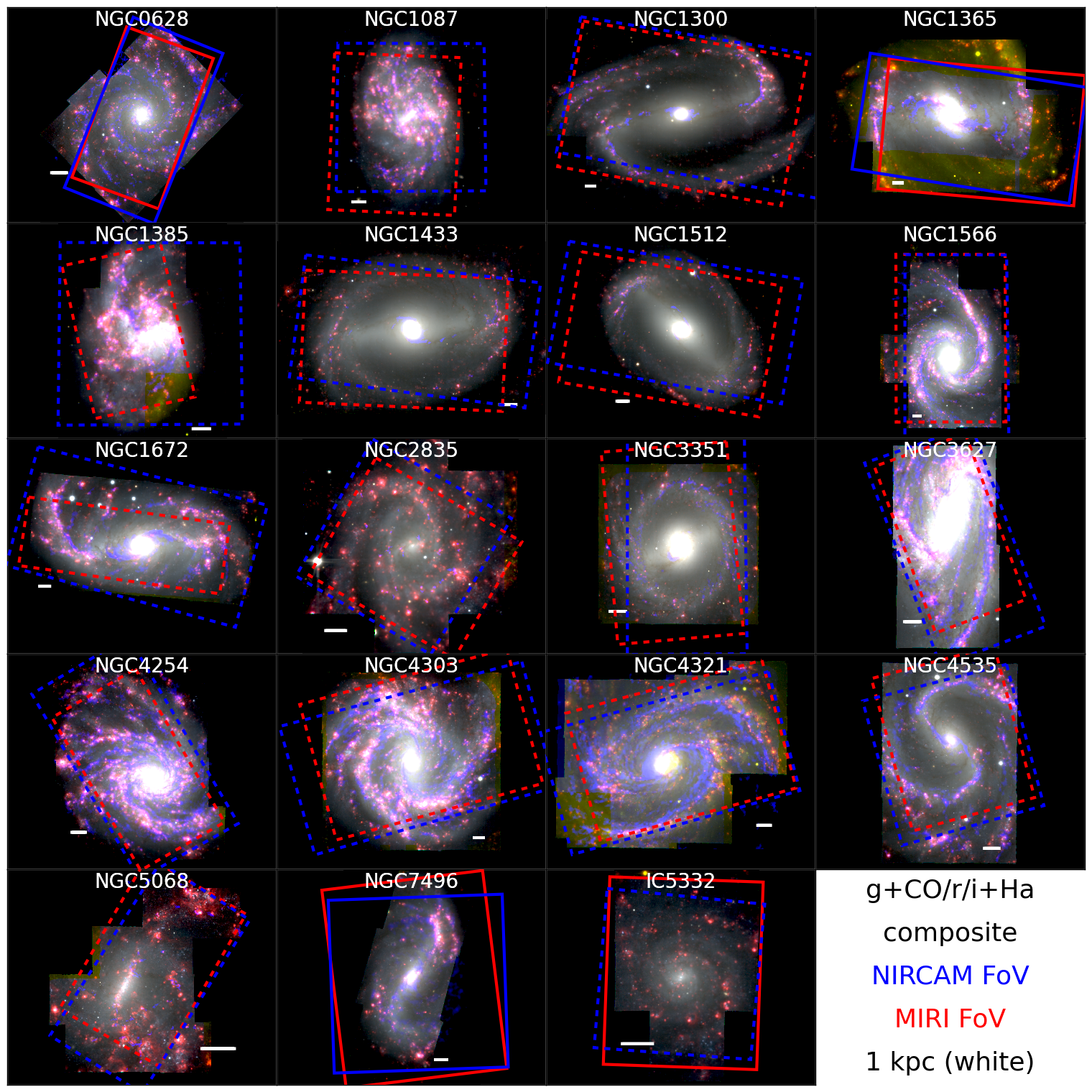}
    \caption{Composite $gri$+H$\alpha$+CO images for the 19 galaxies in the PHANGS-JWST Treasury program. The $gri$+H$\alpha$ images are constructed from VLT MUSE full-field spectral mapping \citep{phangs-muse}, with H$\alpha$ line emission in red, and combined with ALMA CO(2--1) \citep{phangs-alma} maps, with CO(2--1) flux in blue. JWST NIRCam and MIRI footprints have been defined to overlap existing MUSE, ALMA, and HST data. Footprints shown with dashed lines represent observations that have not been executed at time of writing (October 2022), and may rotate slightly due to the range of allowed orient angles specified for the target. The JWST sample spans a factor of $\sim50$ in stellar mass and SFR, and a factor of $\sim100$ in CO surface density (Table \ref{tab:galaxysample2}). It includes galaxies with prominent dust lanes, a full range of stellar bar, bulge, spiral arm morphologies, as well as nuclear starbursts and AGN activity.}
    \label{fig:footprints}
\end{figure*}

Basic properties of the galaxies relevant to their selection are given in Tables \ref{tab:galaxysample1} and \ref{tab:galaxysample2}, and Figure \ref{fig:footprints} shows MUSE+ALMA composite images for the full sample.

Over the distance range of these galaxies (5-20 Mpc; Table \ref{tab:galaxysample1}), JWST's point spread function FWHM subtends 2-6 pc at 2$\mu$m, and 17-65 pc at 21$\mu$m. For the first time, JWST allows for mid-infrared imaging of individual star clusters, HII regions, and molecular clouds across the nearby galaxy population, which are well-matched to observations of these populations from HST, MUSE, and ALMA.  The combination of datasets from these four facilities is essential for achieving our specific science goals (\S \ref{sec:goals}), and the PHANGS program currently provides the largest uniform collection of such data for galaxies in the local volume.

\begin{deluxetable*}{lcccclcccl}
\tablecaption{PHANGS-JWST Target Galaxy Distance, Orientation, and Morphology\label{tab:galaxysample1}}
\centering
\tablehead{
\colhead{Galaxy} & \colhead{$\alpha$} & \colhead{$\delta$} &\colhead{$D$\tablenotemark{a}} & \colhead{$\sigma_D$\tablenotemark{a}} & \colhead{Method\tablenotemark{a}} & \colhead{$i$\tablenotemark{b}} & \colhead{$PA$\tablenotemark{b}} &\colhead{T\tablenotemark{c}} &
\colhead{Morphology\tablenotemark{c}} \\ 
& \colhead{(J2000)} & \colhead{(J2000)} & \colhead{(Mpc)} & \colhead{(Mpc)} & & \colhead{(deg)} & \colhead{(deg)} & & }
\startdata
NGC~0628\tablenotemark{*}    & 01h36m41.75s & +15d47m01.2s        & 9.84                 & 0.63             & TRGB\textsuperscript{1}                   &	9	& 21 & 5.2 & SA(s)c \\ 
NGC~1087    & 02h46m25.16s & -00d29m55.1s        & 15.85                & 2.24             & Group\textsuperscript{2}                  &	43	& 359 & 5.2	& SB(rs)c{\underline d} pec\\ 
NGC~1300    & 03h19m41.08s & -19d24m40.9s        & 18.99                & 2.85             & NAM\textsuperscript{3,4}                  &	32	& 278 & 4	& (R')SB(s,bl,nrl)b \\ 
NGC~1365\tablenotemark{*}    & 03h33m36.37s & -36d08m25.4s        & 19.57                & 0.78             & TRGB\textsuperscript{1}                   &	55	& 201 & 3.2	& (R')SB(r{\underline s},nr)bc \\ 
NGC~1385    & 03h37m28.85s & -24d30m01.1s        & 17.22                & 2.58             & NAM\textsuperscript{3,4}                  &	44	& 181 & 5.9	& SB(s)dm pec \\ 
NGC~1433    & 03h42m01.55s & -47d13m19.5s        & 18.63\tablenotemark{$\dag$} & 1.86      & PNLF\textsuperscript{5}                   &	29	& 200 & 1.5	& (R\_1\_')SB(r,p,nrl,nb)a \\ 
NGC~1512    & 04h03m54.28s & -43d20m55.9s        & 18.83\tablenotemark{$\dag$} & 1.88      & PNLF\textsuperscript{5}                   &	43	& 262 & 1.2	& (R{\underline L})SB(r,bl,nr)a \\
NGC~1566    & 04h20m00.42s & -54d56m16.1s        & 17.69                & 2.00             & Group\textsuperscript{2}                  &	30	& 215 & 4 &	(R\_1\_')SAB(rs,r{\underline s})b \\ 
NGC~1672    & 04h45m42.50s & -59d14m49.9s        & 19.40                & 2.91             & NAM\textsuperscript{3,4}                  &	43	& 134 & 3.3 &  (R')SA{\underline B}(rs,nr)b \\ 
NGC~2835    & 09h17m52.91s & -22d21m16.8s        & 12.22                & 0.94             & TRGB\textsuperscript{1}                   &	41	& 1 & 5	& SB(rs)c \\ 
NGC~3351    & 10h43m57.70s & +11d42m13.7s        & 9.96                 & 0.33             & TRGB\textsuperscript{1}                   &	45	& 193 & 3.1	& (R')SB(r,bl,nr)a \\ 
NGC~3627    & 11h20m14.96s & +12d59m29.5s        & 11.32                & 0.48             & TRGB\textsuperscript{1}                   &	57	& 173 & 3.1	& SB\_x\_(s)b pec\\ 
NGC~4254    & 12h18m49.60s & +14d24m59.4s        & 13.1                 & 2.8              & SCM\textsuperscript{6}                    &	34	& 68 & 5.2 & SA(s)c pec \\ 
NGC~4303    & 12h21m54.90s & +04d28m25.1s        & 16.99                & 3.04             & Group\textsuperscript{2}                  &	24	& 312 & 4 & SAB(rs,nl)b{\underline c} \\ 
NGC~4321    & 12h22m54.83s & +15d49m18.5s        & 15.21                & 0.49             & Cepheid\textsuperscript{7}                &	39	& 156 & 4 & SAB(rs,nr,nb)bc \\ 
NGC~4535    & 12h34m20.31s & +08d11m51.9s        & 15.77                & 0.37             & Cepheid\textsuperscript{7}                &	45	& 180 & 5 & SAB(s)c \\ 
NGC~5068    & 13h18m54.81s & -21d02m20.8s        & 5.20                 & 0.21             & TRGB\textsuperscript{1}                   &	36	& 342 & 6 & SB(s)d \\ 
NGC~7496\tablenotemark{*}    & 23h09m47.29s & -43d25m40.6s        & 18.72                & 2.81             & NAM\textsuperscript{3,4}                    &	36	& 194 & 3.2	& (R')SB\_x\_(rs)b \\
IC~5332\tablenotemark{*}    & 23h34m27.49s & -36d06m03.9s        & 9.01                 & 0.41             & TRGB\textsuperscript{1}                   &	27	& 74 & 6.8	& S{\underline A}B(s)cd
\enddata
\tablenotetext{a}{Compilation of galaxy distances ($D$) and uncertainties ($\sigma_D$). Distance methods include TRGB (Tip of the Red Giant Branch), Group, NAM (Numerical action model), PNLF (Planetary Nebula Luminosity Function), SCM (standardizable candle method), Cepheid (Cepheid period-luminosity relation). Superscripts indicate the primary reference given in the compilation by \citet{anand21}, and are as follows: 1) \citet{edd21}, 2) \citet{kourkchi17}, 3) \citet{shaya17}, 4) \citet{kourkchi20}, 5) \citet{scheuermann22}, 6) \citet{nugent2006}, and 7) \citet{freedman2001}.}
\tablenotetext{b}{Galaxy inclination and position angle adopted from \citet{lang20} where available and \citet{phangs-alma} otherwise.}
\tablenotetext{c}{ Morphological T-type from HyperLEDA \citep{hyperleda}. Morphological classification from \citet{buta15} except for NGC~2835, which comes from \citet{rc3}.}
\tablenotetext{$\dag$}{The PNLF distances to NGC 1433 and 1512 varied from \citet{anand21}, which was used in \citet{phangs-muse} and \citet{phangs-alma}, and \citet{scheuermann22}.}
\tablenotetext{*}{One of first four galaxies observed which form the basis of the PHANGS-JWST First Results presented in this volume.}
\label{tab:galaxysample}
\end{deluxetable*}
\begin{deluxetable*}{lccccccc}
\tablecaption{PHANGS-JWST Target Galaxy Physical Properties\label{tab:galaxysample2}}
\centering
\tablehead{
\colhead{Galaxy} & 
\colhead{log M$_{\star}$\tablenotemark{a}} & 
\colhead{SFR$_\mathrm{tot}$\tablenotemark{b}} & 
\colhead{$R_e$\tablenotemark{c}} & 
\colhead{$\log_{10} L_{\rm CO}$\tablenotemark{d}} &
\colhead{$M_{\rm HI}$\tablenotemark{e}} &
\colhead{$12+\log_{10} {\rm O/H}$\tablenotemark{f}} &
\colhead{$f_{\rm JWST}$\tablenotemark{g}} \\
& 
($M_\odot$) & 
($M_\odot$~yr$^{-1}$) & 
(kpc) & 
(K~km~s$^{-1}$~pc$^2$) &
($M_\odot$) &
&
}
\startdata
NGC~0628\tablenotemark{*}& 10.3 & 1.7 & 3.9 & 8.4 & 9.7  & 8.53 & 0.53 \\
NGC~1087                 & 9.9  & 1.3 & 3.2 & 8.3 & 9.1  & 8.48 & 0.88 \\
NGC~1300                 & 10.6 & 1.2 & 6.5 & 8.5 & 9.4  & 8.62 & 0.87 \\ 
NGC~1365\tablenotemark{*}& 11.0 & 17  & 2.8 & 9.5 & 9.9  & 8.67 & 0.76 \\ 
NGC~1385                 & 10.0 & 2.1 & 3.4 & 8.4 & 9.2  & 8.46 & 0.94 \\ 
NGC~1433                 & 10.9 & 1.1 & 4.3 & 8.5 & 9.4  & 8.57 & 0.50 \\ 
NGC~1512                 & 10.7 & 1.3 & 4.8 & 8.3 & 9.9  & 8.58 & 0.46 \\
NGC~1566                 & 10.8 & 4.6 & 3.2 & 8.9 & 9.8  & 8.61 & 0.70 \\ 
NGC~1672                 & 10.7 & 7.6 & 3.4 & 9.1 & 10.2 & 8.57 & 0.87 \\ 
NGC~2835                 & 10.0 & 1.3 & 3.3 & 7.7 & 9.5  & 8.56 & 0.46 \\ 
NGC~3351                 & 10.4 & 1.3 & 3.0 & 8.1 & 8.9  & 8.58 & 0.66 \\ 
NGC~3627                 & 10.8 & 3.9 & 3.6 & 9.0 & 9.8  & 8.54 & 0.86 \\ 
NGC~4254                 & 10.4 & 3.1 & 2.4 & 8.9 & 9.5  & 8.59 & 0.90 \\ 
NGC~4303                 & 10.5 & 5.4 & 3.4 & 9.0 & 9.7  & 8.58 & 0.80 \\ 
NGC~4321                 & 10.8 & 3.5 & 5.5 & 9.0 & 9.4  & 8.56 & 0.68 \\ 
NGC~4535                 & 10.5 & 2.2 & 6.3 & 8.6 & 9.6  & 8.54 & 0.57 \\ 
NGC~5068                 & 9.4  & 0.3 & 2.0 & 7.3 & 8.8  & 8.32 & 0.71 \\ 
NGC~7496\tablenotemark{*}& 10.0 & 2.2 & 3.8 & 8.3 & 9.1  & 8.51 & 0.88 \\
IC~5332\tablenotemark{*} & 9.7  & 0.4 & 3.6 & 7.1 & 9.3  & 8.30 & 0.27 \\ 
\enddata
\tablenotetext{a}{Galaxy stellar mass.  Following \citet{phangs-alma}, based on Spitzer IRAC 3.6 $\mu$m when available, or WISE 3.4 $\mu$m, and mass-to-light ratio prescription of \citet{leroy19} calculated as a function of radius in the galaxy.}
\tablenotetext{d}{SFR$_\mathrm{tot}$ is the total galaxy star formation rate. Based on GALEX FUV and WISE W4 imaging with SFR prescription calibrated to match results from population synthesis  modeling  of \citet{salim16,salim18} as in \citet{phangs-alma}.}
\tablenotetext{c}{Stellar mass effective radius, from \citet{phangs-alma} and closely resembling near-IR effective radius \citep{munozmateos15}.}
\tablenotetext{d}{Integrated CO (2-1) luminosity. Scale by $\alpha_{\rm CO} \approx 6.7$~M$_\odot$~pc$^{-2}$ (K~km~s$^{-1}$)$^{-1}$ to estimate $M_{\rm mol}$ including helium and metals using a fixed CO-to-H$_2$ conversion factor.}
\tablenotetext{e}{Atomic gas mass, not including helium, from HyperLEDA \citep{hyperleda}.}
\tablenotetext{f}{Gas phase metallicity on the $S$-cal system \citep{scal16} estimated at $R_e$ by \citet{GROVES_HIICAT}.}
\tablenotetext{g}{Fraction of the SFR estimated from the UV+IR that lies within the PHANGS--HST field of view from \citet{phangs-hst}. This can be applied with reasonable precision to aperture correct the SFR, $L_{\rm CO}$, or $M_\star$ to estimate the quantity inside the JWST field of view.}
\tablenotetext{*}{One of first four galaxies observed which form the basis of the PHANGS-JWST First Results presented in this volume.}
\end{deluxetable*}


The PHANGS surveys focus on nearby, star-forming, relatively face-on spiral galaxies. The parent sample (N=90), described in \citet{phangs-alma}, aimed to select galaxies visible from the southern hemisphere (i.e., observable by ALMA and the VLT) that have inclination less that $\sim 75^\circ$, distances less than $\sim 17$~Mpc, stellar masses $\gtrsim 5 \times 10^9$~M$_\odot$, and specific star formation rates, SFR/$M_\star$, above $\sim 10^{-11}$~yr$^{-1}$. The inclination criterion minimizes source blending within the galaxies and line-of-sight dust attenuation. The distance requirement ensures that the galaxies are close enough that HST and JWST can resolve individual clusters and associations, MUSE can identify and characterize individual \textsc{Hii} regions, and ALMA resolves the molecular ISM into individual massive molecular clouds and complexes. Meanwhile the mass and specific star formation rate criteria select galaxies on the ``main sequence'' of star-forming galaxies, ensuring that the surveys capture environments that reflect where most stars form at $z=0$ \citep[e.g.,][]{brinchmann04,salim07,Saintonge17}.

As typical for surveys of nearby galaxies, our best estimates of the properties of galaxies in the sample have evolved since they were first selected, in particular due to the continuing improvement in distance estimates \citep{anand21, phangs-hst}. As a result, the properties of parent sample have broadened slightly beyond the original selection criteria, with the result that the PHANGS-JWST sample
includes galaxies with distances out to $\sim 20$~Mpc and stellar masses as low as $\sim 2.5\times 10^9$~M$_\odot$.

In practice, the requirement that PHANGS-JWST galaxies have the full complement of PHANGS ALMA, MUSE, and HST data means that the PHANGS-JWST selection will be identical to that of the precursor survey with the smallest sample, which is PHANGS-MUSE \citep{phangs-muse}.  PHANGS-MUSE targeted the first 19 galaxies from the PHANGS parent sample for which ALMA CO(2--1) mapping was obtained.  The resulting sample satisfies the general goal of providing a reasonably representative selection of massive star-forming galaxies --  spanning a factor of $\sim50$ in stellar mass and SFR \citep[Figure 3 in][]{phangs-muse}; morphological types from Sa-Sd with a diversity of bar and ring structures; and a factor of $\sim100$ in CO surface densities \citep[Figure 1 in][]{phangs-hst}.

\section{JWST Observations}
\label{sec:observations}

\begin{figure}
\centering
\includegraphics[width=3.2in]{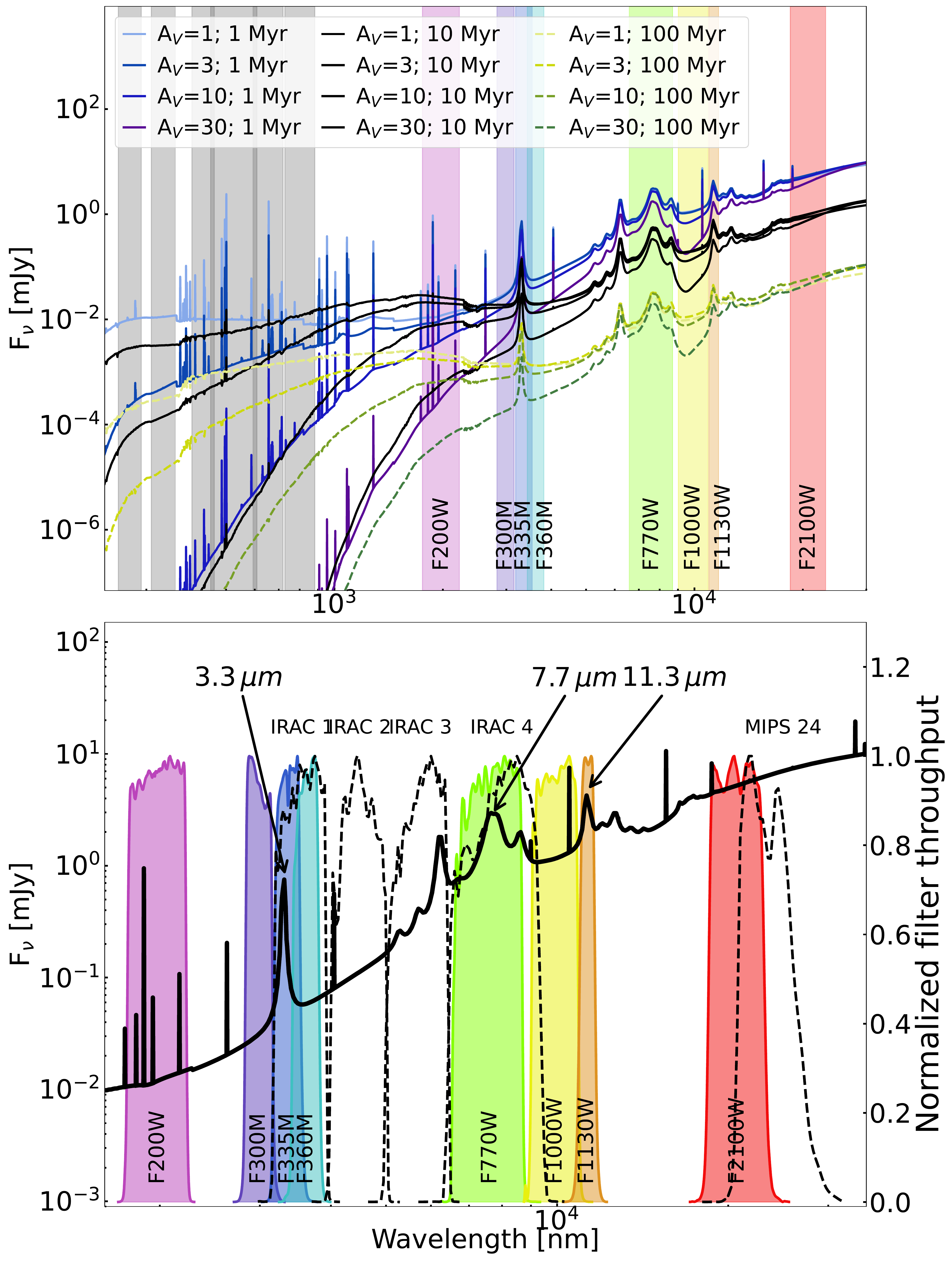}

\caption{\textbf{top}: Model SEDs of dust-enshrouded young star clusters with filter coverage of the PHANGS-HST and JWST Treasury programs. Models are of a solar metallicity $10^4$~M$_\odot$ young star cluster at $d=10$~Mpc with varying ages and levels of extinction generated with CIGALE \citep[][]{boquien19,turner21}. Colored bands show the 8 JWST NIRCAM and MIRI filters, spanning from 2$-21\mu$m, selected to probe stellar photospheric emission, PAH features, and the dust continuum. The 5 UV-visible filters used by PHANGS-HST \citep{phangs-hst} appear in grey. \textbf{bottom}: Normalized throughput for each of the eight filters used in the PHANGS-JWST survey.  For comparison, the four Spitzer IRAC filters and MIPS 24 $\micron$ filter are also shown.  A model dust SED is overplotted to illustrated the features probed by each filter.  The three PAH features targeted by our survey are marked.}
\label{fig:filters}
\end{figure}

PHANGS-JWST observations for our sample of 19 galaxies are obtained under JWST program 2107 (PI: J.~Lee) with a total allocation of \obstime~hrs.\footnote{Accounting for changes in overhead calculations as of 19 November 2022. Approximately 70\% of this allocation is spent on observational overheads (e.g., slew times)}  As mentioned in the Introduction, data for four galaxies (NGC~7496, IC~5332, NGC~628, NGC~1365) were obtained within the first two months of science operations (Table \ref{tab:obslog}), and provide the basis for the first PHANGS-JWST results presented in this Issue.\footnote{Due to a guide star acquisition failure, the NIRCam observations of IC\,5332 were not completed and have been rescheduled for later in JWST Cycle 1. Since our observation strategy uses the time when the NIRCam observations are targeting the galaxy to observe the MIRI off-source position ``in parallel,'' our IC\,5332 dataset also currently lacks an off-source observation for MIRI.}  Six more targets are currently scheduled for observations during 2022, and program completion is anticipated in mid-2023.

\begin{table}[h!]
    \begin{center}
    \caption{PHANGS-JWST First Results Observations\label{tab:obslog}}
    \begin{tabular}{lll}
        \hline 
        \hline
        Galaxy &  MIRI Obs. Date. & NIRCam Obs. Date \\
         & (UT: 2022) & (UT: 2022) \\
        \hline
        NGC\,7496 & 07/06 18:02 - 19:45 & 07/06 19:50 - 20:51 \\
        IC\,5332  & 07/06 21:05 - 22:48 & \nodata \\
        NGC\,0628 & 07/17 12:14 - 14:47 & 07/17 15:01 - 17:03 \\
        NGC\,1365 & 08/13 15:00 - 16:43 & 08/13 18:33 - 20:35 \\
        \hline
    \end{tabular}
    \end{center}
\end{table}

Imaging is conducted with NIRCam and MIRI in eight filters from 2 to 21 $\mu$m (Figure \ref{fig:filters}).  This set of filters is chosen to support our broad science goals (see Section \ref{sec:goals}) as well as to optimize observing efficiency (more below).

PHANGS-JWST coverage is designed to maximize overlap with PHANGS data already obtained by HST, VLT-MUSE, and ALMA, all of which targeted the main star-forming area of the galaxy disk.  This requires 1-2 pointings with NIRCam (Module B only) and 2-4 pointings with MIRI for each galaxy. 
For both NIRCam and MIRI, we use 10\% overlap between rows and columns of mosaic tiles. With this set-up, 13 galaxies required 2 pointings with NIRCam (Module B only) while 9 were observed with a single pointing.  With MIRI, up to four pointings were required, with 1, 6, 3, 9 galaxies requiring 1, 2, 3, 4 pointings respectively.  The resultant PHANGS-JWST NIRCam and MIRI footprints are shown in Figure \ref{fig:footprints}.

\begin{table*}[h]
    \begin{center}
    \caption{Total Exposure Times, Resolution, and Sensitivities for PHANGS-JWST NIRCAM and MIRI Imaging. \label{tab:obsparam1}}
    \begin{tabular}{l c c c c c c c}
    \hline
    \hline
    \multicolumn{7}{c}{NIRCam} \\
    \hline
    Filter & FWHM  &FWHM & $N_\mathrm{exp}$ & Exp. Time & $\Omega_\mathrm{PSF}$\tablenotemark{a} & $\sigma_\mathrm{I}$\tablenotemark{b} & $5\sigma_{f}$\tablenotemark{c}  \\
          & (arcsec) & (pixel) &  & (s) & ($10^{-12}$~sr) & (MJy~sr$^{-1}$) & ($\mu$Jy)\\
    \hline
    F200W & 0.066 & 2.141 & 16 & 1202.5 & 0.167  & 0.068 &  0.079\\
    F300M & 0.100 & 1.585 & 4 & 386.5   & 0.365  & 0.045 &  0.077\\
    F335M & 0.111 & 1.760 & 4 & 386.5   & 0.458  & 0.042 &  0.10\phn\\
    F360M & 0.120 & 1.901 & 8 & 429.5   & 0.530  & 0.058 &  0.17\phn\\
    \hline
    \multicolumn{7}{c}{MIRI}\\
    \hline
    F770W  & 0.25 &  2.27 & 4 & 88.8  & 2.18\phn & 0.11 &  0.87\\
    F1000W & 0.32 &  2.91 & 4 & 122.1 & 3.77\phn & 0.12 &  1.9\phn\\
    F1130W & 0.36 &  3.27 & 4 & 310.8 & 4.95\phn & 0.15 &  2.4\phn\\
    F2100W & 0.67 &  6.09 & 8 & 321.9 & 16.1\phn & 0.25 &  8.0\phn\\
    \hline    
    \hline
    \end{tabular}
    \end{center}
    \tablenotetext{a}{The solid angle subtended by the PSF, calculated using the \textsc{WebbPSF} package \citep{webbpsf}.}
    \tablenotetext{b}{The empirical $1\sigma$ surface brightness sensitivity $\sigma_{I}$ estimated directly from the image data (Section \ref{sec:empiricalnoise}).  This estimate agrees well with the pipeline-provided noise products.}
    \tablenotetext{c}{The $5\sigma$ point source sensitivity ($5\sigma_{f}$) based on empty apertures in low-brightness regions of the image.  See Section \ref{sec:empiricalnoise}.}
\end{table*}

\begin{table*}
    \begin{center}
    \caption{NIRCam Primary Observation \& MIRI Background Parallel Visit Plan\label{tab:obsparam2}}
    \begin{tabular}{ccccccccc}
    \hline
    \hline
    \multicolumn{9}{c}{Intramodulebox pattern with 4 primary dithers, BRIGHT1 readout pattern} \\
    \hline
    NIRCam & NIRCam & Grp/Int & Int/Exp & Exp.\ Time & MIRI Bkgd Filter & Grp/Int & Int/Exp & Exp.\ Time\\
    Short Filter & Long Filter & & & (s) & (in parallel) & & & (s)\\
    \hline
    F200W & F360M & 3 & 1 & 214.7 &  F770W &  8 & 1 & 88.8  \\
    F200W & F360M & 3 & 1 & 214.7 & F1000W & 11 & 1 & 122.1 \\
    F200W & F300M & 5 & 1 & 386.5 & F1130W & 28 & 1 & 310.8 \\
    F200W & F335M & 5 & 1 & 386.5 & F2100W & 14 & 1 & 321.9 \\
    \hline
    \end{tabular}
    \end{center}
\end{table*}

\begin{table*}
\label{tab:obsparam3}
\begin{center}
\caption{MIRI Primary Observation Visit Plan\label{tab:obsparam3}}
\begin{tabular}{cccc}
    \hline 
    \hline
    \multicolumn{4}{c}{4-point extended source dither pattern} \\
    \multicolumn{4}{c}{FASTR1 readout pattern}\\
    \hline
    MIRI Filter & Groups per & Integs.\ per & Exp. Time\\
     & Integration & Exposure & (s)\\
    \hline
    F770W  &  8 &  1 & 88.8 \\
    F1000W & 11 &  1 & 122.1 \\
    F1130W & 28 &  1 & 310.8 \\
    F2100W & 14 &  2 & 321.9 \\
    \hline    
    \hline
\end{tabular}
\end{center}
\end{table*}


Total exposure times per pointing in each filter are listed in Table~\ref{tab:obsparam1} and range from 6.4-20 min with NIRCam and 1.5-5.4 min with MIRI. Tables~\ref{tab:obsparam2} and \ref{tab:obsparam3} summarize the observing parameters and exposure sequences adopted for each filter in each visit.

\subsubsection{NIRCam Primary Observations}

Our NIRCam primary observations are divided into four separate sequences to maximize observing efficiency (Table~\ref{tab:obsparam2}).  This allows for MIRI sky background observations to be taken in parallel with each NIRCam sequence, with one sequence for each of the four MIRI filters in our program.  With NIRCam, we use Module B (FULL subarray) and an INTRAMODULEBOX-4 primary dither pattern with no additional subpixel dither.\footnote{The primary dither pattern itself has very minimally-sampled subpixel dithering.}  We  obtain four F200W  exposures, one at each INTRAMODULEBOX-4 dither position (for a total of 16 exposures), using the BRIGHT1 readout pattern with one integration per exposure.   
Of these independent sequences, two have 5 groups/integration, contributing 386.5s each to the total F200W exposure time, and the other two have 3 groups/integration, yielding 214.7s each. The total F200W exposure time is thus 1202.5s per pointing. 

Simultaneously with F200W primary imaging in the short wavelength channel (and coordinated parallel MIRI imaging of sky background), we observe the target with F300M, F335M, and F360M in the long wavelength channel. The F300M and F335M observations are paired with the two longer F200W exposure sequences.  The exposure time for F300M and F335M is consequently 386.5s (each). During both of the two shorter F200W exposure sequences, F360M imaging is obtained, for a total exposure time of 429.5s.

\subsubsection{MIRI Primary Observations}

For our MIRI primary observations, we use a 4-point dither pattern optimized for extended sources. We observe with the F770W, F1000W, F1130W and F2100W filters and the FASTR1 readout pattern yielding total exposure times per pointing of 88.8s, 122.1s, 310.8s, 321.9s respectively (Table~\ref{tab:obsparam3}). 
All of the galaxies are larger than the MIRI field of view, so we obtain a background measurement using this same MIRI imaging sequence for a blank region of sky near the galaxy in parallel with NIRCam, as just discussed.

\subsubsection{Special Requirements}

We include the following special requirements in our observation plan: 1) a timing requirement (Sequence Observations, Non-interruptible) that ensures the MIRI background imaging 
is obtained directly following the MIRI observation of the target; 2) position angle requirements to maximize the overlap of the primary MIRI observations with existing HST, MUSE, ALMA coverage, while placing the parallel MIRI background observation on blank sky.  In some of the galaxies with largest angular extents in our sample, the parallel sky background measurement may not be completely beyond the outskirts of the galaxy disk. Finally, whenever possible, we attempted to widen the orient constraints to provide a minimum observation window of 2 weeks to provide greater opportunities for scheduling.  
It should be noted that the range in allowable orient angles can result in imaging that is not optimally aligned with previous data or astrophysically interesting features in the galaxy.  For example, in NGC~1365, the observed orientation led to the omission of portions of the eastern spiral arm.

\section{Initial Image Processing}
\label{sec:redux}
In this section, we describe how the STScI JWST Science Calibration Pipeline\footnote{\url{https://github.com/spacetelescope/jwst}} 
is used with  additional processing to enable a first set of analyses by the PHANGS team. 
We completed processing of the images for the papers presented in this Issue by early September 2022. The latest versions of the pipeline and calibration files available at the time of processing were used: version 1.7.0 (for MIRI) and 1.7.1 (for NIRCam) of the JWST pipeline and Calibration Reference Data System\footnote{\url{https://jwst-crds.stsci.edu/}} (CRDS context number 0968 for both NIRCam and MIRI, though for MIRI we used slightly updated flats provided by K.~Gordon).

\subsection{Level One Processing}
\label{sec:level1}
We start with the uncalibrated raw (``uncal'') files obtained from MAST, 
and use Detector1Pipeline, which converts from ``ramps'' to ``slopes'' and applies basic detector-level corrections to the data, including the flagging of saturated pixels and corrections for chip persistence\footnote{\url{https://jwst-pipeline.readthedocs.io/en/latest/jwst/pipeline/calwebb_detector1.html}}.   We ran the Detector1Pipeline with the default parameters, with one exception. To attempt to recover some of the saturated pixels, we remove the restriction that during ramp fitting at least two groups must not reach saturation. These saturated pixels are primarily an issue in the centers of NGC~1365 and NGC~7496, which both host bright AGN. Removing the restriction did not recover the very central pixels coincident with the AGN themselves, but it did allow us to recover a number of the brighter pixels surrounding the central point source.  However, some of these sources, especially the brightest compact, massive star-forming regions in the starburst ring of NGC~1365, still show artifacts that have structure suggesting that the source brightness remains underestimated by the imaging.

\subsection{Level Two Processing}
Stage two of the pipeline produces calibrated individual exposures\footnote{\url{https://jwst-pipeline.readthedocs.io/en/latest/jwst/pipeline/calwebb_image2.html}}. During this stage, the pixel coordinates are translated into WCS coordinates, which includes application of distortion corrections. This stage also involves flat-fielding and background subtraction (the background subtraction is used only for the MIRI data). Because of our parallel observing visit sequence (\S \ref{sec:observations}), the pipeline does not automatically associate the appropriate MIRI off (sky background) observations with the on-galaxy images. Instead, we manually implement this association during this processing step. For IC~5332, only the on-galaxy MIRI observations were obtained during our initial visit. Therefore, we used the off-galaxy observation from NGC~7496 for this purpose in IC~5332. This appears to work reasonably well, but does result in a few visible imperfections in the background subtraction for IC~5332. Also for MIRI, we masked the part of the image associated with the Lyot coronograph. In this early reduction, the coronagraph portions of the image consistently showed imperfect background subtraction that caused issues with our mosaicking.

\subsection{Principal Component Analysis Destriping}
After stage two, we applied additional processing to suppress $1/f$ noise in the NIRCam data. This correlated read noise primarily manifests as striping across the NIRCam images, along detector rows (in the fast-read direction).
This can severely degrade the quality of the images, particularly at shorter wavelengths. \citet{2020Schlawin} studied this effect using simulated data and provided a number of suggestions for how to deal with this $1/f$ noise. We experimented with simple row-by-row median subtraction, but found that it was often unfeasible to calculate a robust background median in the small range covered by each amplifier due to the presence of widespread diffuse emission in most of our images. Moreover, we found that though this median subtraction produced visually better individual images, it yielded poor results when multiple images were combined to estimate 3.3$\mu$m PAH maps \citep{SANDSTROM1_PHANGSJWST}. In those cases, the $1/f$ noise-driven stripes reappeared and were severe in the star-subtracted PAH images. 

\begin{figure*}[ht]
\plotone{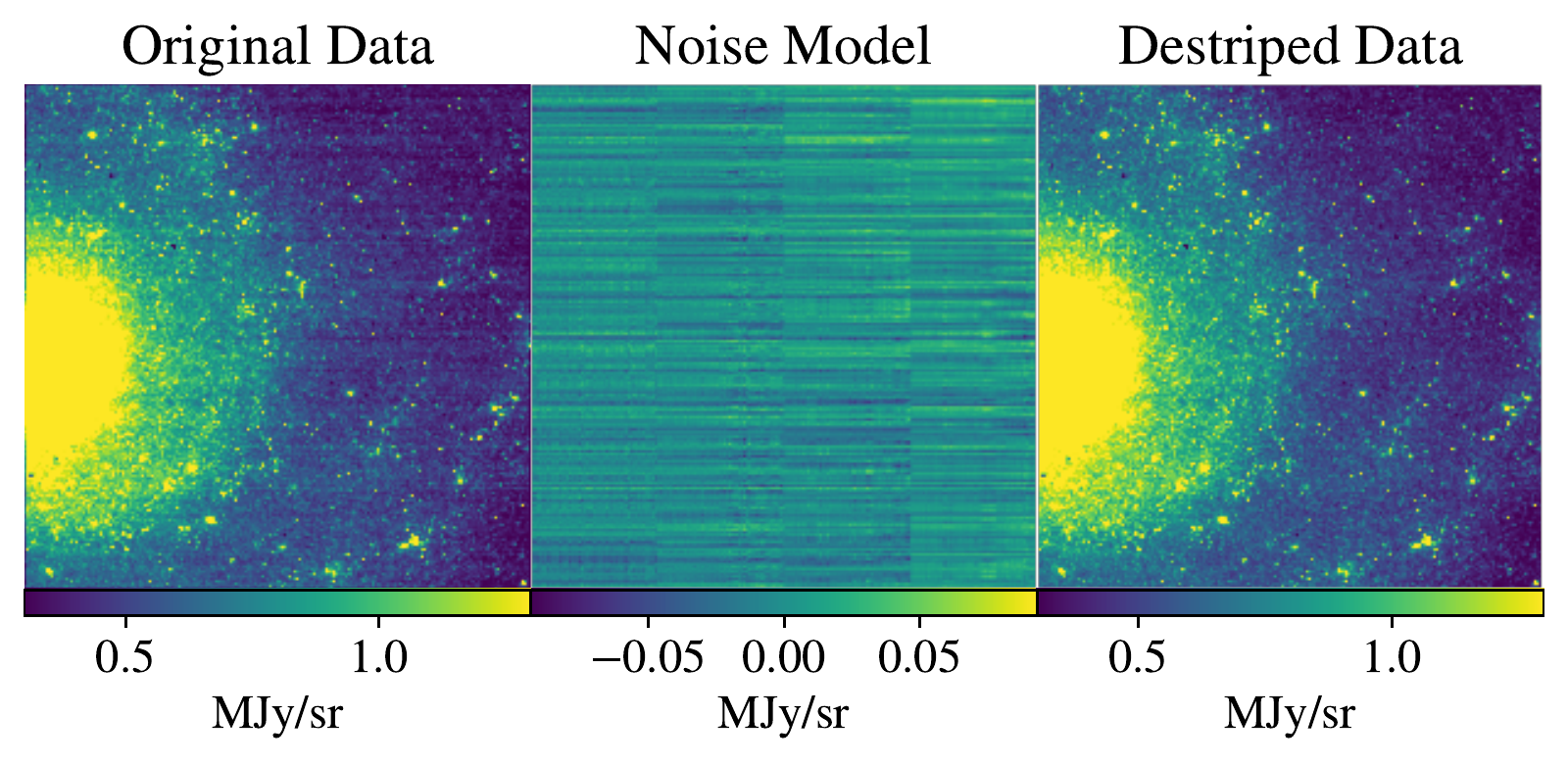}
\caption{{\it Left:} Original calibrated image, presented with a strong intensity transfer function to highlight stripes due to $1/f$ noise. {\it Middle}: Principal component analysis noise model, with the displayed intensity range now centered on zero and the intensity scale boosted by a factor 5 compared to the original image.  {\it Right:} Final destriped data.}
\label{fig:redux_destripe}
\end{figure*}

Instead, as shown in Fig.~\ref{fig:redux_destripe}, we had success modeling the $1/f$ noise using robust principal component analysis (PCA) based on the technique developed in \citet{2005Wild} and \citet{2009Budavari} for SDSS spectra. We treat each column of the data as an individual noise spectrum and fit results for each amplifier separately, because the noise properties differ between amplifiers. We mask out bright sources and then apply a \citet{Butterworth1930} filter to remove large-scale structure but retain the small-scale noise properties. We also shuffle in the $x$ direction and apply a different offset to each column, effectively rolling the column along the $y$-axis by a random amount. These shifts prevent the PCA simply learning where the mask is and responding to the location of bright emission. After this processing, we use PCA to fit 50 components. Then we reconstruct a model describing the $1/f$ noise-induced stripes for that column using the 5 components with the largest eigenvalues. Using only five components limits processing time and we found that increasing the number of components beyond 5 did not significantly improve the noise model. Finally we subtracted the model of the $1/f$ noise-induced stripes from the data.

Figure~\ref{fig:redux_destripe} illustrates this process. The left image shows visible artifacts, mostly horizontal striping. The PCA model successfully captures these features without including the extended bright emission from the source. Occasionally corrupted columns and other features arise, which the PCA also robustly handles. The right image shows visibly less striping and fewer artifacts than the original image. We find that the PCA reduces the RMS noise in the images between around 10\% to 30\%, with the amount of reduction depending mainly on the amount of area filled by emission in the image. For images with more source-free sky background, the PCA produces a more significant decrease in the noise level.

\subsection{Level Three Processing}

The level three stage of the pipeline\footnote{\url{https://jwst-pipeline.readthedocs.io/en/latest/jwst/pipeline/calwebb_image3.html}} is intended to combine individual tiles to produce a final rectified mosaic. This is achieved through relative alignment between tiles, background matching, and the final drizzling to the output pixel grid. 

In our early science reduction, this step involved the most changes from the default pipeline parameters. 
For MIRI, we ran the level three pipeline on individual tiles, i.e., the individual pointings, which we later mosaicked together by hand. For NIRCam, we used the JWST pipeline to stitch the tiles.

We adjusted the stage three pipeline as follows. First, in the {\tt tweakreg} step, which provides relative alignment between tiles, we found it necessary to limit the `roundness' of sources used for alignment to $-0.5 < {\rm roundness} < 0.5$, to reject diffraction spikes from stars and the numerous extended sources in our field. We also turned off the {\tt 2dhist} option, as our relative corrections were generally $\lesssim 1\arcsec$, and leaving this on degraded the quality of the alignment. We fit the per-tile offsets using only ($x, y$) shifts, as we found allowing for rotation produced negligible rotation angles and degraded the quality of the fits.  For data acquired in July and August 2022, we found relative corrections as large as $\sim$1\arcsec. This primarily reflects the quality of the guide star catalog during that period. The size of the correction will likely decrease as JWST observations continue and the guide star catalog is updated and improved.

For the NIRCam data, we calculate and subtract the background level for each detector individually.
By default, the pipeline will subtract a single background for the four short-wavelength NIRCam detectors, however, we found that this produced clear `steps' in our mosaics.

In the final redrizzling step during stage three, we produced mosaics with the conventional north-up orientation to allow for more straightforward comparison with data at other wavelengths.

\subsection{Absolute Astrometric Alignment}
\label{sec:astrometry}
The level three pipeline produced mosaics that have good relative alignment between NIRCam tiles, but the alignment of the JWST images relative to other data with well-established absolute astrometry tended to be poorer, as high as $\sim 1''$ in some cases. After stage three processing, we therefore performed an additional correction to improve the absolute astrometry of the NIRCam and MIRI images.  We adopt the general strategy of bootstraping the astrometric solution from images that are close in wavelength instead of directly using astrometric catalogs, as was followed for PHANGS--HST \citep{phangs-hst}.

Our JWST images lack enough \textit{Gaia} \citep{gaia16} sources to anchor our astrometry directly to GAIA \citep{gaiadr2,gaia2astrometry}. However our NIRCam images detect many AGB stars that are also visible in the PHANGS--HST images, which have astrometric solutions accurate to better than 10 mas.
Thus, we extracted a set of AGB stars from the PHANGS--HST DOLPHOT catalog \citep{thilker22, phangs-hst}, and cross-matched them to sources detected in the NIRCam bands in the level 3 \texttt{source\_catalog} step (the final level 3 step) 
using the \texttt{XYXYMatch} function in \texttt{tweakwcs}. We 
solve for a linear offset and rotation to match the images (though typically the rotation was very small). Typically, we identified a few hundred AGB stars per target from the PHANGS--HST catalogs and the pipeline found 50-100 good matches in the NIRCam imaging.  
After applying the solutions 
comparisons between the NIRCam and HST data suggest that the rms scatter in the astrometric accuracy across the NIRCam images is about $\pm 0.1$ NIRCam pixel.

The MIRI images have fewer point sources than those from NIRCam and HST, so we derive the astrometric solution for MIRI using a cross-correlation approach. To do this, we take advantage of the fact that the F335M NIRCam band, which has been anchored using AGB stars to the PHANGS--HST and \textit{Gaia} frame, contains both stellar and ISM-like emission because of the strong PAH band at $3.3\mu$m \citep[see][]{SANDSTROM1_PHANGSJWST}. The stellar emission allows us to pin the astrometry to the AGB stars and the PHANGS--HST frame, while the morphology of the ISM at $3.3\mu$m resembles the ISM emission in the longer wavelength MIRI bands. This allows us to use a cross-correlation analysis to solve for the relative astrometry of NIRCam and MIRI bands. Specifically, we used the \texttt{image-registration} package to run a cross-correlation and solve for the linear shift (by identifying the maximum of the cross-correlation) needed to match the shortest MIRI band, F770W, to the F335M astrometry. Then, with the F770W astrometry established, we solved for the shift needed to align F1000W to F770W, F1130W to F1000W, and F2100W to F1130W. We adopted this stepwise approach to reflect the increasing dominance of the ISM emission over stellar emission as the wavelength increases. The cross-correlation technique generally produces `good' results but can be sensitive to artifacts, e.g., diffraction spikes around the AGN. By eye, this seems to yield images that are aligned at the level of $\pm 1$ MIRI pixel, i.e., about $0.1''$. While this is sufficient for the first results presented in this Issue (except where noted in specific papers), the alignment is not as good as would be expected based on alignment based on a more limited set of carefully selected, bright, isolated point sources \citep[e.g.,][]{phangs-hst}. We expect to significantly improve this in future work.

\subsection{By-Hand MIRI Mosaicking for Early Science}

After level three processing and alignment, we combined the individual MIRI tiles into a single image for each galaxy at each band. To do this, we defined one of the tiles for each galaxy as the reference tile. Then we reprojected the other tiles onto the astrometric grid for the reference tile and compared the intensity of the two tiles in the region where they overlap. We solved for and applied the additive offset needed to yield a one-to-one match in intensities between the other tile and the reference file. These offsets were typically $\lesssim 0.05$~MJy~sr$^{-1}$ for F770W and F1000W, $\lesssim 0.15$ MJy~sr$^{-1}$ for F1130W, and as high as $\sim 0.3{-}0.8$~MJy~sr$^{-1}$ for F2100W. This yielded a set of individual tiles with the same background level, but did not anchor that background level to the correct absolute value.

Next we defined a new astrometric grid with a larger size but the same pixel scale and orientation as that of the reference field. We then reprojected all tiles onto this new grid, averaging intensities from different tiles with equal weighting where tiles overlapped. The result at this stage was a single combined MIRI image registered via cross-correlation to the NIRCam and other MIRI images, but with an absolute background level that was still uncertain.

Note that we expect to deprecate most of these manual steps as the pipeline processing improves in the future. However, for this first processing during these early months we found that this simple, by-hand approach yielded smoother backgrounds and a better match to previous infrared imaging of our targets than the pipeline.

\subsection{Refinements to the Pipeline Background Subtraction}
\label{sec:background}
As a final step, we adjust the overall background level of the MIRI images to match previous wide-field mid-infrared imaging of our targets. The JWST field-of-view for most of our targets contain relatively little empty sky and the match between the off-source and the on-source image was not sufficient to yield a precise background level. Fortunately, all of our sources have previous observations at least by WISE at $12\mu$m and $22\mu$m \citep{wise} and for NGC~628, NGC~1365, and IC~5332 by \textit{Spitzer} at $8\mu$m and $24\mu$m \citep{kennicutt03,dale09,armus09}. Though these data have much worse resolution than the JWST images they cover a much larger area and extend to empty sky and so have well-established background levels.

We perform this background homogenization in two stages following a procedure detailed in the Appendices of \citet{LEROY2_PHANGSJWST}. First we establish a common background system across all JWST bands. Then we anchor the overall system to an external band with a well-established background. To do this, we make a version of the MIRI images that all share the PSF of the F2100W image and a version of all JWST, \textit{Spitzer}, and WISE data that share a common $15''$ FWHM Gaussian PSF. 
That is, we match the backgrounds among the JWST bands at common resolution. 
These procedures leverage the fact that mid-infrared emission at different wavelengths shows strong, often nearly linear correlations even though the exact band ratios do vary with filter combination \citep[see Appendix in][]{LEROY2_PHANGSJWST}.
This procedure appears to yield good results with an uncertainty in the overall background level of better than $\pm 0.1$~MJy~sr$^{-1}$ across all MIRI bands.

\subsection{Limitations, Caveats, and Expected Improvements}
\label{sec:caveats}

The reduction procedures described above yield data that are useful for the initial work presented in this Issue. Over the longer term, improvements will be made to address known limitations that we summarize here (as well as possible additional issues found during future analysis). In addition to refining our own procedures, we expect to take full advantage of the monthly refinements to the pipeline being released by STScI\footnote{\url{https://jwst-docs.stsci.edu/jwst-science-calibration-pipeline-overview/jwst-operational-pipeline-build-information}}. The pipeline documentation already identifies some of these issues as areas of future development.

Where these limitations affect our First Results scientific papers they are noted, but most of our analyses are conducted to mitigate their impact.  For example, aperture photometry with local background subtraction avoids uncertainties in measurement of the background \citep[e.g.,][]{RODRIGUEZ_PHANGSJWST, WHITMORE_PHANGSJWST}.

\subsubsection{Astrometric Uncertainties}
The relative alignment of the NIRCam images is excellent, with sufficient numbers of common sources available in adjacent bands to align frames and to tie to the astrometry established in the PHANGS-HST imaging\footnote{\url{https://archive.stsci.edu/hlsp/phangs-hst}}.  Using AGB stars, we can 
achieve $<10$~mas astrometric accuracy for the NIRCam data (Section~\ref{sec:astrometry}).  The MIRI images are currently aligned by cross correlation and have larger uncertainty in their astrometric solutions ($\pm 0\farcs1$).

We expect that a combination of better constraints from cross-correlation and the use of AGB stars will increase the astrometric accuracy of the MIRI images. Astrometric solutions are also expected to continue to improve as distortion models are updated by STScI.

Our science images are not currently drizzled onto a matched pixel grid, but adopting and drizzling to the grid adopted for the PHANGS-HST images would significantly facilitate joint HST-JWST analysis. However, the different pixel scales spanned from 2-21$\mu$m make this exercise a trade-off between retaining information and limiting data volume. 

\subsubsection{Flux calibration \& backgrounds}

The zero points (i.e., absolute flux calibration) for the JWST instruments are being refined. 
These changes will primarily primarily affect the NIRCam data \citep[][]{2022Boyer}, and shortly after the completion of the analyses presented in this Volume, these zero-points were updated by STScI. 

The background levels of the MIRI data have been inferred from band-to-band correlations and comparison with existing lower resolution infrared imaging from previous missions.  These are currently uncertain at the $\pm 0.1$~MJy~sr$^{-1}$ level. Comparison of the MIRI data with previous data indicates that, after the subtraction of a background, the intensity measurements agree to better than 10\%.  

We expect that the mosaicking and tiling in the pipeline will improve and replace our by-hand approach to the MIRI data. In addition to yielding a better astrometric solution, this will allow rigorous testing of the pipeline noise estimates and move us towards an accurate noise model for MIRI.  As improved flats and better sky subtraction techniques become available they should also improve the local behavior of the MIRI backgrounds.

\subsubsection{Saturation and Poor Ramp Fitting}

Some of the brightest sources in the MIRI imaging, particularly in NGC\,1365, are saturated, show artifacts resulting from poor ramp fitting, and the surrounding areas are affected by diffraction spikes.  The photometry of sources with poor ramp fits is likely underestimated (Section \ref{sec:level1}).  

\citet{HASSANI_PHANGSJWST} produced bespoke masks to flag diffraction spikes due to the AGN in NGC~1365 and NGC~7496, but as better PSF estimates become available, subtracting such features from the image should become an option. \citet{LIU_PHANGSJWST} present the first efforts at recovering the fluxes of sources with saturated pixels and poor ramp fits by using PSF matching in the outskirts of the source. 

\subsubsection{Angular resolution}

Finally, while not strictly an image processing issue, we caution that the large span of wavelengths of our JWST images implies a similar variation in their angular resolution. Analysis that involves comparisons across multiple JWST bands, or between the JWST and other data, requires care to ensure that relative flux measurements are robust.  In general, the analyses in this Volume rely on flux density measurements from aperture photometry or on surface brightness comparisons, generally at matched resolution.  Where appropriate, we have generated convolution kernels following the method of \citet{aniano2011} to degrade the resolution of short wavelength images to the long wavelength bands, achieving a common resolution match. In the future, PSF-fitting packages like \textsc{dolphot} \citep{dolphot} will be used, and methods to employ joint SED/image deconvolution based on priors from the higher-resolution short-wavelength imaging will be implemented.  However, the results presented in this Volume rely on simpler methods, and should be interpreted accordingly.

\section{Overview of First PHANGS-JWST Images}
\label{sec:results}

\subsection{Comparison to Spitzer}

\begin{figure*}
\centering
\includegraphics[height=8in]{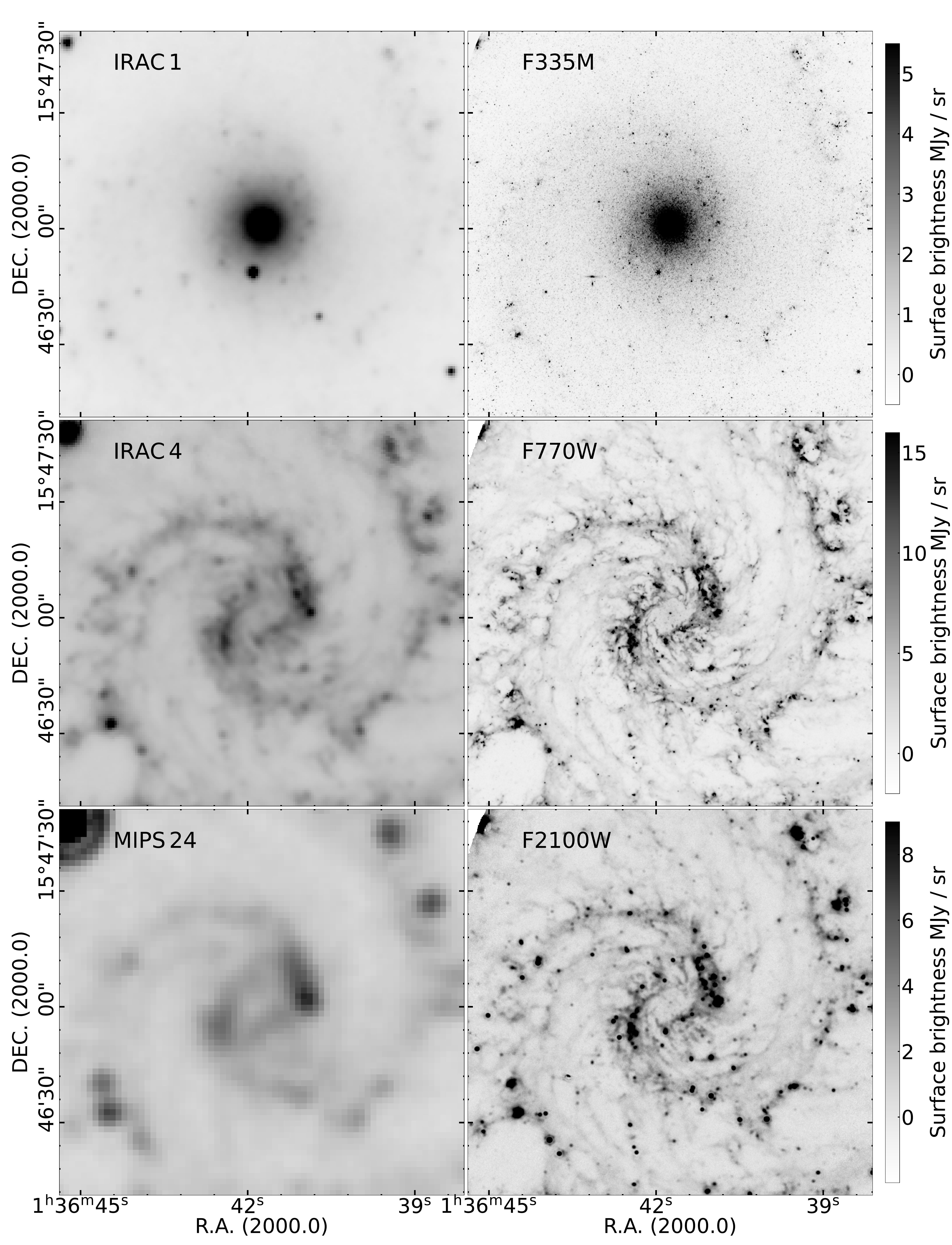}
\caption{New PHANGS-JWST imaging for NCG~628 (right panels) compared with data previously taken with Spitzer (left panels), illustrating the order of magnitude gains in resolution and sensitivity.  Comparisons at three different wavelengths are shown: \textbf{Top panels:} Spitzer IRAC 3.6 $\mu$m (Channel 1) vs.\ JWST NIRCam F335M; \textbf{Middle panels:} Spitzer IRAC 8 $\mu$m (Channel 4) vs.\ JWST MIRI F770W; \textbf{Bottom panels:} Spitzer MIPS 24 $\mu$m vs.\ JWST MIRI F2100W.  The Spitzer IRAC and MIPS data are taken from the SINGS program \citep{kennicutt03, dale05}. The JWST image is formed from a mosaic of two pointings with NIRCAM and three pointings with MIRI.}
\label{fig:spitzer}
\end{figure*}

\begin{figure*}
\centering
\includegraphics[width=\textwidth]{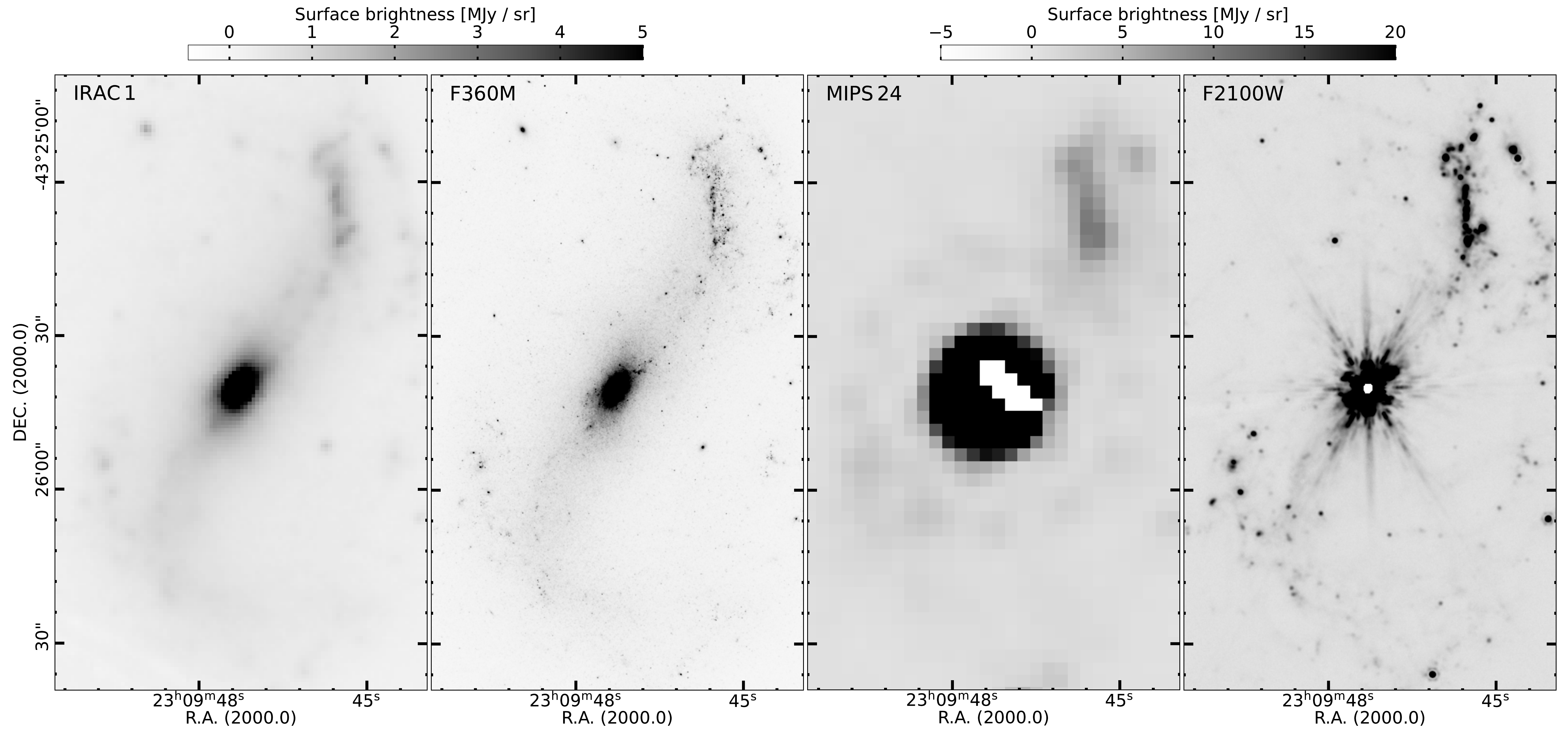}
\caption{Comparison of new PHANGS-JWST imaging with data previously taken by Spitzer, now for NGC~7496. Here comparisons at two wavelengths are shown: (left two panels) Spitzer IRAC 3.6 $\mu$m (Channel 1) vs.\ JWST NIRCam F335M; (right two panels) Spitzer IRAC 8 $\mu$m (Channel 4) vs.\ JWST MIRI F770W.}
\label{fig:spitzer2}
\end{figure*}

\begin{figure*}
\includegraphics[width=3.5in]{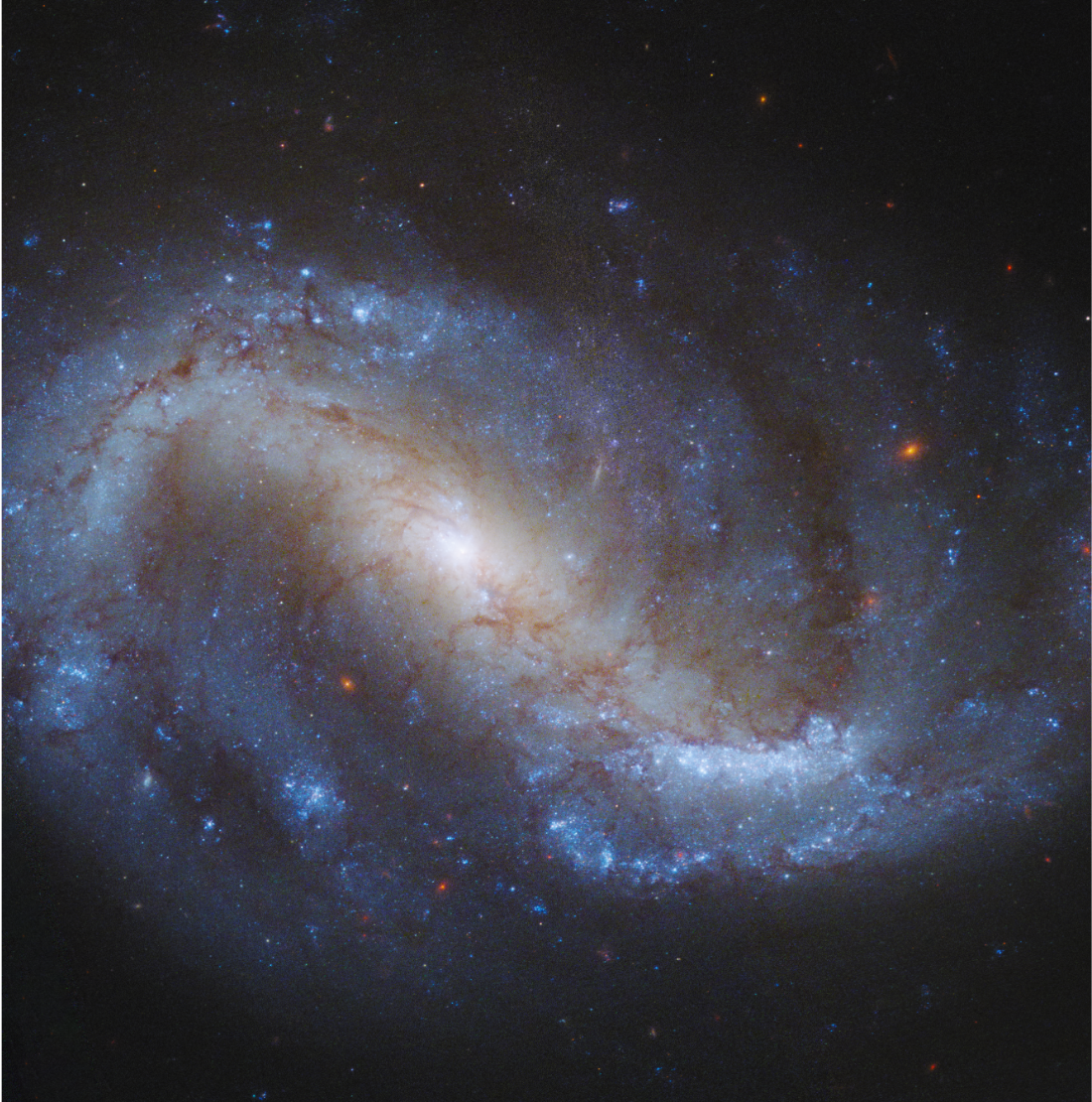}
\includegraphics[width=3.5in]{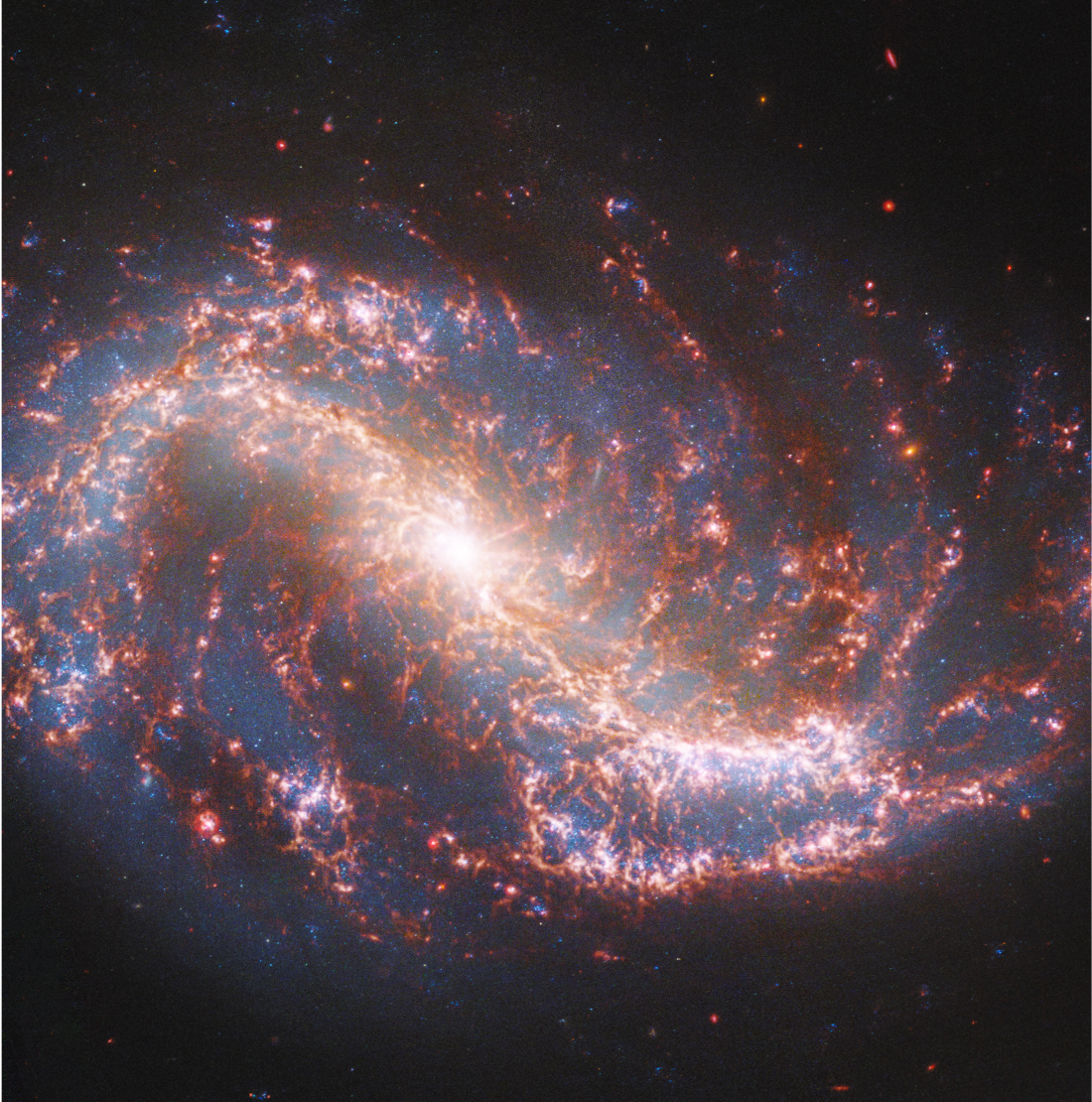}
\caption{Still images from animation (\url{https://app.box.com/s/zpviwjgtv14ozlrc0z1yi4dt9s9mtrmk}) showing a color composite image of NGC~7496 created from HST UV-optical imaging (red: F814W/F555W/F438W; green: F336W; blue:F275W) that fades into the same image combined with a JWST MIRI composite in red hue (with red: F2100W; green: F1130W; blue: F1000W). Both images produced by J. Schmidt. }
\label{fig:n7496_movie}
\end{figure*}

While past infrared missions have provided imaging of the nearby galaxies in the PHANGS-JWST sample, JWST's vast improvement in sensitivity and resolution reveals an extraordinary new view of the complex organization of the ISM.  In Figures~\ref{fig:spitzer} and \ref{fig:spitzer2}, we present our JWST imaging of NGC~628 and NGC~7496 in selected filters, alongside previous Spitzer observations, to illustrate data quality and the new science enabled by the observations. The two galaxies are at 9.84 Mpc and 18.7 Mpc respectively, and span most of the distance range (5-20 Mpc) covered by our sample.  In Figure~\ref{fig:filters}, we show the filter curves of the Spitzer bands which overlap JWST wavelength coverage together with the eight PHANGS-JWST filters. 

Perhaps the most remarkable feature in the PHANGS-JWST imaging is the network of filaments, shells, bubbles, and the star-forming populations nested in these structures, as traced by the dust in emission.  While the highest surface brightness features are apparent in the Spitzer imaging, the JWST imaging shows the pervasiveness of the network deep into the interarm regions, and morphological details that suggest that the multi-scale impact of star formation feedback on the ISM is ubiquitous. With the PHANGS-JWST imaging, we will now be able to develop catalogs of these structures and characterize their ensemble properties \citep[e.g.,][]{WATKINS_PHANGSJWST, RODRIGUEZ_PHANGSJWST, HASSANI_PHANGSJWST, THILKER_PHANGSJWST} even with the lowest resolution image at 21$\mu$m.

For both NGC~628 and NGC~7496, we show the NIRCam F335M and Spitzer/IRAC 3.6 $\mu$m images.  The NIRCam F335M filter better isolates the 3.3 $\mu$m PAH feature, and observations in the flanking F300M and F360M bands (not shown) enable continuum subtraction.  It is clear that the increase in resolution (PSF FWHM 1\farcs6 vs. 0\farcs11), which now samples 5 pc at NGC~628 and 10 pc at NGC~7496, enables the study of individual star clusters and associations, and allows embedded stellar populations to be identified.  A slightly extended point source southeast of the galactic center in NGC~628 is resolved into two background galaxies, and illustrates how JWST significantly improves the characterization of extended stellar structures and the identification of background ``interlopers.''

For NGC~628, the MIRI F770W and Spitzer/IRAC 8 $\mu$m images are shown.  The bandpasses, which are similar, capture the complex of PAH features in the $\sim$8 $\mu$m spectral region (Figure~\ref{fig:filters}).  Overall the same structures are apparent in both images, with MIRI F770W image being eight times sharper (PSF FWHM 2\farcs0 vs. 0\farcs24), sampling 12 pc at NGC~628.

Finally, the MIRI F2100W and Spitzer/MIPS 24 $\mu$m images, which capture flux from the warm dust continuum, are shown for both galaxies.  Here, although the relative increase in the angular resolution is similar as for the previous two sets of images, the improvement in the fidelity of the JWST data is the most dramatic as the MIPS 24 $\mu$m PSF was only 6\arcsec, and the majority of emission appeared diffuse.  With the MIRI F2100W PSF of 0\farcs67 (30 pc at NGC~628 and 60 pc at NGC~7496), the dust emission is resolved into compact sources, mostly along the spiral arms, and shows the same intricate ISM structure seen in the MIRI F770W image.

In terms of surface brightness, the Spitzer images achieve slightly better sensitivities than the JWST data, as might be expected.  The NGC~628 images, from the Spitzer SINGS program \citep{kennicutt03}, are mosaics with 1$\sigma$ surface brightness limits of $\sim$0.02 MJy~sr$^{-1}$ at 3.6$\mu$m and $\sim$0.2 MJy~sr$^{-1}$ at 24$\mu$m.  The corresponding sensitivities are $0.04~\mathrm{MJy~sr^{-1}}$ for NIRCam F335M and $0.25~\mathrm{MJy~sr^{-1}}$ for MIRI F2100W (see Table \ref{tab:obsparam1} and discussion in the next section).  Of course, the principle gain is in the $\sim$10-fold improvement in resolution and corresponding decrease in the solid angle subtended by the PSF, so that the JWST point source sensitivity is $\sim 50$ times better than Spitzer.

\subsection{Empirical Noise Estimates}
\label{sec:empiricalnoise}

Table~\ref{tab:obsparam1} provides sensitivity limits based on our initial reduction of PHANGS-JWST data.  We evaluated the depth of our data with respect to detection of point sources at the resolution achieved in each filter, and measurement of diffuse emission on a fixed larger scale.  

For the point source limit, we first identified the faintest area mapped that was also fully contained within the footprint of maximum exposure time.  Inside each area, aperture photometry was performed for a set of 50 independent circular apertures sized to match the 50\% encircled energy radius of each filter. 
Background estimation for each aperture was also filter-dependent, using an annulus spanning a radial range from 2-3 times the aperture radius.  The sigma-clipped standard deviation of the background-subtracted flux for these `empty' apertures was then scaled by a factor of ten to estimate the 5-sigma point source detection limit in each filter ($5\sigma_{f}$), which accounts for a factor 2 for the choice of 50\% enclosed energy radius and a factor of 5 for the choice of significance level.  No correction was made for the contribution of the point source flux to the background annulus flux, which is expected to be negligible for an unresolved source.  Averages of these galaxy-specific limits per filter are tabulated.  It is important to note that this method, does not include loss of point source sensitivity due to source crowding in more complex, brighter regions of each JWST target.  Such loss of completeness due to crowding and associated photometric bias will be quantified later when PSF-fitting photometry is undertaken for our entire set of targets.  

To estimate the surface brightness sensitivity ($\sigma_{I}$) for the imaging data, we consider the distribution of pixel values in an unsharp masked image.  For each image, we compute a median filter with a circular tophat filter with a diameter 6 times the PSF FWHM in the band.  We subtract off the median smoothed version from the original map to remove large scale structure and produce an unsharp masked image.  We then use a median absolute deviation estimator to measure the local noise in the unsharp masked image.  We iteratively reject all values larger than the local 3$\sigma$ noise measurement in each filter.  After the data rejection converges, the local noise estimates are smoothed with a tophat filter that has a size of 15$\times$ the PSF FWHM and interpolating over missing values.  We compare this local estimate of the noise to the noise estimates provided by the JWST imaging pipeline. 
The resulting noise maps are within 10\% of the pipeline generated noise maps.  We also estimated the local noise by calculating the median autocorrelation of data as a function of scale and by examining the distributions of data in signal-free parts of images (though some bands show emission across the field). All methods arrive at similar noise estimates and reflect the spatial structure in the mosaicked fields.  

Note that because of the extended JWST PSFs, there is less encircled light at a PSF FWHM than for a Gaussian source.  Thus, the simple scaling of surface brightness noise to obtain a point source sensitivity leads to an overestimate of the point source sensitivity: $\sigma_{I} \Omega_\mathrm{PSF} \gtrsim \sigma_f$,  where $\Omega_\mathrm{PSF}$ is the solid angle subtended by the PSF.

\subsection{Color Composites \& Public Outreach}

Since the first observations were taken for the PHANGS-JWST Treasury survey in 2022 July, color composites that combine JWST imaging from multiple filters have been developed and released for all of our targets by astronomers, professional public outreach teams, the press, as well as the public.  The color images have broadly captured the attention of both the science community and general public not only because the structures portrayed are aesthetically stunning, but also because the images vividly illustrate and make qualitatively accessible the physics of star formation, feedback, and the ISM -- particularly when the UV-optical imaging from HST is included.  

Figure~\ref{fig:n7496_movie} shows the first composite PHANGS-JWST image published for NGC~7496, which was observed shortly after the end of commissioning and among the first science data to be released from the mission.  The image is a sum of two composites: one based on the HST UV-optical filters (red: F814W/F555W/F438W; green: F336W; blue:F275W) and a second based on the JWST MIRI F1000W, F1130W, F2100W filters in red hue.  The animation compares the emission from the HST and JWST data. The regions and structures that are dark in the optical due to dust obscuration are illuminated in the mid-infrared by the re-radiated emission from small dust grains.  The complex network of filaments, bubbles, shells, and compact sources described earlier can now be seen in the context of the visible young stellar populations that line the peripheries of the network, and have provided the feedback energy that, together with galactic dynamics, shapes the ISM.  Infrared compact sources without optical counterparts can be identified, revealing the sites of the earliest stages of star formation. 

\begin{figure*}
\centering
\includegraphics[width=6in]{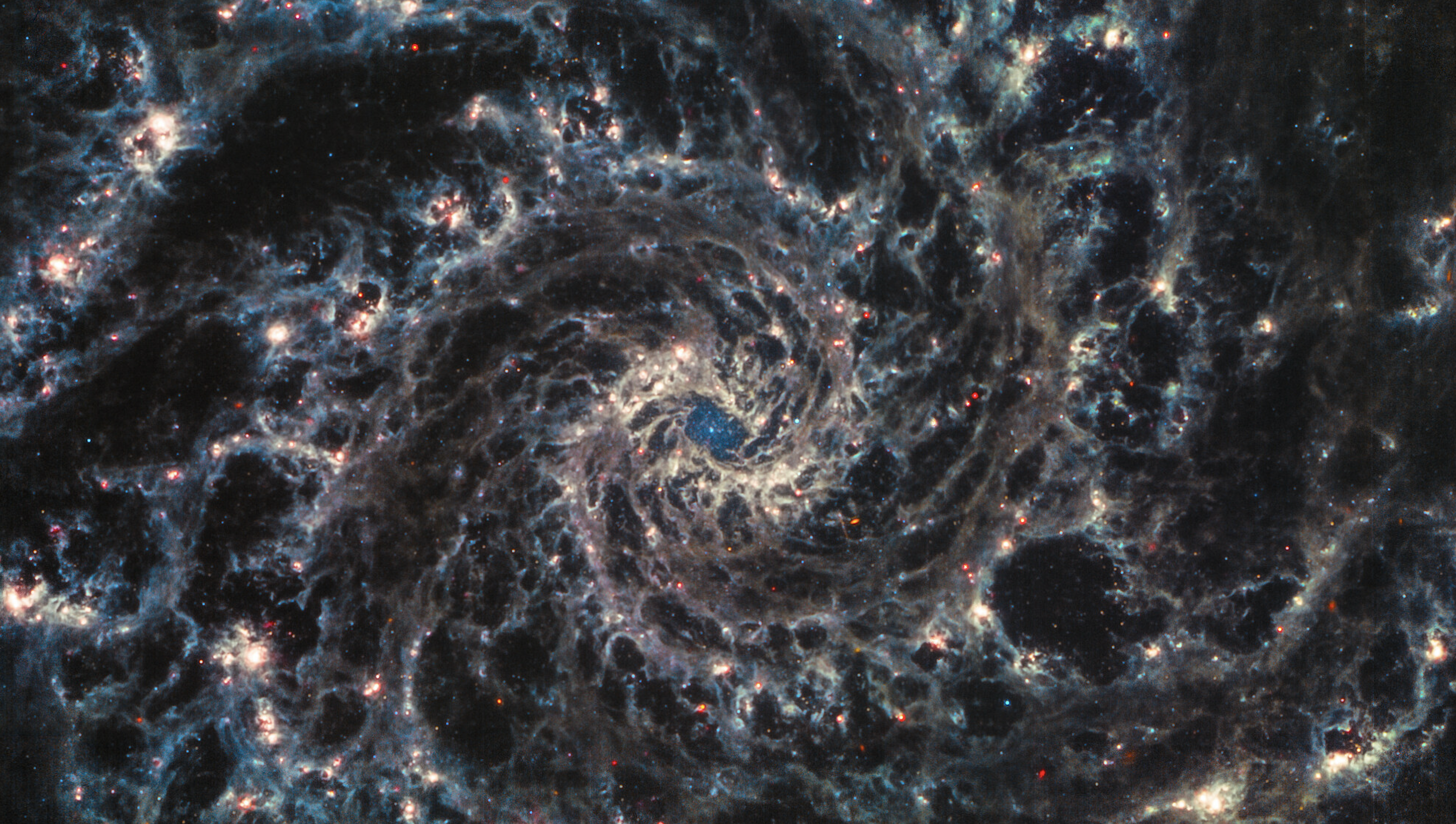}
\includegraphics[width=6in]{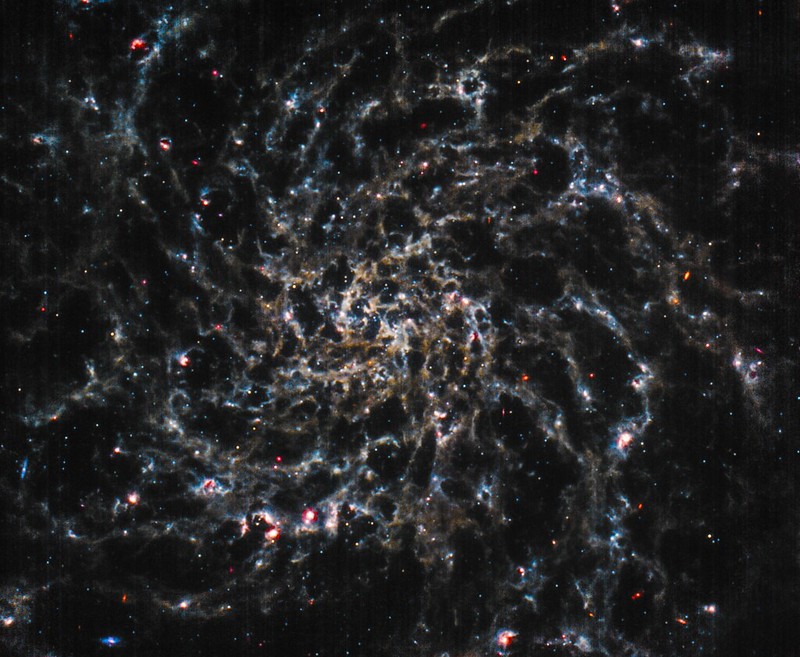}
\caption{Color composite images produced by J. Schmidt from MIRI data of NGC~628 (top) and IC~5332 (bottom). Red: MIRI F2100W; orange: MIRI F1130W; cyan: MIRI F770W and extra overall brightness in grayscale: MIRI F1000W.  The images were used as the models for the first two ESA/JWST Pictures of the Month from the JWST Cycle 1 General Observers Program as featured at \url{https://esawebb.org/images/potm2208a/} and \url{https://esawebb.org/images/comparisons/potm2209a/}.}
 \label{fig:judy1}
\end{figure*}

\begin{figure*}
\centering
\includegraphics[width=6in]{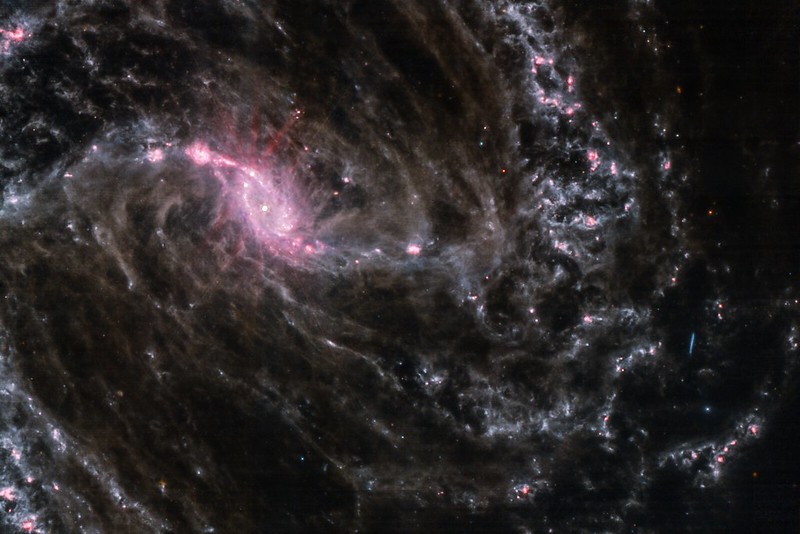}
\includegraphics[width=6in]{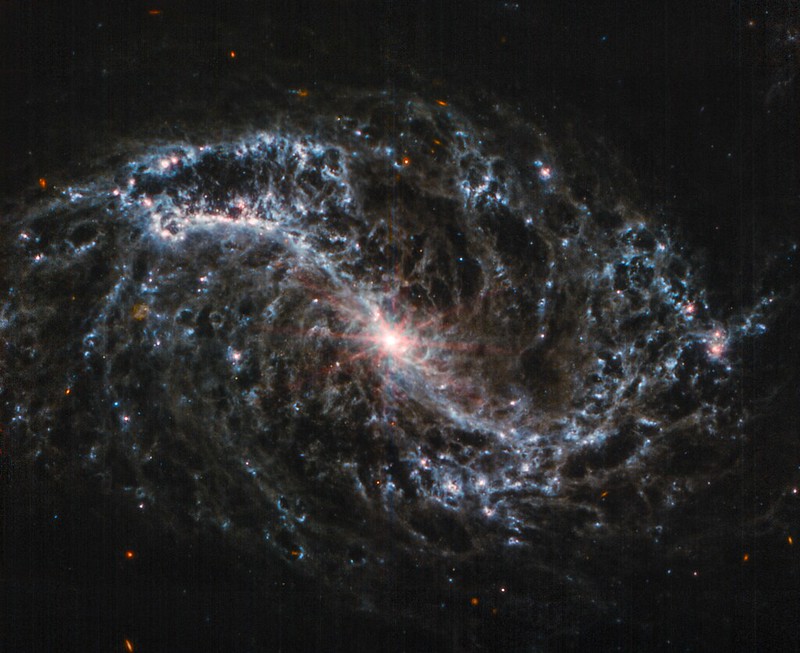}
\caption{Same as Figure~\ref{fig:judy1} but for NGC~1365 (top) and NGC~7496 (bottom).}
 \label{fig:judy2}
\end{figure*}

Figures \ref{fig:judy1} and \ref{fig:judy2} show RGB color composites, now based on only the MIRI imaging, for all four PHANGS-JWST First Results galaxies.  The diversity of the dust emission structure motivates the study of a representative sample of galaxies, and our specific science goals (next section).  The four galaxies bracket the full range of star formation properties in the sample (with NGC~1365 having the highest star formation activity and gas surface density and IC~5332 among the lowest) and feature varied dynamical structures (e.g., bars in NGC~1365 and NGC~7496; strong spiral arms in NGC~628 and weak spiral structure in IC~5332).  The PHANGS-JWST team is collaborating with image processor Judy Schmidt to produce color composites as new data are received, and with the public outreach offices at ESA, STScI, and to support the communication of new results. 

\section{PHANGS-JWST Science Goals}
\label{sec:goals}

In this section, we summarize the science goals that motivate the PHANGS-JWST survey, and highlight how results presented in this Issue begin to make progress toward those goals.  

Our studies of star formation and dust emission on the scales of star clusters, HII regions, and giant molecular clouds are enabled by JWST's new high resolution views of the near- and mid-IR in our selected set of 8 filters. As shown in Figs.~\ref{fig:filters} and ~\ref{fig:sed_embedded}, the F200W, F300M, and F360M filters provide low obscuration views of stellar photospheric emission, with some contribution from nebular emission, and hot dust. The F335M, F770W, F1130W capture PAH emission, tracing a combination of PAH size and charge, with the F300M and F360M allowing for robust continuum subtraction for F335M to isolate the $3.3\micron$ PAH feature. F1000W and F2100W capture the dust continuum, with possible contribution to the F1000W band by silicate absorption.

Together with data from HST, MUSE, and ALMA, the integrated PHANGS dataset allows for multi-scale, multi-phases studies of the star formation cycle across a range of galactic environments.
In Figure~\ref{fig:filters}, the five UV-optical filters used in the PHANGS-HST program \citep{phangs-hst} are also shown; these mainly sample the stellar continuum and provide a view of dust in extinction.  The HST imaging provides high angular resolution (0\farcs8), comparable to JWST NIRCam imaging at 2$\mu$m.  The PHANGS-MUSE program \citep{phangs-muse} provides integral field unit spectroscopy over $0.475 < \lambda /\mu\mathrm{m} < 0.935$, covering many of the same bands as HST but with coarser angular resolution ($\sim 1''$).  PHANGS-ALMA \citep{phangs-alma} shows the distribution and kinematics of the CO that traces the star forming molecular medium, also at $\sim 1''$ resolution.

\subsection{Constraining the natal cluster mass function and cluster formation efficiency} \label{sec:SG1}

\begin{figure*}[t!]
\includegraphics[width=7in]{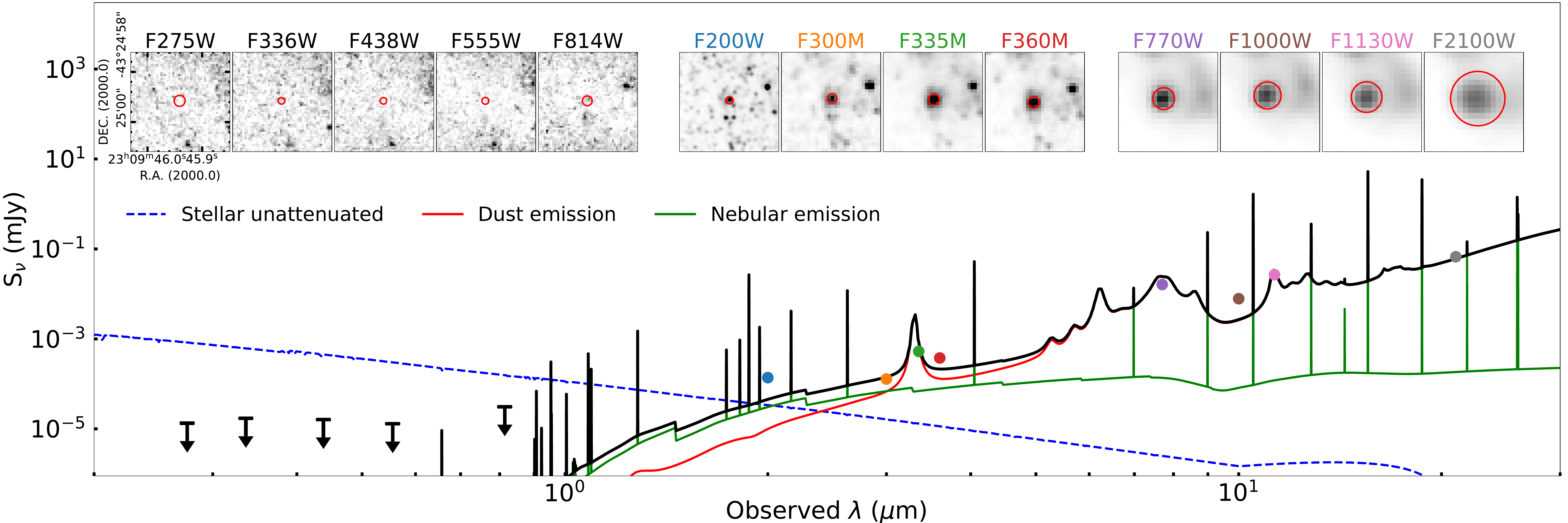}
 \caption{Example of a dust embedded young star cluster---a compact infrared source with no detection in the optical HST images.  Postage stamps from all 13 HST+JWST PHANGS images are shown together with photometry, as an illustration of the SED fitting that will be performed in the future to constrain the physical properties of these objects.}
 \label{fig:sed_embedded}
\end{figure*}

PHANGS-JWST near- and mid-IR imaging will be used to detect dust embedded clusters and associations, to mitigate and measure the effects of extinction, and to constrain other physical properties such as age and mass through SED-fitting (e.g.~Figure~\ref{fig:sed_embedded}).  We will test different selection strategies to systematically identify embedded cluster candidates, and produce the first complete catalogs of these sites of the earliest stages of star formation.  The catalogs will enable us to measure the fraction of stars that are born in compact clusters; characterize the distribution of cluster masses at birth; and investigate dependence on environment.

The ensemble properties of star clusters are highly debated in the literature
\citep[e.g.,][]{larsen09,chandar17,krumholz19,adamo20,adamo20b}. Due to dust extinction, it has been difficult to obtain a truly complete census of clusters across all galactic environments and ages. 
The PHANGS-JWST infrared imaging will now allow us to study the clusters in dusty galaxy centers, spiral arms, and bar-end regions. These are sites where super star clusters tend to be found, the cluster formation efficiency is expected to be highest \citep[e.g.,][]{kruijssen12,krumholz14,adamo15,ginsburg18,grudic21}, and molecular cloud populations show clear enhancements in both density and turbulence \citep[e.g.,][]{longmore14,beuther17,sun20,henshaw22}. Extrapolating from the visible cluster population, we expect to detect $\gtrsim$100 new embedded clusters in each galaxy or ${\sim}2{,}000$ across the sample, with the greatest gain in the dustiest and most active parts of the sample. Analysis combining JWST with HST data will also allow us to break the age-reddening degeneracy for optically detected clusters, and derive robust constraints on age, mass, and reddening for both embedded and visible cluster populations (e.g.~Figure~\ref{fig:filters}).

These investigations use the full set of PHANGS-JWST data for measuring the properties of embedded clusters. In particular, the near-IR bands provide key constraints on stellar properties.  The F200W observations provide good visiblity through the dust in the region of the SED still dominated by stellar photospheric emission.  The excellent angular resolution (0\farcs07) at 2$\mu$m is required to resolve star clusters, which have effective radii between $0.5$~pc to about $10$~pc \citep{pz10,ryon17}.   

\citet{RODRIGUEZ_PHANGSJWST} \citet{WHITMORE_PHANGSJWST} take the first steps in identifying embedded clusters, estimating their physical properties, and examining the incompleteness in optical samples.  \citet{HOYER_PHANGSJWST} study the nuclear star cluster in NGC628 and the nature of the mid-infrared structure surrounding it. \citet{SCHINNERER_PHANGSJWST} show that massive, dusty clusters can provide a sensitive diagnostic of the very dusty star-forming environments in nuclear regions when comparing ALMA data and simulations.  Some of the clusters detected in the PHANGS-JWST survey push current stellar and dust models used by CIGALE to their limits due to their very young ages, high dust content, and/or low mass.  Including dust models computed in the optically thick case and stellar populations with a stochastically sampled Initial Mass Function may prove necessary to model such clusters.

\subsection{Measuring the Duration of Embedded Star Formation}

PHANGS-JWST data enable characterization of the poorly studied embedded phase of star-formation before feedback disrupts the natal cloud. 
Observations that resolve galaxies into discrete star-forming regions, clusters, and molecular clouds have enabled statistical measurements of the timescales associated with the major phases of the star formation cycle \citep[e.g.,][]{kawamura09, kruijssen14, meidt15, corbelli17, grasha18, grasha19, kruijssen18, kruijssen19, hannon19, chevance20, ward22, kim22, hannon22}. Over the last decade, the (anti-)correlation between clouds and \ion{H}{2} regions or optically-identified clusters has constrained the total lifetimes of clouds to 10-30~Myr \citep[where the range represents physical variation rather than the uncertainty, see e.g.,][]{chevance20b,chevance22b}. However, both the duration of the earliest, embedded phase and the length of time that a cloud remains truly devoid of star formation activity remain poorly known. 

Measuring these timescales is essential for understanding the dominant physics of star formation. The time between cloud formation and the onset of star formation constrains the (im)balance between the driving and decay of turbulence \citep[e.g.,][]{klessen16,padoan17}. The duration of the embedded phase reflects the time that it takes feedback to start dispersing a cloud \citep[e.g.,][]{olivier21}.  Measuring the embedded timescale can also distinguish between models in which star and cluster formation ``accelerate'' over a cloud's lifetime \citep{murray11, lee16} and those that favor a more steady process \citep{krumholz05, krumholz20, krumholz07}. Timescales for the embedded and optically visible stages can be used to infer which feedback mechanisms quench star formation \citep[e.g.,][]{lopez14, kruijssen19, chevance22, kim22, hannon22}, and whether clouds are magnetically sub- or super-critical \citep{jgkim21}.

Accessing the embedded phase of star formation requires a tracer of obscured star formation that reaches the scale of individual clouds. The limited resolution of \textit{Spitzer} ($\sim 6''$) limited the number of galaxies where such measurements were possible to five \citep{kim21}. With the increased spatial resolution and sensitivity in the mid-infrared achieved with MIRI, the time evolution of the early stages of star formation can now be characterised in galaxies out to $\sim 30\,$Mpc. 

By combining some of the first PHANGS-JWST MIRI observations with PHANGS-ALMA CO observations \citep{phangs-alma}, \citet{KIM_PHANGSJWST} determine the relative duration of the early phases of star formation, from truly quiescent clouds, to embedded star formation, revealed star formation, and cold gas-free \ion{H}{2} regions in NGC~628. In combination with the stellar ``clocks'' offered by star clusters (see \S\ref{sec:SG1}) and H$\alpha$ emission \citep[e.g.,][]{leroy2013, haydon20}, these relative measurements are converted into absolute timescales. PHANGS-JWST data will enable this measurement across the wide range of environments that are present in the sample. These observations will constrain how the physical mechanisms that regulate and drive the successive phases of star formation map to environment.

\subsection{Characterizing How Dust Properties Respond to Local Environment}

Most of the imaging bands used in PHANGS-JWST show emission features from dust continuum and PAHs. The PAHs reprocess as much as $20$\% of all absorbed stellar UV/optical light \citep{smith07}. As they do so, they eject electrons that heat the ISM and produce bright mid-IR emission features \citep[e.g.,][]{tielens08} that are among the most highly detectable star formation tracers across cosmic time \citep[e.g.,][]{riechers14}. However, PAH properties and abundance vary substantially within and between galaxies. Fewer PAHs are present in \ion{H}{2} regions or at low metallicities \citep[e.g.,][]{engelbracht05,chastenet19} but they can make up $>$5\% of the dust mass in the diffuse ISM of solar metallicity galaxies \citep{draine07,aniano20}. This variability hampers the use of PAH emission as a star formation tracer \citep{calzetti07,whitcomb20} and leads to critical uncertainties when modeling their role in the ISM. The PHANGS-JWST Treasury data will yield the first highly resolved atlas of PAH properties---abundance, size distribution, and charge inferred from mid-IR band ratios---in the nearby galaxy population. These data can be used to build an empirical model to predict the abundance and characteristics of PAHs as a function of local metallicity, ISM phase, density, and radiation field (from HST, MUSE, and ALMA). The measurements underpinning this model must be highly resolved, because radiation field, density, and ISM phase can vary by factors of ${\sim}1000$ on 10-100~pc scales, altering the PAHs \citep{chastenet19,draine21}.  PHANGS-JWST First Results in this Issue from \citet{CHASTENET1_PHANGSJWST}, \citet{CHASTENET2_PHANGSJWST}, \citet{DALE_PHANGSJWST}, and \citet{EGOROV_PHANGSJWST} all show that PAH emission properties do indeed vary systematically with the local galactic environment.

\subsection{A High Resolution View of the Neutral ISM}
\label{sec:highresview}
PAH emission offers a promising tracer for the structure of the neutral ISM. Traditionally, \ion{H}{1} 21-cm emission and the lowest energy rotational lines of CO ($\lambda = 2.6$ and 1.3~mm) have been used to map out the atomic and molecular phases of the ISM.  These tracers suffer some limitations: the very long wavelength of 21-cm emission makes it intrinsically faint, and current interferomers can typically only achieve resolutions of 5-10'' for non-prohibitive integration times.  The (sub-)millimeter CO emission lines are at shorter wavelengths and thus significantly better resolution of 1'' can be achieved with ALMA \citep{phangs-alma}, but they are an imperfect proxy for the H$_2$ that constitutes the molecular ISM.  With JWST, the mid-IR PAH emission potentially provides a much higher resolution view ($<0\farcs37$) of the same interstellar gas that is traced by \ion{H}{1} and CO.  While the emissivity of PAHs likely varies with environment \citep{draine07,hensley22}, the spatial distribution of the mid-IR PAH emission yields an exceptionally detailed, high dynamic range image of the structure of neutral gas in a galaxy.  Based on studies in the Milky Way and the nearest galaxies, the older ``spherical cloud'' view of the ISM has been integrated with a newer filamentary description of ISM structure \citep{hacar22}.  The enthusiasm for such a filamentary view has been driven by infrared imaging by the previous generations of instruments, but it was difficult to extend this view to a galactic scale in all but the nearest galaxies.  JWST imaging of PAH emission yields such high resolution maps of neutral gas throughout the Local Volume. 

Figure~\ref{fig:judy1} shows the filamentary organization of the gas in NGC\,628, which prompted several initial investigations of ISM structure using PAHs as ISM tracers. \citet{SANDSTROM2_PHANGSJWST} demonstrates a new JWST band combination that produces high quality maps of the PAH feature at 3.3 $\mu$m with excellent stellar continuum subtraction.
These maps offer the highest resolution ($0\farcs11$) wide-area maps of the ISM in nearby galaxies. \citet{SANDSTROM1_PHANGSJWST} verifies that PAH features are intimately connected to the neutral ISM and uses the genus statistic for characterizing the emission topology.  \citet{THILKER_PHANGSJWST} show how the filamentary network of 3.3 $\mu$m PAH emission relates to extinction lanes identified in the optical HST imaging, and to the young cluster population of a galaxy.  \citet{MEIDT_PHANGSJWST} show that the spatial separation scales between filaments is consistent with large scale gravitational fragmentation models.  \citet{WILLIAMS_PHANGSJWST} leverage the high resolution images to make a careful measurement of evolutionary timescales in a spiral arm spur.

PAH imaging also shows us where the neutral ISM is {\it not} found.  Rings of PAH emission with faint centers can be used to define the wall of ``bubbles'' or feedback-driven shells.  While bubbles from feedback have been long studied \citep[e.g.,][and references therein]{oey97, weisz09}, the exquisite resolution, the sensitivity to structure on many scales, and the wide dynamic range of the PHANGS-JWST images make them an excellent tracer of bubble walls.  PHANGS-JWST data can be used to produce inventories of bubbles and shell features. These structures can be connected to the sources of feedback identified from PHANGS-MUSE and PHANGS-HST catalogs of star clusters and associations, and to the kinematics of the cold ISM as traced by PHANGS-ALMA.  As an early demonstration of such analyses, \citet{WATKINS_PHANGSJWST} create a bubble catalog for NGC~628, and examine their properties, including the size distribution for the ensemble population. \citet{BARNES_PHANGSJWST} demonstrate the capabilities of the PHANGS multi-wavelength dataset for measuring the expansion rates and understanding the origins of large kpc-scale cavities, focusing on the ``Phantom Void'' in the southeast quadrant of NGC~628.

\subsection{Robust mid-IR Star Formation Diagnostics From Cloud to Galaxy Scales}

Mid-IR emission from hot dust is widely used to trace star formation at both low and high $z$ \citep[e.g.,][]{kennicutt12}. Unfortunately, the calibration of mid-IR star formation tracers has a strong dependence on both spatial scale and environment \citep[e.g.,][]{boquien15}, which likely stems from a variable heating contribution by old stars and uncertain dust geometry \citep[e.g.,][]{groves12,nersesian19}. 
PHANGS-JWST imaging, combined with supporting PHANGS datasets, enables a data-driven description of the origin and variations in mid-IR emission. Mid-IR emission is due to the heating of dust and PAHs by both young and old stars, but the bright thermal dust emission in the mid-IR continuum bands (F2100W and F1000W) is expected to respond differently to the local dust-heating radiation field than the PAH emission \citep{draine07, hensley22}, offering a potentially powerful diagnostic of local interstellar conditions. With the insights obtained by comparing mid-IR band ratios and physical dust models, we will build on previous work that calibrated the mid-IR emission as a star formation rate indicator on galaxy-wide and sub-kpc scales \citep[e.g.][]{calzetti07}. 

Combining robust high resolution measurements of the embedded star formation  from PHANGS-JWST 
with PHANGS-ALMA and PHANGS-MUSE imaging of the molecular gas and optically visible tracers of star formation and feedback will enable the first cloud-scale inventory of all stages in the star formation cycle across the 19 galaxies in our sample. Previous work on $\lesssim\!100$~pc scales suggested that up to $\sim60$\% of molecular gas in galaxies may be devoid of H$\alpha$ emission \citep[e.g.,][]{schruba10,onodera10,kreckel18,kruijssen19,schinnerer19b,chevance20,kim22,pan22}. Yet highly resolved IR observations show that truly non-star-forming clouds are scarce in our Galaxy \citep{mooney88,ginsburg12} and \textit{Spitzer} results suggest that molecular clouds in Local Group galaxies spend up to half their lifetime hosting only embedded massive star formation \citep{kim21}.  With PHANGS-JWST, we will confirm whether there is a large reservoir of molecular gas without star formation in nearby star-forming disk galaxies and, if so, whether it due to a delayed onset \citep[e.g.][]{klessen16,padoan17} or sustained suppression \citep[e.g.,][]{federrath16,semenov17,meidt18,meidt20,kruijssen19b} of star formation.

PHANGS-JWST high resolution measurements will also enable important empirical tests of cloud-scale models of star formation. These models predict the integrated and per free-fall time star formation rate and efficiency of individual clouds based on their density, turbulence, and dynamical state \citep[e.g.,][]{padoan17,jgkim21}, quantities that have already been measured using the PHANGS-ALMA data \citep[e.g.,][]{sun20,sun22}. Cloud-scale star formation models are widely used to interpret observations \citep[e.g.,][]{barnes17,utomo18} and implemented as sub-grid models in galaxy simulations \citep[e.g.,][]{gensior20,kretschmer20}.  The models have never been rigorously tested beyond the Local Group, however, since the rapid action of feedback prohibits linking optical SFR estimates to the progenitor cloud's gas mass, and hence the integrated star formation rate per unit gas mass.  
Lacking cloud-by-cloud estimates of the embedded star formation activity, studies in external galaxies have resorted to using population averages, obtaining results that suggest---but cannot currently prove---important inconsistencies with the models, such as inferred star formation efficiencies that decrease with the gravitational boundedness of the gas \citep[e.g.,][]{leroy17,schruba19}. The new PHANGS-JWST data are essential for unambiguous cloud-scale tests of these predictions.

The PHANGS-JWST papers in this Issue have begun to explore these questions.  \citet{HASSANI_PHANGSJWST} study compact sources in F2100W, finding that nearly all sources are compact star-forming regions.  These regions show strong association both with the molecular gas and H$\alpha$ emission. \citet{LIU_PHANGSJWST} evaluate how strong heating in the IR from dusty star-forming regions is connected to the density and temperature of the molecular ISM as seen by ALMA. 
\citet{LEROY2_PHANGSJWST} use a large set of archival data to show that the mid-IR has excellent capacity to trace the properties of the star-forming ISM. This retrospective analysis provides context for the high resolution view provided by PHANGS-JWST.  Using the new data, \citet{LEROY1_PHANGSJWST} conduct a correlation analysis of the mid-IR data, showing strong correlations with both the H$\alpha$ and the CO in galaxies and these correlations are stronger than the correlation between H$\alpha$ and CO at high resolution.  
\section{Data Products}
\label{sec:products}
Preliminary JWST image products for NGC~7496, IC~5332, NGC~628, NGC~1365, used as the basis for the PHANGS-JWST First Results presented in this Volume, are available at the Canadian Astronomy Data Centre as part of the PHANGS archive\footnote{\url{https://www.canfar.net/storage/vault/list/phangs/RELEASES} -- Note to be removed in proof: JWST data will be posted on acceptance.}.  This data release includes stacked, mosaicked, calibrated images based on all PHANGS-JWST data obtained through 2022 August: imaging in four NIRCam and four MIRI filters for NGC~7496, NGC~628, NGC~1365, and in four MIRI filter for IC~5332.$^3$  The processing of these data are subject to the caveats discussed in Section~\ref{sec:caveats}.

After the completion of the PHANGS-JWST observations, we will provide an expanded set of image products for all 19 galaxies through the Mikulski Archive for Space Telescopes (MAST) as well as at the CADC PHANGS archive.  We anticipate significant improvements in the processing and calibration to address currently known issues (Sections \ref{sec:redux} and \ref{sec:caveats}) as well as those that arise in future analyses.  These later data releases will include the full PHANGS-JWST imaging stack at native resolution together with products convolved to match the angular resolution of the lowest resolution PHANGS-JWST image (MIRI 21$\micron$) and other PHANGS datasets (e.g., maps from ALMA and MUSE).  

The PHANGS-JWST papers in this Issue give examples of the high-level science products that would be valuable to be developed from JWST data in combination with the rest of the PHANGS surveys.  

PHANGS-JWST First Results on stellar clusters illustrate the potential (and challenges) in using the data at their highest resolution (which at 2$\mu$m surpasses the resolution of HST optical imaging) to develop catalogs of embedded star clusters, to augment the PHANGS-HST star cluster and association catalogs with photometry from 2-21$\mu$m, and to measure their physical properties through SED fitting \citep{RODRIGUEZ_PHANGSJWST, WHITMORE_PHANGSJWST, HOYER_PHANGSJWST}.  Catalogs of compact sources in the MIRI bands can be produced to identify larger star-forming regions on scales comparable to the HII regions observed with MUSE, and the giant molecular clouds resolved by ALMA \citep{HASSANI_PHANGSJWST, SCHINNERER_PHANGSJWST}. As PHANGS-JWST analysis matures, it will be possible to develop complete catalogs of dust-enshrouded and visible stellar populations across multiple physics scales \citep{larson22}, which include physical properties such as mass, age and dust reddening, via SED fitting which is able to break degeneracies between reddening and age \citep{whitmore22}.  

PHANGS-JWST First Results on feedback show how catalogs of ISM bubbles and shells can be produced \citep{WATKINS_PHANGSJWST, BARNES_PHANGSJWST} and used in combination with catalogs of the young stellar populations to identify the contribution of feedback and dynamical processes in the formation and evolution of these structures. The importance of complete inventories of the youngest stellar populations and feedback-driven ISM structures and of access to the earliest embedded phase of star formation via mid-IR JWST imaging for quantitative estimates of timescales, star formation efficiencies and feedback parameters is also highlighted by \citet{KIM_PHANGSJWST,WATKINS_PHANGSJWST, BARNES_PHANGSJWST}.

 As shown in Section~\ref{sec:results}, JWST's high resolution capabilities have provided the most detailed maps of the diffuse ISM possible in nearby galaxies to-date \citep{SANDSTROM1_PHANGSJWST} to enable studies of local influences on heating \citep{LEROY1_PHANGSJWST}.  Dust emission and absorption maps can be produced from the combination of the PHANGS JWST, HST, VLT-MUSE, and ALMA datasets \citep[e.g.,][]{THILKER_PHANGSJWST,MEIDT_PHANGSJWST,WILLIAMS_PHANGSJWST}.  

Finally, PHANGS-JWST First Results show how PAH features can be sensitive diagnostics of ISM conditions through changing flux ratios.  Using empirical methods, \citet{SANDSTROM2_PHANGSJWST} demonstrate how to isolate PAH features from the stellar continuum using JWST bands, which can be developed into a full set of PAH-only maps of the galaxies.  \citet{CHASTENET1_PHANGSJWST, CHASTENET2_PHANGSJWST} use JWST data to find variation in PAH properties in the broader ISM, with \citet{EGOROV_PHANGSJWST} and \citet{DALE_PHANGSJWST} exploring \ion{H}{2} regions and star clusters respectively.  These ratio measurements can be refined into environmental diagnostics of properties like the metallicity and radiation field.

\section{Summary}
We present the PHANGS-JWST Cycle 1 Treasury Program, one of the key observational campaigns conducted by the Physics at High Angular resolution in Nearby GalaxieS (PHANGS) collaboration \citep{schinnerer19}.  PHANGS-JWST is a \obstime~ hour program to image 19 nearby galaxies ($D<20~\mathrm{Mpc}$) in eight filters from 2$\mu$m to 21$\mu$m using NIRCam (F200W, F300M, F335M, F360M) and MIRI (F770W, F1000W, F1130W, and F2100W). 
\begin{enumerate}
    \item We summarize the properties of the PHANGS-JWST nearby spiral galaxy sample, which all have overlapping coverage from ALMA, HST, and VLT-MUSE and other facilities (Section \ref{sec:sample}), and present our NIRCam and MIRI observing strategy (Section \ref{sec:observations}).
    \item In Section \ref{sec:redux}, we present the modifications to the standard pipeline and additional processing steps used to produce the data products for the First Results analysis in this Issue. In particular, we describe changes to the default astrometric alignment, a mitigation strategy for the $1/f$ noise in NIRCam using Principal Component Analysis, and a background validation strategy for MIRI data.  Current limitations and uncertainties in this first reduction are summarized.
    \item In Section \ref{sec:results} we provide an overview of the imaging of the first four targets (NGC\,0628, NGC\,1365, NGC\,7496, and IC\,5332), compare our new JWST data to existing Spitzer imaging, and compute limiting point source and surface brightness sensitivities of the data.  We highlight the value of the PHANGS-JWST imaging for engaging the public in JWST discovery.
    \item In Section \ref{sec:goals} we describe the science goals of our program and preview the Letters in this Issue, while in Section \ref{sec:products} we describe currently available and planned future image products, as well as higher level science products that would be valuable to develop for the community.
\end{enumerate}
As a public Cycle 1 Treasury program, the PHANGS-JWST dataset is designed to broadly enable community science beyond the team science objectives outlined here. 
Open access to such standard imaging observations early in the mission is essential for learning how  to  best  study  star formation, stellar populations, and dust in the local universe at high resolution in the infrared with JWST.  The early publication of our first results, initial lessons-learned with data processing, and the challenges to be addressed as we expand analysis to the full PHANGS-JWST sample, is intended to serve as a pathfinder to facilitate forthcoming work.

\section*{Software}

\software{Python associated: Astropy (and its affiliated packages) \citep{astropy_collaboration_astropy_2018,astropy_collaboration_astropy_2022}. SAOImage DS9 \citep{smithsonian_astrophysical_observatory_saoimage_2000,joye_new_2003}. CIGALE \citep{cigale}.}

\section*{Acknowledgements}
This work is conducted as part of the PHANGS Collaboration\footnote{\url{http://phangs.org/}}.

This work is based on observations made with the NASA/ESA/CSA James Webb Space Telescope. The data were obtained from the Mikulski Archive for Space Telescopes at the Space Telescope Science Institute, which is operated by the Association of Universities for Research in Astronomy, Inc., under NASA contract NAS 5-03127 for JWST. These observations are associated with program 2107.

JCL acknowledges the W.M. Keck Institute for Space Studies (KISS) for providing collaboration space, and its support of workshops which helped to seed the work in this paper.  In particular, we have benefited from discussions at the KISS workshops: ``Bridging the Gap: Observations and Theory of Star Formation Meet on Large and Small Scales'' in 2014 and ``Star Formation in Nearby Galaxies with JWST'' in 2017.  

This paper makes use of the following ALMA data: ADS/JAO.ALMA\#2012.1.00650.S, 
ADS/JAO.ALMA\#2015.1.00956.S, 
ADS/JAO.ALMA\#2017.1.00886.L, 
ADS/JAO.ALMA\#2018.1.01651.S. 
ALMA is a partnership of ESO (representing its member states), NSF (USA) and NINS (Japan), together with NRC (Canada), MOST and ASIAA (Taiwan), and KASI (Republic of Korea), in cooperation with the Republic of Chile. The Joint ALMA Observatory is operated by ESO, AUI/NRAO and NAOJ.

Based on observations collected at the European Southern Observatory under ESO programmes 094.C-0623 (PI: Kreckel), 095.C-0473,  098.C-0484 (PI: Blanc), 1100.B-0651 (PHANGS-MUSE; PI: Schinnerer), as well as 094.B-0321 (MAGNUM; PI: Marconi), 099.B-0242, 0100.B-0116, 098.B-0551 (MAD; PI: Carollo) and 097.B-0640 (TIMER; PI: Gadotti).

MB acknowledges support from FONDECYT regular grant 1211000 and by the ANID BASAL project FB210003.
G.A.B. acknowledges the support from ANID Basal project FB210003.
E.C. acknowledges support from ANID Basal projects ACE210002 and FB210003.
KG is supported by the Australian Research Council through the Discovery Early Career Researcher Award (DECRA) Fellowship DE220100766 funded by the Australian Government. 

MC gratefully acknowledges funding from the DFG through an Emmy Noether Research Group (grant number CH2137/1-1).
COOL Research DAO is a Decentralized Autonomous Organization supporting research in astrophysics aimed at uncovering our cosmic origins.
JMDK gratefully acknowledges funding from the ERC under the European Union's Horizon 2020 research and innovation programme via the ERC Starting Grant MUSTANG (grant agreement number 714907).

ADB acknowledges support by NSF-AST2108140 and award JWST-GO-02107.008-A.

ATB, IB, JdB, FB would like to acknowledge funding from the European Research Council (ERC) under the European Union’s Horizon 2020 research and innovation programme (grant agreement No.726384/Empire)

CE gratefully acknowledges funding from the Deutsche Forschungsgemeinschaft (DFG) Sachbeihilfe, grant number BI1546/3-1. 

RSK, SCOG,  MS and EJW acknowledge the funding provided by DFG in SFB 881 ``The Milky Way System'', subprojects A1, B1, B2,  B8, and P1 -- funding-ID 138713538). RSK is thankful for  support from the ERC via the ERC Synergy Grant ``ECOGAL'' (project ID 855130), from the Heidelberg Cluster of Excellence (EXC 2181 - 390900948) ``STRUCTURES'', funded by the German Excellence Strategy, and from the German Ministry for Economic Affairs and Climate Action in project ``MAINN'' (funding ID 50OO2206).

JK gratefully acknowledges funding from the Deutsche Forschungsgemeinschaft (DFG, German Research Foundation) through the DFG Sachbeihilfe (grant number KR4801/2-1).

KK, OE, JL and FS gratefully acknowledge funding from the Deutsche Forschungsgemeinschaft (DFG, German Research Foundation) in the form of an Emmy Noether Research Group (grant number KR4598/2-1, PI Kreckel). 

JS acknowledges support by the Natural Sciences and Engineering Research Council of Canada (NSERC) through a Canadian Institute for Theoretical Astrophysics (CITA) National Fellowship.

SD is supported by funding from the European Research Council (ERC) under the European Union’s Horizon 2020 research and innovation programme (grant agreement no. 101018897 CosmicExplorer).

HAP acknowledges support by the National Science and Technology Council of Taiwan under grant 110-2112-M-032-020-MY3.

\bibliographystyle{aasjournal} 
\bibliography{all,phangsjwst} 

\begin{thebibliography}{}
\expandafter\ifx\csname natexlab\endcsname\relax\def\natexlab#1{#1}\fi
\providecommand{\url}[1]{\href{#1}{#1}}
\providecommand{\dodoi}[1]{doi:~\href{http://doi.org/#1}{\nolinkurl{#1}}}
\providecommand{\doeprint}[1]{\href{http://ascl.net/#1}{\nolinkurl{http://ascl.net/#1}}}
\providecommand{\doarXiv}[1]{\href{https://arxiv.org/abs/#1}{\nolinkurl{https://arxiv.org/abs/#1}}}

\bibitem[{{Adamo} {et~al.}(2015){Adamo}, {Kruijssen}, {Bastian}, {Silva-Villa},
  \& {Ryon}}]{adamo15}
{Adamo}, A., {Kruijssen}, J.~M.~D., {Bastian}, N., {Silva-Villa}, E., \&
  {Ryon}, J. 2015, \mnras, 452, 246, \dodoi{10.1093/mnras/stv1203}

\bibitem[{{Adamo} {et~al.}(2020{\natexlab{a}}){Adamo}, {Hollyhead}, {Messa},
  {Ryon}, {Bajaj}, {Runnholm}, {Aalto}, {Calzetti}, {Gallagher}, {Hayes},
  {Kruijssen}, {K{\"o}nig}, {Larsen}, {Melinder}, {Sabbi}, {Smith}, \&
  {{\"O}stlin}}]{adamo20}
{Adamo}, A., {Hollyhead}, K., {Messa}, M., {et~al.} 2020{\natexlab{a}}, \mnras,
  499, 3267, \dodoi{10.1093/mnras/staa2380}

\bibitem[{{Adamo} {et~al.}(2020{\natexlab{b}}){Adamo}, {Zeidler}, {Kruijssen},
  {Chevance}, {Gieles}, {Calzetti}, {Charbonnel}, {Zinnecker}, \&
  {Krause}}]{adamo20b}
{Adamo}, A., {Zeidler}, P., {Kruijssen}, J.~M.~D., {et~al.} 2020{\natexlab{b}},
  \ssr, 216, 69, \dodoi{10.1007/s11214-020-00690-x}

\bibitem[{{Anand} {et~al.}(2021{\natexlab{a}}){Anand}, {Lee}, {Van Dyk},
  {Leroy}, {Rosolowsky}, {Schinnerer}, {Larson}, {Kourkchi}, {Kreckel},
  {Scheuermann}, {Rizzi}, {Thilker}, {Tully}, {Bigiel}, {Blanc}, {Boquien},
  {Chandar}, {Dale}, {Emsellem}, {Deger}, {Glover}, {Grasha}, {Groves}, {S.
  Klessen}, {Kruijssen}, {Querejeta}, {S{\'a}nchez-Bl{\'a}zquez}, {Schruba},
  {Turner}, {Ubeda}, {Williams}, \& {Whitmore}}]{anand21}
{Anand}, G.~S., {Lee}, J.~C., {Van Dyk}, S.~D., {et~al.} 2021{\natexlab{a}},
  \mnras, 501, 3621, \dodoi{10.1093/mnras/staa3668}

\bibitem[{{Anand} {et~al.}(2021{\natexlab{b}}){Anand}, {Rizzi}, {Tully},
  {Shaya}, {Karachentsev}, {Makarov}, {Makarova}, {Wu}, {Dolphin}, \&
  {Kourkchi}}]{edd21}
{Anand}, G.~S., {Rizzi}, L., {Tully}, R.~B., {et~al.} 2021{\natexlab{b}}, \aj,
  162, 80, \dodoi{10.3847/1538-3881/ac0440}

\bibitem[{{Aniano} {et~al.}(2011){Aniano}, {Draine}, {Gordon}, \&
  {Sandstrom}}]{aniano2011}
{Aniano}, G., {Draine}, B.~T., {Gordon}, K.~D., \& {Sandstrom}, K. 2011, \pasp,
  123, 1218, \dodoi{10.1086/662219}

\bibitem[{{Aniano} {et~al.}(2020){Aniano}, {Draine}, {Hunt}, {Sandstrom},
  {Calzetti}, {Kennicutt}, {Dale}, {Galametz}, {Gordon}, {Leroy}, {Smith},
  {Roussel}, {Sauvage}, {Walter}, {Armus}, {Bolatto}, {Boquien}, {Crocker}, {De
  Looze}, {Donovan Meyer}, {Helou}, {Hinz}, {Johnson}, {Koda}, {Miller},
  {Montiel}, {Murphy}, {Rela{\~n}o}, {Rix}, {Schinnerer}, {Skibba}, {Wolfire},
  \& {Engelbracht}}]{aniano20}
{Aniano}, G., {Draine}, B.~T., {Hunt}, L.~K., {et~al.} 2020, \apj, 889, 150,
  \dodoi{10.3847/1538-4357/ab5fdb}

\bibitem[{{Armus} {et~al.}(2009){Armus}, {Mazzarella}, {Evans}, {Surace},
  {Sanders}, {Iwasawa}, {Frayer}, {Howell}, {Chan}, {Petric}, {Vavilkin},
  {Kim}, {Haan}, {Inami}, {Murphy}, {Appleton}, {Barnes}, {Bothun}, {Bridge},
  {Charmandaris}, {Jensen}, {Kewley}, {Lord}, {Madore}, {Marshall},
  {Melbourne}, {Rich}, {Satyapal}, {Schulz}, {Spoon}, {Sturm}, {U}, {Veilleux},
  \& {Xu}}]{armus09}
{Armus}, L., {Mazzarella}, J.~M., {Evans}, A.~S., {et~al.} 2009, \pasp, 121,
  559, \dodoi{10.1086/600092}

\bibitem[{{Astropy Collaboration} {et~al.}(2018){Astropy Collaboration},
  {Price-Whelan}, Sip{\H o}cz, G{\"u}nther, Lim, Crawford, Conseil, Shupe,
  Craig, Dencheva, Ginsburg, VanderPlas, Bradley, {P{\'e}rez-Su{\'a}rez}, {de
  Val-Borro}, Aldcroft, Cruz, Robitaille, Tollerud, Ardelean, Babej, Bach,
  Bachetti, Bakanov, Bamford, Barentsen, Barmby, Baumbach, Berry, Biscani,
  Boquien, Bostroem, Bouma, Brammer, Bray, Breytenbach, Buddelmeijer, Burke,
  Calderone, Cano~Rodr{\'i}guez, Cara, Cardoso, Cheedella, Copin, Corrales,
  Crichton, D'Avella, Deil, Depagne, Dietrich, Donath, Droettboom, Earl, Erben,
  Fabbro, Ferreira, Finethy, Fox, Garrison, Gibbons, Goldstein, Gommers, Greco,
  Greenfield, Groener, Grollier, Hagen, Hirst, Homeier, Horton, Hosseinzadeh,
  Hu, Hunkeler, Ivezi{\'c}, Jain, Jenness, Kanarek, Kendrew, Kern, Kerzendorf,
  Khvalko, King, Kirkby, Kulkarni, Kumar, Lee, Lenz, Littlefair, Ma, Macleod,
  Mastropietro, McCully, Montagnac, Morris, Mueller, Mumford, Muna, Murphy,
  Nelson, Nguyen, Ninan, N{\"o}the, Ogaz, Oh, Parejko, Parley, Pascual, Patil,
  Patil, Plunkett, Prochaska, Rastogi, Reddy~Janga, Sabater, Sakurikar,
  Seifert, Sherbert, {Sherwood-Taylor}, Shih, Sick, Silbiger, Singanamalla,
  Singer, Sladen, Sooley, Sornarajah, Streicher, Teuben, Thomas, Tremblay,
  Turner, Terr{\'o}n, {van Kerkwijk}, {de la Vega}, Watkins, Weaver, Whitmore,
  Woillez, Zabalza, \& {Astropy
  Contributors}}]{astropy_collaboration_astropy_2018}
{Astropy Collaboration}, {Price-Whelan}, A.~M., Sip{\H o}cz, B.~M., {et~al.}
  2018, The Astronomical Journal, 156, 123, \dodoi{10.3847/1538-3881/aabc4f}

\bibitem[{{Astropy Collaboration} {et~al.}(2022){Astropy Collaboration},
  {Price-Whelan}, Lim, Earl, Starkman, Bradley, Shupe, Patil, Corrales,
  Brasseur, N{\"o}the, Donath, Tollerud, Morris, Ginsburg, Vaher, Weaver,
  Tocknell, Jamieson, {van Kerkwijk}, Robitaille, Merry, Bachetti, G{\"u}nther,
  Aldcroft, {Alvarado-Montes}, Archibald, B{\'o}di, Bapat, Barentsen,
  Baz{\'a}n, Biswas, Boquien, Burke, Cara, Cara, Conroy, Conseil, Craig, Cross,
  Cruz, D'Eugenio, Dencheva, Devillepoix, Dietrich, Eigenbrot, Erben, Ferreira,
  {Foreman-Mackey}, Fox, Freij, Garg, Geda, Glattly, Gondhalekar, Gordon,
  Grant, Greenfield, Groener, Guest, Gurovich, Handberg, Hart,
  {Hatfield-Dodds}, Homeier, Hosseinzadeh, Jenness, Jones, Joseph, Kalmbach,
  Karamehmetoglu, Ka{\l}uszy{\'n}ski, Kelley, Kern, Kerzendorf, Koch, Kulumani,
  Lee, Ly, Ma, MacBride, Maljaars, Muna, Murphy, Norman, O'Steen, Oman,
  Pacifici, Pascual, {Pascual-Granado}, Patil, Perren, Pickering, Rastogi,
  Roulston, Ryan, Rykoff, Sabater, Sakurikar, Salgado, Sanghi, Saunders,
  Savchenko, Schwardt, {Seifert-Eckert}, Shih, Jain, Shukla, Sick, Simpson,
  Singanamalla, Singer, Singhal, Sinha, Sip{\H o}cz, Spitler, Stansby,
  Streicher, {\v S}umak, Swinbank, Taranu, Tewary, Tremblay, de~{Val-Borro},
  Van~Kooten, Vasovi{\'c}, Verma, {de Miranda Cardoso}, Williams, Wilson,
  Winkel, {Wood-Vasey}, Xue, Yoachim, Zhang, Zonca, \& {Astropy Project
  Contributors}}]{astropy_collaboration_astropy_2022}
{Astropy Collaboration}, {Price-Whelan}, A.~M., Lim, P.~L., {et~al.} 2022, The
  Astrophysical Journal, 935, 167, \dodoi{10.3847/1538-4357/ac7c74}

\bibitem[{{Barnes} {et~al.}(subm.){Barnes}, {Watkins},
  {et~al.}}]{BARNES_PHANGSJWST}
{Barnes}, A., {Watkins}, E., {et~al.} subm., \apjl

\bibitem[{{Barnes} {et~al.}(2017){Barnes}, {Longmore}, {Battersby}, {Bally},
  {Kruijssen}, {Henshaw}, \& {Walker}}]{barnes17}
{Barnes}, A.~T., {Longmore}, S.~N., {Battersby}, C., {et~al.} 2017, \mnras,
  469, 2263, \dodoi{10.1093/mnras/stx941}

\bibitem[{{Becklin} \& {Neugebauer}(1967)}]{becklin67}
{Becklin}, E.~E., \& {Neugebauer}, G. 1967, \apj, 147, 799,
  \dodoi{10.1086/149055}

\bibitem[{{Beuther} {et~al.}(2017){Beuther}, {Meidt}, {Schinnerer}, {Paladino},
  \& {Leroy}}]{beuther17}
{Beuther}, H., {Meidt}, S., {Schinnerer}, E., {Paladino}, R., \& {Leroy}, A.
  2017, \aap, 597, A85, \dodoi{10.1051/0004-6361/201526749}

\bibitem[{{Boquien} {et~al.}(2019{\natexlab{a}}){Boquien}, {Burgarella},
  {Roehlly}, {Buat}, {Ciesla}, {Corre}, {Inoue}, \& {Salas}}]{boquien19}
{Boquien}, M., {Burgarella}, D., {Roehlly}, Y., {et~al.} 2019{\natexlab{a}},
  \aap, 622, A103, \dodoi{10.1051/0004-6361/201834156}

\bibitem[{{Boquien} {et~al.}(2019{\natexlab{b}}){Boquien}, {Burgarella},
  {Roehlly}, {Buat}, {Ciesla}, {Corre}, {Inoue}, \& {Salas}}]{cigale}
---. 2019{\natexlab{b}}, \aap, 622, A103, \dodoi{10.1051/0004-6361/201834156}

\bibitem[{{Boquien} {et~al.}(2015){Boquien}, {Calzetti}, {Aalto}, {Boselli},
  {Braine}, {Buat}, {Combes}, {Israel}, {Kramer}, {Lord}, {Rela{\~n}o},
  {Rosolowsky}, {Stacey}, {Tabatabaei}, {van der Tak}, {van der Werf},
  {Verley}, \& {Xilouris}}]{boquien15}
{Boquien}, M., {Calzetti}, D., {Aalto}, S., {et~al.} 2015, \aap, 578, A8,
  \dodoi{10.1051/0004-6361/201423518}

\bibitem[{{Boyer} {et~al.}(2022){Boyer}, {Anderson}, {Gennaro}, {Geha},
  {Wingfield McQuinn}, {Tollerud}, {Correnti}, {Brenner Newman}, {Cohen},
  {Kallivayalil}, {Beaton}, {Cole}, {Dolphin}, {Kalirai}, {Sandstrom},
  {Savino}, {Skillman}, {Weisz}, \& {Williams}}]{2022Boyer}
{Boyer}, M.~L., {Anderson}, J., {Gennaro}, M., {et~al.} 2022, arXiv e-prints,
  arXiv:2209.03348.
\newblock \doarXiv{2209.03348}

\bibitem[{{Brinchmann} {et~al.}(2004){Brinchmann}, {Charlot}, {White},
  {Tremonti}, {Kauffmann}, {Heckman}, \& {Brinkmann}}]{brinchmann04}
{Brinchmann}, J., {Charlot}, S., {White}, S.~D.~M., {et~al.} 2004, \mnras, 351,
  1151, \dodoi{10.1111/j.1365-2966.2004.07881.x}

\bibitem[{{Budav{\'a}ri} {et~al.}(2009){Budav{\'a}ri}, {Wild}, {Szalay},
  {Dobos}, \& {Yip}}]{2009Budavari}
{Budav{\'a}ri}, T., {Wild}, V., {Szalay}, A.~S., {Dobos}, L., \& {Yip}, C.-W.
  2009, \mnras, 394, 1496, \dodoi{10.1111/j.1365-2966.2009.14415.x}

\bibitem[{{Buta} {et~al.}(2015){Buta}, {Sheth}, {Athanassoula}, {Bosma},
  {Knapen}, {Laurikainen}, {Salo}, {Elmegreen}, {Ho}, {Zaritsky}, {Courtois},
  {Hinz}, {Mu{\~n}oz-Mateos}, {Kim}, {Regan}, {Gadotti}, {Gil de Paz}, {Laine},
  {Men{\'e}ndez-Delmestre}, {Comer{\'o}n}, {Erroz Ferrer}, {Seibert},
  {Mizusawa}, {Holwerda}, \& {Madore}}]{buta15}
{Buta}, R.~J., {Sheth}, K., {Athanassoula}, E., {et~al.} 2015, \apjs, 217, 32,
  \dodoi{10.1088/0067-0049/217/2/32}

\bibitem[{Butterworth(1930)}]{Butterworth1930}
Butterworth, S. 1930, Experimental Wireless \& the Wireless Engineer, 7, 536

\bibitem[{{Calzetti} {et~al.}(2007){Calzetti}, {Kennicutt}, {Engelbracht},
  {Leitherer}, {Draine}, {Kewley}, {Moustakas}, {Sosey}, {Dale}, {Gordon},
  {Helou}, {Hollenbach}, {Armus}, {Bendo}, {Bot}, {Buckalew}, {Jarrett}, {Li},
  {Meyer}, {Murphy}, {Prescott}, {Regan}, {Rieke}, {Roussel}, {Sheth}, {Smith},
  {Thornley}, \& {Walter}}]{calzetti07}
{Calzetti}, D., {Kennicutt}, R.~C., {Engelbracht}, C.~W., {et~al.} 2007, \apj,
  666, 870, \dodoi{10.1086/520082}

\bibitem[{{Chandar} {et~al.}(2017){Chandar}, {Fall}, {Whitmore}, \&
  {Mulia}}]{chandar17}
{Chandar}, R., {Fall}, S.~M., {Whitmore}, B.~C., \& {Mulia}, A.~J. 2017, \apj,
  849, 128, \dodoi{10.3847/1538-4357/aa92ce}

\bibitem[{{Chastenet} {et~al.}(2019){Chastenet}, {Sandstrom}, {Chiang},
  {Leroy}, {Utomo}, {Bot}, {Gordon}, {Draine}, {Fukui}, {Onishi}, \&
  {Tsuge}}]{chastenet19}
{Chastenet}, J., {Sandstrom}, K., {Chiang}, I.~D., {et~al.} 2019, \apj, 876,
  62, \dodoi{10.3847/1538-4357/ab16cf}

\bibitem[{{Chastenet} {et~al.}(subm.{\natexlab{a}})}]{CHASTENET1_PHANGSJWST}
{Chastenet}, J., {et~al.} subm.{\natexlab{a}}, \apjl

\bibitem[{{Chastenet} {et~al.}(subm.{\natexlab{b}})}]{CHASTENET2_PHANGSJWST}
---. subm.{\natexlab{b}}, \apjl

\bibitem[{{Chevance} {et~al.}(2022{\natexlab{a}}){Chevance}, {Krumholz},
  {McLeod}, {Ostriker}, {Rosolowsky}, \& {Sternberg}}]{chevance22b}
{Chevance}, M., {Krumholz}, M.~R., {McLeod}, A.~F., {et~al.}
  2022{\natexlab{a}}, arXiv e-prints, arXiv:2203.09570.
\newblock \doarXiv{2203.09570}

\bibitem[{{Chevance} {et~al.}(2020{\natexlab{a}}){Chevance}, {Kruijssen},
  {Hygate}, {Schruba}, {Longmore}, {Groves}, {Henshaw}, {Herrera}, {Hughes},
  {Jeffreson}, {Lang}, {Leroy}, {Meidt}, {Pety}, {Razza}, {Rosolowsky},
  {Schinnerer}, {Bigiel}, {Blanc}, {Emsellem}, {Faesi}, {Glover}, {Haydon},
  {Ho}, {Kreckel}, {Lee}, {Liu}, {Querejeta}, {Saito}, {Sun}, {Usero}, \&
  {Utomo}}]{chevance20}
{Chevance}, M., {Kruijssen}, J.~M.~D., {Hygate}, A. P.~S., {et~al.}
  2020{\natexlab{a}}, \mnras, 493, 2872, \dodoi{10.1093/mnras/stz3525}

\bibitem[{{Chevance} {et~al.}(2020{\natexlab{b}}){Chevance}, {Kruijssen},
  {Vazquez-Semadeni}, {Nakamura}, {Klessen}, {Ballesteros-Paredes}, {Inutsuka},
  {Adamo}, \& {Hennebelle}}]{chevance20b}
{Chevance}, M., {Kruijssen}, J.~M.~D., {Vazquez-Semadeni}, E., {et~al.}
  2020{\natexlab{b}}, \ssr, 216, 50, \dodoi{10.1007/s11214-020-00674-x}

\bibitem[{{Chevance} {et~al.}(2022{\natexlab{b}}){Chevance}, {Kruijssen},
  {Krumholz}, {Groves}, {Keller}, {Hughes}, {Glover}, {Henshaw}, {Herrera},
  {Kim}, {Leroy}, {Pety}, {Razza}, {Rosolowsky}, {Schinnerer}, {Schruba},
  {Barnes}, {Bigiel}, {Blanc}, {Dale}, {Emsellem}, {Faesi}, {Grasha},
  {Klessen}, {Kreckel}, {Liu}, {Longmore}, {Meidt}, {Querejeta}, {Saito},
  {Sun}, \& {Usero}}]{chevance22}
{Chevance}, M., {Kruijssen}, J.~M.~D., {Krumholz}, M.~R., {et~al.}
  2022{\natexlab{b}}, \mnras, 509, 272, \dodoi{10.1093/mnras/stab2938}

\bibitem[{{Corbelli} {et~al.}(2017){Corbelli}, {Braine}, {Bandiera},
  {Brouillet}, {Combes}, {Druard}, {Gratier}, {Mata}, {Schuster}, {Xilouris},
  \& {Palla}}]{corbelli17}
{Corbelli}, E., {Braine}, J., {Bandiera}, R., {et~al.} 2017, \aap, 601, A146,
  \dodoi{10.1051/0004-6361/201630034}

\bibitem[{{Dale} {et~al.}(subm.)}]{DALE_PHANGSJWST}
{Dale}, D., {et~al.} subm., \apjl

\bibitem[{{Dale} {et~al.}(2005){Dale}, {Bendo}, {Engelbracht}, {Gordon},
  {Regan}, {Armus}, {Cannon}, {Calzetti}, {Draine}, {Helou}, {Joseph},
  {Kennicutt}, {Li}, {Murphy}, {Roussel}, {Walter}, {Hanson}, {Hollenbach},
  {Jarrett}, {Kewley}, {Lamanna}, {Leitherer}, {Meyer}, {Rieke}, {Rieke},
  {Sheth}, {Smith}, \& {Thornley}}]{dale05}
{Dale}, D.~A., {Bendo}, G.~J., {Engelbracht}, C.~W., {et~al.} 2005, \apj, 633,
  857, \dodoi{10.1086/491642}

\bibitem[{{Dale} {et~al.}(2009){Dale}, {Cohen}, {Johnson}, {Schuster},
  {Calzetti}, {Engelbracht}, {Gil de Paz}, {Kennicutt}, {Lee}, {Begum},
  {Block}, {Dalcanton}, {Funes}, {Gordon}, {Johnson}, {Marble}, {Sakai},
  {Skillman}, {van Zee}, {Walter}, {Weisz}, {Williams}, {Wu}, \& {Wu}}]{dale09}
{Dale}, D.~A., {Cohen}, S.~A., {Johnson}, L.~C., {et~al.} 2009, \apj, 703, 517,
  \dodoi{10.1088/0004-637X/703/1/517}

\bibitem[{{de Vaucouleurs} {et~al.}(1991){de Vaucouleurs}, {de Vaucouleurs},
  {Corwin}, {Buta}, {Paturel}, \& {Fouqu{\'e}}}]{rc3}
{de Vaucouleurs}, G., {de Vaucouleurs}, A., {Corwin}, Jr., H.~G., {et~al.}
  1991, {Third Reference Catalogue of Bright Galaxies. Volume I: Explanations
  and references. Volume II: Data for galaxies between 0$^{h}$ and 12$^{h}$.
  Volume III: Data for galaxies between 12$^{h}$ and 24$^{h}$.}

\bibitem[{{Deger} {et~al.}(2022){Deger}, {Lee}, {Whitmore}, {Thilker},
  {Boquien}, {Chandar}, {Dale}, {Ubeda}, {White}, {Grasha}, {Glover},
  {Schruba}, {Barnes}, {Klessen}, {Kruijssen}, {Rosolowsky}, \&
  {Williams}}]{deger22}
{Deger}, S., {Lee}, J.~C., {Whitmore}, B.~C., {et~al.} 2022, \mnras, 510, 32,
  \dodoi{10.1093/mnras/stab3213}

\bibitem[{{Dolphin}(2016)}]{dolphot}
{Dolphin}, A. 2016, {DOLPHOT: Stellar photometry}, Astrophysics Source Code
  Library, record ascl:1608.013.
\newblock \doeprint{1608.013}

\bibitem[{{Draine} {et~al.}(2021){Draine}, {Li}, {Hensley}, {Hunt},
  {Sandstrom}, \& {Smith}}]{draine21}
{Draine}, B.~T., {Li}, A., {Hensley}, B.~S., {et~al.} 2021, \apj, 917, 3,
  \dodoi{10.3847/1538-4357/abff51}

\bibitem[{{Draine} {et~al.}(2007){Draine}, {Dale}, {Bendo}, {Gordon}, {Smith},
  {Armus}, {Engelbracht}, {Helou}, {Kennicutt}, {Li}, {Roussel}, {Walter},
  {Calzetti}, {Moustakas}, {Murphy}, {Rieke}, {Bot}, {Hollenbach}, {Sheth}, \&
  {Teplitz}}]{draine07}
{Draine}, B.~T., {Dale}, D.~A., {Bendo}, G., {et~al.} 2007, \apj, 663, 866,
  \dodoi{10.1086/518306}

\bibitem[{{Egorov} {et~al.}(subm.)}]{EGOROV_PHANGSJWST}
{Egorov}, O., {et~al.} subm., \apjl

\bibitem[{{Emsellem} {et~al.}(2022){Emsellem}, {Schinnerer}, {Santoro},
  {Belfiore}, {Pessa}, {McElroy}, {Blanc}, {Congiu}, {Groves}, {Ho}, {Kreckel},
  {Razza}, {Sanchez-Blazquez}, {Egorov}, {Faesi}, {Klessen}, {Leroy}, {Meidt},
  {Querejeta}, {Rosolowsky}, {Scheuermann}, {Anand}, {Barnes},
  {Be{\v{s}}li{\'c}}, {Bigiel}, {Boquien}, {Cao}, {Chevance}, {Dale},
  {Eibensteiner}, {Glover}, {Grasha}, {Henshaw}, {Hughes}, {Koch}, {Kruijssen},
  {Lee}, {Liu}, {Pan}, {Pety}, {Saito}, {Sandstrom}, {Schruba}, {Sun},
  {Thilker}, {Usero}, {Watkins}, \& {Williams}}]{phangs-muse}
{Emsellem}, E., {Schinnerer}, E., {Santoro}, F., {et~al.} 2022, \aap, 659,
  A191, \dodoi{10.1051/0004-6361/202141727}

\bibitem[{{Engelbracht} {et~al.}(2005){Engelbracht}, {Gordon}, {Rieke},
  {Werner}, {Dale}, \& {Latter}}]{engelbracht05}
{Engelbracht}, C.~W., {Gordon}, K.~D., {Rieke}, G.~H., {et~al.} 2005, \apjl,
  628, L29, \dodoi{10.1086/432613}

\bibitem[{{Federrath} {et~al.}(2016){Federrath}, {Rathborne}, {Longmore},
  {Kruijssen}, {Bally}, {Contreras}, {Crocker}, {Garay}, {Jackson}, {Testi}, \&
  {Walsh}}]{federrath16}
{Federrath}, C., {Rathborne}, J.~M., {Longmore}, S.~N., {et~al.} 2016, \apj,
  832, 143, \dodoi{10.3847/0004-637X/832/2/143}

\bibitem[{{Freedman} {et~al.}(2001){Freedman}, {Madore}, {Gibson}, {Ferrarese},
  {Kelson}, {Sakai}, {Mould}, {Kennicutt}, {Ford}, {Graham}, {Huchra},
  {Hughes}, {Illingworth}, {Macri}, \& {Stetson}}]{freedman2001}
{Freedman}, W.~L., {Madore}, B.~F., {Gibson}, B.~K., {et~al.} 2001, \apj, 553,
  47, \dodoi{10.1086/320638}

\bibitem[{{Gaia Collaboration} {et~al.}(2016){Gaia Collaboration}, {Prusti},
  {de Bruijne}, {Brown}, {Vallenari}, {Babusiaux}, {Bailer-Jones}, {Bastian},
  {Biermann}, {Evans}, {Eyer}, {Jansen}, {Jordi}, {Klioner}, {Lammers},
  {Lindegren}, {Luri}, {Mignard}, {Milligan}, {Panem}, {Poinsignon},
  {Pourbaix}, {Randich}, {Sarri}, {Sartoretti}, {Siddiqui}, {Soubiran},
  {Valette}, {van Leeuwen}, {Walton}, {Aerts}, {Arenou}, {Cropper}, {Drimmel},
  {H{\o}g}, {Katz}, {Lattanzi}, {O'Mullane}, {Grebel}, {Holland}, {Huc},
  {Passot}, {Bramante}, {Cacciari}, {Casta{\~n}eda}, {Chaoul}, {Cheek}, {De
  Angeli}, {Fabricius}, {Guerra}, {Hern{\'a}ndez}, {Jean-Antoine-Piccolo},
  {Masana}, {Messineo}, {Mowlavi}, {Nienartowicz}, {Ord{\'o}{\~n}ez-Blanco},
  {Panuzzo}, {Portell}, {Richards}, {Riello}, {Seabroke}, {Tanga},
  {Th{\'e}venin}, {Torra}, {Els}, {Gracia-Abril}, {Comoretto},
  {Garcia-Reinaldos}, {Lock}, {Mercier}, {Altmann}, {Andrae}, {Astraatmadja},
  {Bellas-Velidis}, {Benson}, {Berthier}, {Blomme}, {Busso}, {Carry},
  {Cellino}, {Clementini}, {Cowell}, {Creevey}, {Cuypers}, {Davidson}, {De
  Ridder}, {de Torres}, {Delchambre}, {Dell'Oro}, {Ducourant}, {Fr{\'e}mat},
  {Garc{\'\i}a-Torres}, {Gosset}, {Halbwachs}, {Hambly}, {Harrison}, {Hauser},
  {Hestroffer}, {Hodgkin}, {Huckle}, {Hutton}, {Jasniewicz}, {Jordan},
  {Kontizas}, {Korn}, {Lanzafame}, {Manteiga}, {Moitinho}, {Muinonen},
  {Osinde}, {Pancino}, {Pauwels}, {Petit}, {Recio-Blanco}, {Robin}, {Sarro},
  {Siopis}, {Smith}, {Smith}, {Sozzetti}, {Thuillot}, {van Reeven}, {Viala},
  {Abbas}, {Abreu Aramburu}, {Accart}, {Aguado}, {Allan}, {Allasia},
  {Altavilla}, {{\'A}lvarez}, {Alves}, {Anderson}, {Andrei}, {Anglada Varela},
  {Antiche}, {Antoja}, {Ant{\'o}n}, {Arcay}, {Atzei}, {Ayache}, {Bach},
  {Baker}, {Balaguer-N{\'u}{\~n}ez}, {Barache}, {Barata}, {Barbier}, {Barblan},
  {Baroni}, {Barrado y Navascu{\'e}s}, {Barros}, {Barstow}, {Becciani},
  {Bellazzini}, {Bellei}, {Bello Garc{\'\i}a}, {Belokurov}, {Bendjoya},
  {Berihuete}, {Bianchi}, {Bienaym{\'e}}, {Billebaud}, {Blagorodnova},
  {Blanco-Cuaresma}, {Boch}, {Bombrun}, {Borrachero}, {Bouquillon}, {Bourda},
  {Bouy}, {Bragaglia}, {Breddels}, {Brouillet}, {Br{\"u}semeister},
  {Bucciarelli}, {Budnik}, {Burgess}, {Burgon}, {Burlacu}, {Busonero}, {Buzzi},
  {Caffau}, {Cambras}, {Campbell}, {Cancelliere}, {Cantat-Gaudin}, {Carlucci},
  {Carrasco}, {Castellani}, {Charlot}, {Charnas}, {Charvet}, {Chassat},
  {Chiavassa}, {Clotet}, {Cocozza}, {Collins}, {Collins}, {Costigan}, {Crifo},
  {Cross}, {Crosta}, {Crowley}, {Dafonte}, {Damerdji}, {Dapergolas}, {David},
  {David}, {De Cat}, {de Felice}, {de Laverny}, {De Luise}, {De March}, {de
  Martino}, {de Souza}, {Debosscher}, {del Pozo}, {Delbo}, {Delgado},
  {Delgado}, {di Marco}, {Di Matteo}, {Diakite}, {Distefano}, {Dolding}, {Dos
  Anjos}, {Drazinos}, {Dur{\'a}n}, {Dzigan}, {Ecale}, {Edvardsson}, {Enke},
  {Erdmann}, {Escolar}, {Espina}, {Evans}, {Eynard Bontemps}, {Fabre},
  {Fabrizio}, {Faigler}, {Falc{\~a}o}, {Farr{\`a}s Casas}, {Faye}, {Federici},
  {Fedorets}, {Fern{\'a}ndez-Hern{\'a}ndez}, {Fernique}, {Fienga}, {Figueras},
  {Filippi}, {Findeisen}, {Fonti}, {Fouesneau}, {Fraile}, {Fraser}, {Fuchs},
  {Furnell}, {Gai}, {Galleti}, {Galluccio}, {Garabato}, {Garc{\'\i}a-Sedano},
  {Gar{\'e}}, {Garofalo}, {Garralda}, {Gavras}, {Gerssen}, {Geyer}, {Gilmore},
  {Girona}, {Giuffrida}, {Gomes}, {Gonz{\'a}lez-Marcos},
  {Gonz{\'a}lez-N{\'u}{\~n}ez}, {Gonz{\'a}lez-Vidal}, {Granvik}, {Guerrier},
  {Guillout}, {Guiraud}, {G{\'u}rpide}, {Guti{\'e}rrez-S{\'a}nchez}, {Guy},
  {Haigron}, {Hatzidimitriou}, {Haywood}, {Heiter}, {Helmi}, {Hobbs},
  {Hofmann}, {Holl}, {Holland}, {Hunt}, {Hypki}, {Icardi}, {Irwin}, {Jevardat
  de Fombelle}, {Jofr{\'e}}, {Jonker}, {Jorissen}, {Julbe}, {Karampelas},
  {Kochoska}, {Kohley}, {Kolenberg}, {Kontizas}, {Koposov}, {Kordopatis},
  {Koubsky}, {Kowalczyk}, {Krone-Martins}, {Kudryashova}, {Kull}, {Bachchan},
  {Lacoste-Seris}, {Lanza}, {Lavigne}, {Le Poncin-Lafitte}, {Lebreton},
  {Lebzelter}, {Leccia}, {Leclerc}, {Lecoeur-Taibi}, {Lemaitre}, {Lenhardt},
  {Leroux}, {Liao}, {Licata}, {Lindstr{\o}m}, {Lister}, {Livanou}, {Lobel},
  {L{\"o}ffler}, {L{\'o}pez}, {Lopez-Lozano}, {Lorenz}, {Loureiro},
  {MacDonald}, {Magalh{\~a}es Fernandes}, {Managau}, {Mann}, {Mantelet},
  {Marchal}, {Marchant}, {Marconi}, {Marie}, {Marinoni}, {Marrese},
  {Marschalk{\'o}}, {Marshall}, {Mart{\'\i}n-Fleitas}, {Martino}, {Mary},
  {Matijevi{\v{c}}}, {Mazeh}, {McMillan}, {Messina}, {Mestre}, {Michalik},
  {Millar}, {Miranda}, {Molina}, {Molinaro}, {Molinaro}, {Moln{\'a}r},
  {Moniez}, {Montegriffo}, {Monteiro}, {Mor}, {Mora}, {Morbidelli}, {Morel},
  {Morgenthaler}, {Morley}, {Morris}, {Mulone}, {Muraveva}, {Musella},
  {Narbonne}, {Nelemans}, {Nicastro}, {Noval}, {Ord{\'e}novic},
  {Ordieres-Mer{\'e}}, {Osborne}, {Pagani}, {Pagano}, {Pailler}, {Palacin},
  {Palaversa}, {Parsons}, {Paulsen}, {Pecoraro}, {Pedrosa}, {Pentik{\"a}inen},
  {Pereira}, {Pichon}, {Piersimoni}, {Pineau}, {Plachy}, {Plum}, {Poujoulet},
  {Pr{\v{s}}a}, {Pulone}, {Ragaini}, {Rago}, {Rambaux}, {Ramos-Lerate},
  {Ranalli}, {Rauw}, {Read}, {Regibo}, {Renk}, {Reyl{\'e}}, {Ribeiro},
  {Rimoldini}, {Ripepi}, {Riva}, {Rixon}, {Roelens}, {Romero-G{\'o}mez},
  {Rowell}, {Royer}, {Rudolph}, {Ruiz-Dern}, {Sadowski}, {Sagrist{\`a}
  Sell{\'e}s}, {Sahlmann}, {Salgado}, {Salguero}, {Sarasso}, {Savietto},
  {Schnorhk}, {Schultheis}, {Sciacca}, {Segol}, {Segovia}, {Segransan},
  {Serpell}, {Shih}, {Smareglia}, {Smart}, {Smith}, {Solano}, {Solitro},
  {Sordo}, {Soria Nieto}, {Souchay}, {Spagna}, {Spoto}, {Stampa}, {Steele},
  {Steidelm{\"u}ller}, {Stephenson}, {Stoev}, {Suess}, {S{\"u}veges}, {Surdej},
  {Szabados}, {Szegedi-Elek}, {Tapiador}, {Taris}, {Tauran}, {Taylor},
  {Teixeira}, {Terrett}, {Tingley}, {Trager}, {Turon}, {Ulla}, {Utrilla},
  {Valentini}, {van Elteren}, {Van Hemelryck}, {van Leeuwen}, {Varadi},
  {Vecchiato}, {Veljanoski}, {Via}, {Vicente}, {Vogt}, {Voss}, {Votruba},
  {Voutsinas}, {Walmsley}, {Weiler}, {Weingrill}, {Werner}, {Wevers},
  {Whitehead}, {Wyrzykowski}, {Yoldas}, {{\v{Z}}erjal}, {Zucker}, {Zurbach},
  {Zwitter}, {Alecu}, {Allen}, {Allende Prieto}, {Amorim},
  {Anglada-Escud{\'e}}, {Arsenijevic}, {Azaz}, {Balm}, {Beck}, {Bernstein},
  {Bigot}, {Bijaoui}, {Blasco}, {Bonfigli}, {Bono}, {Boudreault}, {Bressan},
  {Brown}, {Brunet}, {Bunclark}, {Buonanno}, {Butkevich}, {Carret}, {Carrion},
  {Chemin}, {Ch{\'e}reau}, {Corcione}, {Darmigny}, {de Boer}, {de Teodoro}, {de
  Zeeuw}, {Delle Luche}, {Domingues}, {Dubath}, {Fodor}, {Fr{\'e}zouls},
  {Fries}, {Fustes}, {Fyfe}, {Gallardo}, {Gallegos}, {Gardiol}, {Gebran},
  {Gomboc}, {G{\'o}mez}, {Grux}, {Gueguen}, {Heyrovsky}, {Hoar}, {Iannicola},
  {Isasi Parache}, {Janotto}, {Joliet}, {Jonckheere}, {Keil}, {Kim},
  {Klagyivik}, {Klar}, {Knude}, {Kochukhov}, {Kolka}, {Kos}, {Kutka}, {Lainey},
  {LeBouquin}, {Liu}, {Loreggia}, {Makarov}, {Marseille}, {Martayan},
  {Martinez-Rubi}, {Massart}, {Meynadier}, {Mignot}, {Munari}, {Nguyen},
  {Nordlander}, {Ocvirk}, {O'Flaherty}, {Olias Sanz}, {Ortiz}, {Osorio},
  {Oszkiewicz}, {Ouzounis}, {Palmer}, {Park}, {Pasquato}, {Peltzer}, {Peralta},
  {P{\'e}turaud}, {Pieniluoma}, {Pigozzi}, {Poels}, {Prat}, {Prod'homme},
  {Raison}, {Rebordao}, {Risquez}, {Rocca-Volmerange}, {Rosen}, {Ruiz-Fuertes},
  {Russo}, {Sembay}, {Serraller Vizcaino}, {Short}, {Siebert}, {Silva},
  {Sinachopoulos}, {Slezak}, {Soffel}, {Sosnowska}, {Strai{\v{z}}ys}, {ter
  Linden}, {Terrell}, {Theil}, {Tiede}, {Troisi}, {Tsalmantza}, {Tur},
  {Vaccari}, {Vachier}, {Valles}, {Van Hamme}, {Veltz}, {Virtanen}, {Wallut},
  {Wichmann}, {Wilkinson}, {Ziaeepour}, \& {Zschocke}}]{gaia16}
{Gaia Collaboration}, {Prusti}, T., {de Bruijne}, J.~H.~J., {et~al.} 2016,
  \aap, 595, A1, \dodoi{10.1051/0004-6361/201629272}

\bibitem[{{Gensior} {et~al.}(2020){Gensior}, {Kruijssen}, \&
  {Keller}}]{gensior20}
{Gensior}, J., {Kruijssen}, J.~M.~D., \& {Keller}, B.~W. 2020, \mnras, 495,
  199, \dodoi{10.1093/mnras/staa1184}

\bibitem[{{Ginsburg} {et~al.}(2012){Ginsburg}, {Bressert}, {Bally}, \&
  {Battersby}}]{ginsburg12}
{Ginsburg}, A., {Bressert}, E., {Bally}, J., \& {Battersby}, C. 2012, \apjl,
  758, L29, \dodoi{10.1088/2041-8205/758/2/L29}

\bibitem[{{Ginsburg} \& {Kruijssen}(2018)}]{ginsburg18}
{Ginsburg}, A., \& {Kruijssen}, J.~M.~D. 2018, \apjl, 864, L17,
  \dodoi{10.3847/2041-8213/aada89}

\bibitem[{{Grasha} {et~al.}(2018){Grasha}, {Calzetti}, {Bittle}, {Johnson},
  {Donovan Meyer}, {Kennicutt}, {Elmegreen}, {Adamo}, {Krumholz}, {Fumagalli},
  {Grebel}, {Gouliermis}, {Cook}, {Gallagher}, {Aloisi}, {Dale}, {Linden},
  {Sacchi}, {Thilker}, {Walterbos}, {Messa}, {Wofford}, \& {Smith}}]{grasha18}
{Grasha}, K., {Calzetti}, D., {Bittle}, L., {et~al.} 2018, \mnras, 481, 1016,
  \dodoi{10.1093/mnras/sty2154}

\bibitem[{{Grasha} {et~al.}(2019){Grasha}, {Calzetti}, {Adamo}, {Kennicutt},
  {Elmegreen}, {Messa}, {Dale}, {Fedorenko}, {Mahadevan}, {Grebel},
  {Fumagalli}, {Kim}, {Dobbs}, {Gouliermis}, {Ashworth}, {Gallagher}, {Smith},
  {Tosi}, {Whitmore}, {Schinnerer}, {Colombo}, {Hughes}, {Leroy}, \&
  {Meidt}}]{grasha19}
{Grasha}, K., {Calzetti}, D., {Adamo}, A., {et~al.} 2019, \mnras, 483, 4707,
  \dodoi{10.1093/mnras/sty3424}

\bibitem[{{Groves} {et~al.}(2012){Groves}, {Krause}, {Sandstrom}, {Schmiedeke},
  {Leroy}, {Linz}, {Kapala}, {Rix}, {Schinnerer}, {Tabatabaei}, {Walter}, \&
  {da Cunha}}]{groves12}
{Groves}, B., {Krause}, O., {Sandstrom}, K., {et~al.} 2012, \mnras, 426, 892,
  \dodoi{10.1111/j.1365-2966.2012.21696.x}

\bibitem[{{Groves} {et~al.}(subm.)}]{GROVES_HIICAT}
{Groves}, B., {et~al.} subm., \mnras

\bibitem[{{Grudi{\'c}} {et~al.}(2021){Grudi{\'c}}, {Kruijssen},
  {Faucher-Gigu{\`e}re}, {Hopkins}, {Ma}, {Quataert}, \&
  {Boylan-Kolchin}}]{grudic21}
{Grudi{\'c}}, M.~Y., {Kruijssen}, J.~M.~D., {Faucher-Gigu{\`e}re}, C.-A.,
  {et~al.} 2021, \mnras, 506, 3239, \dodoi{10.1093/mnras/stab1894}

\bibitem[{{Hacar} {et~al.}(2022){Hacar}, {Clark}, {Heitsch}, {Kainulainen},
  {Panopoulou}, {Seifried}, \& {Smith}}]{hacar22}
{Hacar}, A., {Clark}, S., {Heitsch}, F., {et~al.} 2022, arXiv e-prints,
  arXiv:2203.09562.
\newblock \doarXiv{2203.09562}

\bibitem[{{Hannon} {et~al.}(2019){Hannon}, {Lee}, {Whitmore}, {Chandar},
  {Adamo}, {Mobasher}, {Aloisi}, {Calzetti}, {Cignoni}, {Cook}, {Dale},
  {Deger}, {Della Bruna}, {Elmegreen}, {Gouliermis}, {Grasha}, {Grebel},
  {Herrero}, {Hunter}, {Johnson}, {Kennicutt}, {Kim}, {Sacchi}, {Smith},
  {Thilker}, {Turner}, {Walterbos}, \& {Wofford}}]{hannon19}
{Hannon}, S., {Lee}, J.~C., {Whitmore}, B.~C., {et~al.} 2019, \mnras, 490,
  4648, \dodoi{10.1093/mnras/stz2820}

\bibitem[{{Hannon} {et~al.}(2022){Hannon}, {Lee}, {Whitmore}, {Mobasher},
  {Thilker}, {Chandar}, {Adamo}, {Wofford}, {Orozco-Duarte}, {Calzetti}, {Della
  Bruna}, {Kreckel}, {Groves}, {Barnes}, {Boquien}, {Belfiore}, \&
  {Linden}}]{hannon22}
---. 2022, \mnras, 512, 1294, \dodoi{10.1093/mnras/stac550}

\bibitem[{{Hassani} {et~al.}(subm.)}]{HASSANI_PHANGSJWST}
{Hassani}, H., {et~al.} subm., \apjl

\bibitem[{{Haydon} {et~al.}(2020){Haydon}, {Kruijssen}, {Chevance}, {Hygate},
  {Krumholz}, {Schruba}, \& {Longmore}}]{haydon20}
{Haydon}, D.~T., {Kruijssen}, J.~M.~D., {Chevance}, M., {et~al.} 2020, \mnras,
  498, 235, \dodoi{10.1093/mnras/staa2430}

\bibitem[{{Henshaw} {et~al.}(2022){Henshaw}, {Barnes}, {Battersby}, {Ginsburg},
  {Sormani}, \& {Walker}}]{henshaw22}
{Henshaw}, J.~D., {Barnes}, A.~T., {Battersby}, C., {et~al.} 2022, arXiv
  e-prints, arXiv:2203.11223.
\newblock \doarXiv{2203.11223}

\bibitem[{{Hensley} \& {Draine}(2022)}]{hensley22}
{Hensley}, B.~S., \& {Draine}, B.~T. 2022, arXiv e-prints, arXiv:2208.12365.
\newblock \doarXiv{2208.12365}

\bibitem[{{Hoyer} {et~al.}(2022){Hoyer}, {Pinna}, {Kamlah}, {Nogueras-Lara},
  {Feldmeier-Krause}, {Neumayer}, {Sormani}, {Boquien}, {Emsellem}, {Seth},
  {Klessen}, {Williams}, {Schinnerer}, {Barnes}, {Leroy}, {Bonoli},
  {Kruijssen}, {Neumann}, {S{\'a}nchez-Bl{\'a}zquez}, {Dale}, {Watkins},
  {Thilker}, {Rosolowsky}, {Bigiel}, {Grasha}, {Egorov}, {Liu}, {Sandstrom},
  {Larson}, {Blanc}, \& {Hassani}}]{HOYER_PHANGSJWST}
{Hoyer}, N., {Pinna}, F., {Kamlah}, A. W.~H., {et~al.} 2022, arXiv e-prints,
  arXiv:2211.13997.
\newblock \doarXiv{2211.13997}

\bibitem[{Joye \& Mandel(2003)}]{joye_new_2003}
Joye, W.~A., \& Mandel, E. 2003, 295, 489

\bibitem[{{Kawamura} {et~al.}(2009){Kawamura}, {Mizuno}, {Minamidani},
  {Filipovi{\'c}}, {Staveley-Smith}, {Kim}, {Mizuno}, {Onishi}, {Mizuno}, \&
  {Fukui}}]{kawamura09}
{Kawamura}, A., {Mizuno}, Y., {Minamidani}, T., {et~al.} 2009, \apjs, 184, 1,
  \dodoi{10.1088/0067-0049/184/1/1}

\bibitem[{{Kennicutt} \& {Evans}(2012)}]{kennicutt12}
{Kennicutt}, R.~C., \& {Evans}, N.~J. 2012, \araa, 50, 531,
  \dodoi{10.1146/annurev-astro-081811-125610}

\bibitem[{{Kennicutt} {et~al.}(2003){Kennicutt}, {Armus}, {Bendo}, {Calzetti},
  {Dale}, {Draine}, {Engelbracht}, {Gordon}, {Grauer}, {Helou}, {Hollenbach},
  {Jarrett}, {Kewley}, {Leitherer}, {Li}, {Malhotra}, {Regan}, {Rieke},
  {Rieke}, {Roussel}, {Smith}, {Thornley}, \& {Walter}}]{kennicutt03}
{Kennicutt}, Jr., R.~C., {Armus}, L., {Bendo}, G., {et~al.} 2003, \pasp, 115,
  928, \dodoi{10.1086/376941}

\bibitem[{{Kessler} {et~al.}(2003){Kessler}, {Mueller}, {Leech}, {Arviset},
  {Garcia-Lario}, {Metcalfe}, {Pollock}, {Prusti}, \& {Salama}}]{iso}
{Kessler}, M.~F., {Mueller}, T.~G., {Leech}, K., {et~al.} 2003, {The ISO
  Handbook, Volume I - Mission \& Satellite Overview}

\bibitem[{{Kim} {et~al.}(2021{\natexlab{a}}){Kim}, {Chevance}, {Kruijssen},
  {Schruba}, {Sandstrom}, {Barnes}, {Bigiel}, {Blanc}, {Cao}, {Dale}, {Faesi},
  {Glover}, {Grasha}, {Groves}, {Herrera}, {Klessen}, {Kreckel}, {Lee},
  {Leroy}, {Pety}, {Querejeta}, {Schinnerer}, {Sun}, {Usero}, {Ward}, \&
  {Williams}}]{kim21}
{Kim}, J., {Chevance}, M., {Kruijssen}, J.~M.~D., {et~al.} 2021{\natexlab{a}},
  \mnras, 504, 487, \dodoi{10.1093/mnras/stab878}

\bibitem[{{Kim} {et~al.}(2022{\natexlab{a}}){Kim}, {Chevance}, {Kruijssen},
  {Leroy}, {Schruba}, {Barnes}, {Bigiel}, {Blanc}, {Cao}, {Congiu}, {Dale},
  {Faesi}, {Glover}, {Grasha}, {Groves}, {Hughes}, {Klessen}, {Kreckel},
  {McElroy}, {Pan}, {Pety}, {Querejeta}, {Razza}, {Rosolowsky}, {Saito},
  {Schinnerer}, {Sun}, {Tomi{\v{c}}i{\'c}}, {Usero}, \& {Williams}}]{kim22}
---. 2022{\natexlab{a}}, \mnras, 516, 3006, \dodoi{10.1093/mnras/stac2339}

\bibitem[{{Kim} {et~al.}(2022{\natexlab{b}}){Kim}, {Chevance}, {Kruijssen},
  {Barnes}, {Bigiel}, {Blanc}, {Boquien}, {Cao}, {Congiu}, {Dale}, {Egorov},
  {Faesi}, {Glover}, {Grasha}, {Groves}, {Hassani}, {Hughes}, {Klessen},
  {Kreckel}, {Larson}, {Lee}, {Leroy}, {Liu}, {Longmore}, {Meidt}, {Pan},
  {Pety}, {Querejeta}, {Rosolowsky}, {Saito}, {Sandstrom}, {Schinnerer},
  {Smith}, {Usero}, {Watkins}, \& {Williams}}]{KIM_PHANGSJWST}
---. 2022{\natexlab{b}}, arXiv e-prints, arXiv:2211.15698.
\newblock \doarXiv{2211.15698}

\bibitem[{{Kim} {et~al.}(2021{\natexlab{b}}){Kim}, {Ostriker}, \&
  {Filippova}}]{jgkim21}
{Kim}, J.-G., {Ostriker}, E.~C., \& {Filippova}, N. 2021{\natexlab{b}}, \apj,
  911, 128, \dodoi{10.3847/1538-4357/abe934}

\bibitem[{{Kleinmann} \& {Low}(1967)}]{kleinmann67}
{Kleinmann}, D.~E., \& {Low}, F.~J. 1967, \apjl, 149, L1,
  \dodoi{10.1086/180039}

\bibitem[{{Klessen} \& {Glover}(2016)}]{klessen16}
{Klessen}, R.~S., \& {Glover}, S. C.~O. 2016, Saas-Fee Advanced Course, 43, 85,
  \dodoi{10.1007/978-3-662-47890-5_2}

\bibitem[{{Kourkchi} {et~al.}(2020){Kourkchi}, {Courtois}, {Graziani},
  {Hoffman}, {Pomar{\`e}de}, {Shaya}, \& {Tully}}]{kourkchi20}
{Kourkchi}, E., {Courtois}, H.~M., {Graziani}, R., {et~al.} 2020, \aj, 159, 67,
  \dodoi{10.3847/1538-3881/ab620e}

\bibitem[{{Kourkchi} \& {Tully}(2017)}]{kourkchi17}
{Kourkchi}, E., \& {Tully}, R.~B. 2017, \apj, 843, 16,
  \dodoi{10.3847/1538-4357/aa76db}

\bibitem[{{Kreckel} {et~al.}(2018){Kreckel}, {Faesi}, {Kruijssen}, {Schruba},
  {Groves}, {Leroy}, {Bigiel}, {Blanc}, {Chevance}, {Herrera}, {Hughes},
  {McElroy}, {Pety}, {Querejeta}, {Rosolowsky}, {Schinnerer}, {Sun}, {Usero},
  \& {Utomo}}]{kreckel18}
{Kreckel}, K., {Faesi}, C., {Kruijssen}, J.~M.~D., {et~al.} 2018, \apjl, 863,
  L21, \dodoi{10.3847/2041-8213/aad77d}

\bibitem[{{Kretschmer} \& {Teyssier}(2020)}]{kretschmer20}
{Kretschmer}, M., \& {Teyssier}, R. 2020, \mnras, 492, 1385,
  \dodoi{10.1093/mnras/stz3495}

\bibitem[{{Kruijssen}(2012)}]{kruijssen12}
{Kruijssen}, J.~M.~D. 2012, \mnras, 426, 3008,
  \dodoi{10.1111/j.1365-2966.2012.21923.x}

\bibitem[{{Kruijssen} \& {Longmore}(2014)}]{kruijssen14}
{Kruijssen}, J.~M.~D., \& {Longmore}, S.~N. 2014, \mnras, 439, 3239,
  \dodoi{10.1093/mnras/stu098}

\bibitem[{{Kruijssen} {et~al.}(2018){Kruijssen}, {Schruba}, {Hygate}, {Hu},
  {Haydon}, \& {Longmore}}]{kruijssen18}
{Kruijssen}, J.~M.~D., {Schruba}, A., {Hygate}, A. P.~S., {et~al.} 2018,
  \mnras, 479, 1866, \dodoi{10.1093/mnras/sty1128}

\bibitem[{{Kruijssen} {et~al.}(2019{\natexlab{a}}){Kruijssen}, {Schruba},
  {Chevance}, {Longmore}, {Hygate}, {Haydon}, {McLeod}, {Dalcanton}, {Tacconi},
  \& {van Dishoeck}}]{kruijssen19}
{Kruijssen}, J.~M.~D., {Schruba}, A., {Chevance}, M., {et~al.}
  2019{\natexlab{a}}, \nat, 569, 519, \dodoi{10.1038/s41586-019-1194-3}

\bibitem[{{Kruijssen} {et~al.}(2019{\natexlab{b}}){Kruijssen}, {Dale},
  {Longmore}, {Walker}, {Henshaw}, {Jeffreson}, {Petkova}, {Ginsburg},
  {Barnes}, {Battersby}, {Immer}, {Jackson}, {Keto}, {Krieger}, {Mills},
  {S{\'a}nchez-Monge}, {Schmiedeke}, {Suri}, \& {Zhang}}]{kruijssen19b}
{Kruijssen}, J.~M.~D., {Dale}, J.~E., {Longmore}, S.~N., {et~al.}
  2019{\natexlab{b}}, \mnras, 484, 5734, \dodoi{10.1093/mnras/stz381}

\bibitem[{{Krumholz}(2014)}]{krumholz14}
{Krumholz}, M.~R. 2014, \physrep, 539, 49,
  \dodoi{10.1016/j.physrep.2014.02.001}

\bibitem[{{Krumholz} \& {McKee}(2005)}]{krumholz05}
{Krumholz}, M.~R., \& {McKee}, C.~F. 2005, \apj, 630, 250,
  \dodoi{10.1086/431734}

\bibitem[{{Krumholz} \& {McKee}(2020)}]{krumholz20}
---. 2020, \mnras, 494, 624, \dodoi{10.1093/mnras/staa659}

\bibitem[{{Krumholz} {et~al.}(2019){Krumholz}, {McKee}, \& {Bland
  -Hawthorn}}]{krumholz19}
{Krumholz}, M.~R., {McKee}, C.~F., \& {Bland -Hawthorn}, J. 2019, \araa, 57,
  227, \dodoi{10.1146/annurev-astro-091918-104430}

\bibitem[{{Krumholz} \& {Tan}(2007)}]{krumholz07}
{Krumholz}, M.~R., \& {Tan}, J.~C. 2007, \apj, 654, 304, \dodoi{10.1086/509101}

\bibitem[{{Lang} {et~al.}(2020){Lang}, {Meidt}, {Rosolowsky}, {Nofech},
  {Schinnerer}, {Leroy}, {Emsellem}, {Pessa}, {Glover}, {Groves}, {Hughes},
  {Kruijssen}, {Querejeta}, {Schruba}, {Bigiel}, {Blanc}, {Chevance},
  {Colombo}, {Faesi}, {Henshaw}, {Herrera}, {Liu}, {Pety}, {Puschnig}, {Saito},
  {Sun}, \& {Usero}}]{lang20}
{Lang}, P., {Meidt}, S.~E., {Rosolowsky}, E., {et~al.} 2020, \apj, 897, 122,
  \dodoi{10.3847/1538-4357/ab9953}

\bibitem[{{Larsen}(2009)}]{larsen09}
{Larsen}, S.~S. 2009, \aap, 494, 539, \dodoi{10.1051/0004-6361:200811212}

\bibitem[{{Larson} {et~al.}(2022){Larson}, {Lee}, {Thilker}, {Whitmore},
  {Lilly}, {Deger}, {Dale}, {Ubeda}, \& {Chandar}}]{larson22}
{Larson}, K., {Lee}, J., {Thilker}, D., {et~al.} 2022

\bibitem[{{Lee} {et~al.}(2016){Lee}, {Miville-Desch{\^e}nes}, \&
  {Murray}}]{lee16}
{Lee}, E.~J., {Miville-Desch{\^e}nes}, M.-A., \& {Murray}, N.~W. 2016, \apj,
  833, 229, \dodoi{10.3847/1538-4357/833/2/229}

\bibitem[{{Lee} {et~al.}(2022){Lee}, {Whitmore}, {Thilker}, {Deger}, {Larson},
  {Ubeda}, {Anand}, {Boquien}, {Chandar}, {Dale}, {Emsellem}, {Leroy},
  {Rosolowsky}, {Schinnerer}, {Schmidt}, {Lilly}, {Turner}, {Van Dyk}, {White},
  {Barnes}, {Belfiore}, {Bigiel}, {Blanc}, {Cao}, {Chevance}, {Congiu},
  {Egorov}, {Glover}, {Grasha}, {Groves}, {Henshaw}, {Hughes}, {Klessen},
  {Koch}, {Kreckel}, {Kruijssen}, {Liu}, {Lopez}, {Mayker}, {Meidt}, {Murphy},
  {Pan}, {Pety}, {Querejeta}, {Razza}, {Saito}, {S{\'a}nchez-Bl{\'a}zquez},
  {Santoro}, {Sardone}, {Scheuermann}, {Schruba}, {Sun}, {Usero}, {Watkins}, \&
  {Williams}}]{phangs-hst}
{Lee}, J.~C., {Whitmore}, B.~C., {Thilker}, D.~A., {et~al.} 2022, \apjs, 258,
  10, \dodoi{10.3847/1538-4365/ac1fe5}

\bibitem[{{Leroy} {et~al.}(subm.{\natexlab{a}})}]{LEROY2_PHANGSJWST}
{Leroy}, A., {et~al.} subm.{\natexlab{a}}, \apjl

\bibitem[{{Leroy} {et~al.}(subm.{\natexlab{b}})}]{LEROY1_PHANGSJWST}
---. subm.{\natexlab{b}}, \apjl

\bibitem[{{Leroy} {et~al.}(2013){Leroy}, {Walter}, {Sandstrom}, {Schruba},
  {Munoz-Mateos}, {Bigiel}, {Bolatto}, {Brinks}, {de Blok}, {Meidt}, {Rix},
  {Rosolowsky}, {Schinnerer}, {Schuster}, \& {Usero}}]{leroy2013}
{Leroy}, A.~K., {Walter}, F., {Sandstrom}, K., {et~al.} 2013, \aj, 146, 19,
  \dodoi{10.1088/0004-6256/146/2/19}

\bibitem[{{Leroy} {et~al.}(2017){Leroy}, {Schinnerer}, {Hughes}, {Kruijssen},
  {Meidt}, {Schruba}, {Sun}, {Bigiel}, {Aniano}, {Blanc}, {Bolatto},
  {Chevance}, {Colombo}, {Gallagher}, {Garcia-Burillo}, {Kramer}, {Querejeta},
  {Pety}, {Thompson}, \& {Usero}}]{leroy17}
{Leroy}, A.~K., {Schinnerer}, E., {Hughes}, A., {et~al.} 2017, \apj, 846, 71,
  \dodoi{10.3847/1538-4357/aa7fef}

\bibitem[{{Leroy} {et~al.}(2019){Leroy}, {Sandstrom}, {Lang}, {Lewis}, {Salim},
  {Behrens}, {Chastenet}, {Chiang}, {Gallagher}, {Kessler}, \&
  {Utomo}}]{leroy19}
{Leroy}, A.~K., {Sandstrom}, K.~M., {Lang}, D., {et~al.} 2019, \apjs, 244, 24,
  \dodoi{10.3847/1538-4365/ab3925}

\bibitem[{{Leroy} {et~al.}(2021){Leroy}, {Schinnerer}, {Hughes}, {Rosolowsky},
  {Pety}, {Schruba}, {Usero}, {Blanc}, {Chevance}, {Emsellem}, {Faesi},
  {Herrera}, {Liu}, {Meidt}, {Querejeta}, {Saito}, {Sandstrom}, {Sun},
  {Williams}, {Anand}, {Barnes}, {Behrens}, {Belfiore}, {Benincasa},
  {Be{\v{s}}li{\'c}}, {Bigiel}, {Bolatto}, {den Brok}, {Cao}, {Chandar},
  {Chastenet}, {Chiang}, {Congiu}, {Dale}, {Deger}, {Eibensteiner}, {Egorov},
  {Garc{\'\i}a-Rodr{\'\i}guez}, {Glover}, {Grasha}, {Henshaw}, {Ho}, {Kepley},
  {Kim}, {Klessen}, {Kreckel}, {Koch}, {Kruijssen}, {Larson}, {Lee}, {Lopez},
  {Machado}, {Mayker}, {McElroy}, {Murphy}, {Ostriker}, {Pan}, {Pessa},
  {Puschnig}, {Razza}, {S{\'a}nchez-Bl{\'a}zquez}, {Santoro}, {Sardone},
  {Scheuermann}, {Sliwa}, {Sormani}, {Stuber}, {Thilker}, {Turner}, {Utomo},
  {Watkins}, \& {Whitmore}}]{phangs-alma}
{Leroy}, A.~K., {Schinnerer}, E., {Hughes}, A., {et~al.} 2021, \apjs, 257, 43,
  \dodoi{10.3847/1538-4365/ac17f3}

\bibitem[{{Lindegren} {et~al.}(2018{\natexlab{a}}){Lindegren}, {Hern{\'a}ndez},
  {Bombrun}, {Klioner}, {Bastian}, {Ramos-Lerate}, {de Torres},
  {Steidelm{\"u}ller}, {Stephenson}, {Hobbs}, {Lammers}, {Biermann}, {Geyer},
  {Hilger}, {Michalik}, {Stampa}, {McMillan}, {Casta{\~n}eda}, {Clotet},
  {Comoretto}, {Davidson}, {Fabricius}, {Gracia}, {Hambly}, {Hutton}, {Mora},
  {Portell}, {van Leeuwen}, {Abbas}, {Abreu}, {Altmann}, {Andrei}, {Anglada},
  {Balaguer-N{\'u}{\~n}ez}, {Barache}, {Becciani}, {Bertone}, {Bianchi},
  {Bouquillon}, {Bourda}, {Br{\"u}semeister}, {Bucciarelli}, {Busonero},
  {Buzzi}, {Cancelliere}, {Carlucci}, {Charlot}, {Cheek}, {Crosta}, {Crowley},
  {de Bruijne}, {de Felice}, {Drimmel}, {Esquej}, {Fienga}, {Fraile}, {Gai},
  {Garralda}, {Gonz{\'a}lez-Vidal}, {Guerra}, {Hauser}, {Hofmann}, {Holl},
  {Jordan}, {Lattanzi}, {Lenhardt}, {Liao}, {Licata}, {Lister}, {L{\"o}ffler},
  {Marchant}, {Martin-Fleitas}, {Messineo}, {Mignard}, {Morbidelli}, {Poggio},
  {Riva}, {Rowell}, {Salguero}, {Sarasso}, {Sciacca}, {Siddiqui}, {Smart},
  {Spagna}, {Steele}, {Taris}, {Torra}, {van Elteren}, {van Reeven}, \&
  {Vecchiato}}]{gaiadr2}
{Lindegren}, L., {Hern{\'a}ndez}, J., {Bombrun}, A., {et~al.}
  2018{\natexlab{a}}, \aap, 616, A2, \dodoi{10.1051/0004-6361/201832727}

\bibitem[{{Lindegren} {et~al.}(2018{\natexlab{b}}){Lindegren}, {Hern{\'a}ndez},
  {Bombrun}, {Klioner}, {Bastian}, {Ramos-Lerate}, {de Torres},
  {Steidelm{\"u}ller}, {Stephenson}, {Hobbs}, {Lammers}, {Biermann}, {Geyer},
  {Hilger}, {Michalik}, {Stampa}, {McMillan}, {Casta{\~n}eda}, {Clotet},
  {Comoretto}, {Davidson}, {Fabricius}, {Gracia}, {Hambly}, {Hutton}, {Mora},
  {Portell}, {van Leeuwen}, {Abbas}, {Abreu}, {Altmann}, {Andrei}, {Anglada},
  {Balaguer-N{\'u}{\~n}ez}, {Barache}, {Becciani}, {Bertone}, {Bianchi},
  {Bouquillon}, {Bourda}, {Br{\"u}semeister}, {Bucciarelli}, {Busonero},
  {Buzzi}, {Cancelliere}, {Carlucci}, {Charlot}, {Cheek}, {Crosta}, {Crowley},
  {de Bruijne}, {de Felice}, {Drimmel}, {Esquej}, {Fienga}, {Fraile}, {Gai},
  {Garralda}, {Gonz{\'a}lez-Vidal}, {Guerra}, {Hauser}, {Hofmann}, {Holl},
  {Jordan}, {Lattanzi}, {Lenhardt}, {Liao}, {Licata}, {Lister}, {L{\"o}ffler},
  {Marchant}, {Martin-Fleitas}, {Messineo}, {Mignard}, {Morbidelli}, {Poggio},
  {Riva}, {Rowell}, {Salguero}, {Sarasso}, {Sciacca}, {Siddiqui}, {Smart},
  {Spagna}, {Steele}, {Taris}, {Torra}, {van Elteren}, {van Reeven}, \&
  {Vecchiato}}]{gaia2astrometry}
---. 2018{\natexlab{b}}, \aap, 616, A2, \dodoi{10.1051/0004-6361/201832727}

\bibitem[{{Liu} {et~al.}(subm.)}]{LIU_PHANGSJWST}
{Liu}, D., {et~al.} subm., \apjl

\bibitem[{{Longmore} {et~al.}(2014){Longmore}, {Kruijssen}, {Bastian}, {Bally},
  {Rathborne}, {Testi}, {Stolte}, {Dale}, {Bressert}, \& {Alves}}]{longmore14}
{Longmore}, S.~N., {Kruijssen}, J.~M.~D., {Bastian}, N., {et~al.} 2014, in
  Protostars and Planets VI, ed. H.~{Beuther}, R.~S. {Klessen}, C.~P.
  {Dullemond}, \& T.~{Henning}, 291,
  \dodoi{10.2458/azu_uapress_9780816531240-ch013}

\bibitem[{{Lopez} {et~al.}(2014){Lopez}, {Krumholz}, {Bolatto}, {Prochaska},
  {Ramirez-Ruiz}, \& {Castro}}]{lopez14}
{Lopez}, L.~A., {Krumholz}, M.~R., {Bolatto}, A.~D., {et~al.} 2014, \apj, 795,
  121, \dodoi{10.1088/0004-637X/795/2/121}

\bibitem[{{Makarov} {et~al.}(2014){Makarov}, {Prugniel}, {Terekhova},
  {Courtois}, \& {Vauglin}}]{hyperleda}
{Makarov}, D., {Prugniel}, P., {Terekhova}, N., {Courtois}, H., \& {Vauglin},
  I. 2014, \aap, 570, A13, \dodoi{10.1051/0004-6361/201423496}

\bibitem[{{Meidt} {et~al.}(subm.)}]{MEIDT_PHANGSJWST}
{Meidt}, S., {et~al.} subm., \apjl

\bibitem[{{Meidt} {et~al.}(2015){Meidt}, {Hughes}, {Dobbs}, {Pety}, {Thompson},
  {Garc{\'\i}a-Burillo}, {Leroy}, {Schinnerer}, {Colombo}, {Querejeta},
  {Kramer}, {Schuster}, \& {Dumas}}]{meidt15}
{Meidt}, S.~E., {Hughes}, A., {Dobbs}, C.~L., {et~al.} 2015, \apj, 806, 72,
  \dodoi{10.1088/0004-637X/806/1/72}

\bibitem[{{Meidt} {et~al.}(2018){Meidt}, {Leroy}, {Rosolowsky}, {Kruijssen},
  {Schinnerer}, {Schruba}, {Pety}, {Blanc}, {Bigiel}, {Chevance}, {Hughes},
  {Querejeta}, \& {Usero}}]{meidt18}
{Meidt}, S.~E., {Leroy}, A.~K., {Rosolowsky}, E., {et~al.} 2018, \apj, 854,
  100, \dodoi{10.3847/1538-4357/aaa290}

\bibitem[{{Meidt} {et~al.}(2020){Meidt}, {Glover}, {Kruijssen}, {Leroy},
  {Rosolowsky}, {Hughes}, {Schinnerer}, {Schruba}, {Usero}, {Bigiel}, {Blanc},
  {Chevance}, {Pety}, {Querejeta}, \& {Utomo}}]{meidt20}
{Meidt}, S.~E., {Glover}, S. C.~O., {Kruijssen}, J.~M.~D., {et~al.} 2020, \apj,
  892, 73, \dodoi{10.3847/1538-4357/ab7000}

\bibitem[{{Mooney} \& {Solomon}(1988)}]{mooney88}
{Mooney}, T.~J., \& {Solomon}, P.~M. 1988, \apjl, 334, L51,
  \dodoi{10.1086/185310}

\bibitem[{{Mu{\~n}oz-Mateos} {et~al.}(2015){Mu{\~n}oz-Mateos}, {Sheth},
  {Regan}, {Kim}, {Laine}, {Erroz-Ferrer}, {Gil de Paz}, {Comeron}, {Hinz},
  {Laurikainen}, {Salo}, {Athanassoula}, {Bosma}, {Bouquin}, {Schinnerer},
  {Ho}, {Zaritsky}, {Gadotti}, {Madore}, {Holwerda}, {Men{\'e}ndez-Delmestre},
  {Knapen}, {Meidt}, {Querejeta}, {Mizusawa}, {Seibert}, {Laine}, \&
  {Courtois}}]{munozmateos15}
{Mu{\~n}oz-Mateos}, J.~C., {Sheth}, K., {Regan}, M., {et~al.} 2015, \apjs, 219,
  3, \dodoi{10.1088/0067-0049/219/1/3}

\bibitem[{{Murray}(2011)}]{murray11}
{Murray}, N. 2011, \apj, 729, 133, \dodoi{10.1088/0004-637X/729/2/133}

\bibitem[{{Nersesian} {et~al.}(2019){Nersesian}, {Xilouris}, {Bianchi},
  {Galliano}, {Jones}, {Baes}, {Casasola}, {Cassar{\`a}}, {Clark}, {Davies},
  {Decleir}, {Dobbels}, {De Looze}, {De Vis}, {Fritz}, {Galametz}, {Madden},
  {Mosenkov}, {Tr{\v{c}}ka}, {Verstocken}, {Viaene}, \& {Lianou}}]{nersesian19}
{Nersesian}, A., {Xilouris}, E.~M., {Bianchi}, S., {et~al.} 2019, \aap, 624,
  A80, \dodoi{10.1051/0004-6361/201935118}

\bibitem[{{Neugebauer} {et~al.}(1984){Neugebauer}, {Habing}, {van Duinen},
  {Aumann}, {Baud}, {Beichman}, {Beintema}, {Boggess}, {Clegg}, {de Jong},
  {Emerson}, {Gautier}, {Gillett}, {Harris}, {Hauser}, {Houck}, {Jennings},
  {Low}, {Marsden}, {Miley}, {Olnon}, {Pottasch}, {Raimond}, {Rowan-Robinson},
  {Soifer}, {Walker}, {Wesselius}, \& {Young}}]{neugebaueretal84}
{Neugebauer}, G., {Habing}, H.~J., {van Duinen}, R., {et~al.} 1984, \apjl, 278,
  L1, \dodoi{10.1086/184209}

\bibitem[{{Nugent} {et~al.}(2006){Nugent}, {Sullivan}, {Ellis}, {Gal-Yam},
  {Leonard}, {Howell}, {Astier}, {Carlberg}, {Conley}, {Fabbro}, {Fouchez},
  {Neill}, {Pain}, {Perrett}, {Pritchet}, \& {Regnault}}]{nugent2006}
{Nugent}, P., {Sullivan}, M., {Ellis}, R., {et~al.} 2006, \apj, 645, 841,
  \dodoi{10.1086/504413}

\bibitem[{{Oey} \& {Clarke}(1997)}]{oey97}
{Oey}, M.~S., \& {Clarke}, C.~J. 1997, \mnras, 289, 570,
  \dodoi{10.1093/mnras/289.3.570}

\bibitem[{{Olivier} {et~al.}(2021){Olivier}, {Lopez}, {Rosen}, {Nayak},
  {Reiter}, {Krumholz}, \& {Bolatto}}]{olivier21}
{Olivier}, G.~M., {Lopez}, L.~A., {Rosen}, A.~L., {et~al.} 2021, \apj, 908, 68,
  \dodoi{10.3847/1538-4357/abd24a}

\bibitem[{{Onodera} {et~al.}(2010){Onodera}, {Kuno}, {Tosaki}, {Kohno},
  {Nakanishi}, {Sawada}, {Muraoka}, {Komugi}, {Miura}, {Kaneko}, {Hirota}, \&
  {Kawabe}}]{onodera10}
{Onodera}, S., {Kuno}, N., {Tosaki}, T., {et~al.} 2010, \apjl, 722, L127,
  \dodoi{10.1088/2041-8205/722/2/L127}

\bibitem[{{Padoan} {et~al.}(2017){Padoan}, {Haugb{\o}lle}, {Nordlund}, \&
  {Frimann}}]{padoan17}
{Padoan}, P., {Haugb{\o}lle}, T., {Nordlund}, {\r{A}}., \& {Frimann}, S. 2017,
  \apj, 840, 48, \dodoi{10.3847/1538-4357/aa6afa}

\bibitem[{{Pan} {et~al.}(2022){Pan}, {Schinnerer}, {Hughes}, {Leroy}, {Groves},
  {Barnes}, {Belfiore}, {Bigiel}, {Blanc}, {Cao}, {Chevance}, {Congiu}, {Dale},
  {Eibensteiner}, {Emsellem}, {Faesi}, {Glover}, {Grasha}, {Herrera}, {Ho},
  {Klessen}, {Kruijssen}, {Lang}, {Liu}, {McElroy}, {Meidt}, {Murphy}, {Pety},
  {Querejeta}, {Razza}, {Rosolowsky}, {Saito}, {Santoro}, {Schruba}, {Sun},
  {Tomi{\v{c}}i{\'c}}, {Usero}, {Utomo}, \& {Williams}}]{pan22}
{Pan}, H.-A., {Schinnerer}, E., {Hughes}, A., {et~al.} 2022, \apj, 927, 9,
  \dodoi{10.3847/1538-4357/ac474f}

\bibitem[{{Perrin} {et~al.}(2014){Perrin}, {Sivaramakrishnan}, {Lajoie},
  {Elliott}, {Pueyo}, {Ravindranath}, \& {Albert}}]{webbpsf}
{Perrin}, M.~D., {Sivaramakrishnan}, A., {Lajoie}, C.-P., {et~al.} 2014, in
  Society of Photo-Optical Instrumentation Engineers (SPIE) Conference Series,
  Vol. 9143, Space Telescopes and Instrumentation 2014: Optical, Infrared, and
  Millimeter Wave, ed. J.~{Oschmann}, Jacobus~M., M.~{Clampin}, G.~G. {Fazio},
  \& H.~A. {MacEwen}, 91433X, \dodoi{10.1117/12.2056689}

\bibitem[{{Pilbratt} {et~al.}(2010){Pilbratt}, {Riedinger}, {Passvogel},
  {Crone}, {Doyle}, {Gageur}, {Heras}, {Jewell}, {Metcalfe}, {Ott}, \&
  {Schmidt}}]{pilbrattetal10}
{Pilbratt}, G.~L., {Riedinger}, J.~R., {Passvogel}, T., {et~al.} 2010, \aap,
  518, L1, \dodoi{10.1051/0004-6361/201014759}

\bibitem[{{Pilyugin} \& {Grebel}(2016)}]{scal16}
{Pilyugin}, L.~S., \& {Grebel}, E.~K. 2016, \mnras, 457, 3678,
  \dodoi{10.1093/mnras/stw238}

\bibitem[{{Portegies Zwart} {et~al.}(2010){Portegies Zwart}, {McMillan}, \&
  {Gieles}}]{pz10}
{Portegies Zwart}, S.~F., {McMillan}, S. L.~W., \& {Gieles}, M. 2010, \araa,
  48, 431, \dodoi{10.1146/annurev-astro-081309-130834}

\bibitem[{{Riechers} {et~al.}(2014){Riechers}, {Pope}, {Daddi}, {Armus},
  {Carilli}, {Walter}, {Hodge}, {Chary}, {Morrison}, {Dickinson},
  {Dannerbauer}, \& {Elbaz}}]{riechers14}
{Riechers}, D.~A., {Pope}, A., {Daddi}, E., {et~al.} 2014, \apj, 786, 31,
  \dodoi{10.1088/0004-637X/786/1/31}

\bibitem[{{Rodriguez} {et~al.}(2022){Rodriguez}, {Lee}, {Whitmore}, {Thilker},
  {Maschmann}, {Chandar}, {Dale}, {Kruijssen}, {Boquien}, {Grasha}, {Watkins},
  {Barnes}, {Sormani}, {Williams}, {Kim}, {Anand}, {Chevance}, {Bigiel},
  {Leroy}, {Klessen}, {Rosolowsky}, {Sandstrom}, {Hassani}, {Kim}, {Larson},
  {Deger}, {Liu}, {Faesi}, {Cao}, {Belfiore}, {Pessa}, {Kreckel}, {Groves},
  {Pety}, {Indebetouw}, {Egorov}, {Blanc}, {Saito}, {Emsellem}, {Hughes}, \&
  {Schinnerer}}]{RODRIGUEZ_PHANGSJWST}
{Rodriguez}, J., {Lee}, J., {Whitmore}, B., {et~al.} 2022, arXiv e-prints,
  arXiv:2211.13426.
\newblock \doarXiv{2211.13426}

\bibitem[{{Rosolowsky} {et~al.}(2021){Rosolowsky}, {Hughes}, {Leroy}, {Sun},
  {Querejeta}, {Schruba}, {Usero}, {Herrera}, {Liu}, {Pety}, {Saito},
  {Be{\v{s}}li{\'c}}, {Bigiel}, {Blanc}, {Chevance}, {Dale}, {Deger}, {Faesi},
  {Glover}, {Henshaw}, {Klessen}, {Kruijssen}, {Larson}, {Lee}, {Meidt}, {Mok},
  {Schinnerer}, {Thilker}, \& {Williams}}]{rosolowsky21}
{Rosolowsky}, E., {Hughes}, A., {Leroy}, A.~K., {et~al.} 2021, \mnras, 502,
  1218, \dodoi{10.1093/mnras/stab085}

\bibitem[{{Ryon} {et~al.}(2017){Ryon}, {Gallagher}, {Smith}, {Adamo},
  {Calzetti}, {Bright}, {Cignoni}, {Cook}, {Dale}, {Elmegreen}, {Fumagalli},
  {Gouliermis}, {Grasha}, {Grebel}, {Kim}, {Messa}, {Thilker}, \&
  {Ubeda}}]{ryon17}
{Ryon}, J.~E., {Gallagher}, J.~S., {Smith}, L.~J., {et~al.} 2017, \apj, 841,
  92, \dodoi{10.3847/1538-4357/aa719e}

\bibitem[{{Saintonge} {et~al.}(2017){Saintonge}, {Catinella}, {Tacconi},
  {Kauffmann}, {Genzel}, {Cortese}, {Dav{\'e}}, {Fletcher},
  {Graci{\'a}-Carpio}, {Kramer}, {Heckman}, {Janowiecki}, {Lutz}, {Rosario},
  {Schiminovich}, {Schuster}, {Wang}, {Wuyts}, {Borthakur}, {Lamperti}, \&
  {Roberts-Borsani}}]{Saintonge17}
{Saintonge}, A., {Catinella}, B., {Tacconi}, L.~J., {et~al.} 2017, \apjs, 233,
  22, \dodoi{10.3847/1538-4365/aa97e0}

\bibitem[{{Salim} {et~al.}(2018){Salim}, {Boquien}, \& {Lee}}]{salim18}
{Salim}, S., {Boquien}, M., \& {Lee}, J.~C. 2018, \apj, 859, 11,
  \dodoi{10.3847/1538-4357/aabf3c}

\bibitem[{{Salim} {et~al.}(2007){Salim}, {Rich}, {Charlot}, {Brinchmann},
  {Johnson}, {Schiminovich}, {Seibert}, {Mallery}, {Heckman}, {Forster},
  {Friedman}, {Martin}, {Morrissey}, {Neff}, {Small}, {Wyder}, {Bianchi},
  {Donas}, {Lee}, {Madore}, {Milliard}, {Szalay}, {Welsh}, \& {Yi}}]{salim07}
{Salim}, S., {Rich}, R.~M., {Charlot}, S., {et~al.} 2007, \apjs, 173, 267,
  \dodoi{10.1086/519218}

\bibitem[{{Salim} {et~al.}(2016){Salim}, {Lee}, {Janowiecki}, {da Cunha},
  {Dickinson}, {Boquien}, {Burgarella}, {Salzer}, \& {Charlot}}]{salim16}
{Salim}, S., {Lee}, J.~C., {Janowiecki}, S., {et~al.} 2016, \apjs, 227, 2,
  \dodoi{10.3847/0067-0049/227/1/2}

\bibitem[{{Sandstrom} {et~al.}(subm.{\natexlab{a}})}]{SANDSTROM1_PHANGSJWST}
{Sandstrom}, K., {et~al.} subm.{\natexlab{a}}, \apjl

\bibitem[{{Sandstrom} {et~al.}(subm.{\natexlab{b}})}]{SANDSTROM2_PHANGSJWST}
---. subm.{\natexlab{b}}, \apjl

\bibitem[{{Santoro} {et~al.}(2022){Santoro}, {Kreckel}, {Belfiore}, {Groves},
  {Congiu}, {Thilker}, {Blanc}, {Schinnerer}, {Ho}, {Diederik Kruijssen},
  {Meidt}, {Klessen}, {Schruba}, {Querejeta}, {Pessa}, {Chevance}, {Kim},
  {Emsellem}, {McElroy}, {Barnes}, {Bigiel}, {Boquien}, {Dale}, {Glover},
  {Grasha}, {Lee}, {Leroy}, {Pan}, {Rosolowsky}, {Saito}, {Sanchez-Blazquez},
  {Watkins}, \& {Williams}}]{santoro22}
{Santoro}, F., {Kreckel}, K., {Belfiore}, F., {et~al.} 2022, \aap, 658, A188,
  \dodoi{10.1051/0004-6361/202141907}

\bibitem[{{Scheuermann} {et~al.}(2022){Scheuermann}, {Kreckel}, {Anand},
  {Blanc}, {Congiu}, {Santoro}, {Van Dyk}, {Barnes}, {Bigiel}, {Glover},
  {Groves}, {Klessen}, {Kruijssen}, {Rosolowsky}, {Schinnerer}, {Schruba},
  {Watkins}, \& {Williams}}]{scheuermann22}
{Scheuermann}, F., {Kreckel}, K., {Anand}, G.~S., {et~al.} 2022, \mnras, 511,
  6087, \dodoi{10.1093/mnras/stac110}

\bibitem[{{Schinnerer} {et~al.}(2019{\natexlab{a}}){Schinnerer}, {Leroy},
  {Blanc}, {Emsellem}, {Hughes}, {Rosolowsky}, {Schruba}, {Bigiel}, {Escala},
  {Groves}, {Kreckel}, {Kruijssen}, {Lee}, {Meidt}, {Pety}, {Sanchez-Blazquez},
  {Sandstrom}, {Usero}, {Barnes}, {Belfiore}, {Be{\v{s}}li{\'c}}, {Chandar},
  {Chatzigiannakis}, {Chevance}, {Congiu}, {Dale}, {Faesi}, {Gallagher},
  {Garcia-Rodriguez}, {Glover}, {Grasha}, {Henshaw}, {Herrera}, {Ho}, {Hygate},
  {Jimenez-Donaire}, {Kessler}, {Kim}, {Klessen}, {Koch}, {Lang}, {Larson}, {Le
  Reste}, {Liu}, {McElroy}, {Nofech}, {Ostriker}, {Pessa Gutierrez},
  {Puschnig}, {Querejeta}, {Razza}, {Saito}, {Santoro}, {Stuber}, {Sun},
  {Thilker}, {Turner}, {Ubeda}, {Utreras}, {Utomo}, {van Dyk}, {Ward}, \&
  {Whitmore}}]{schinnerer19}
{Schinnerer}, E., {Leroy}, A., {Blanc}, G., {et~al.} 2019{\natexlab{a}}, The
  Messenger, 177, 36, \dodoi{10.18727/0722-6691/5151}

\bibitem[{{Schinnerer} {et~al.}(2019{\natexlab{b}}){Schinnerer}, {Hughes},
  {Leroy}, {Groves}, {Blanc}, {Kreckel}, {Bigiel}, {Chevance}, {Dale},
  {Emsellem}, {Faesi}, {Glover}, {Grasha}, {Henshaw}, {Hygate}, {Kruijssen},
  {Meidt}, {Pety}, {Querejeta}, {Rosolowsky}, {Saito}, {Schruba}, {Sun}, \&
  {Utomo}}]{schinnerer19b}
{Schinnerer}, E., {Hughes}, A., {Leroy}, A., {et~al.} 2019{\natexlab{b}}, \apj,
  887, 49, \dodoi{10.3847/1538-4357/ab50c2}

\bibitem[{{Schinnerer} {et~al.}(subm.)}]{SCHINNERER_PHANGSJWST}
{Schinnerer}, E., {et~al.} subm., \apjl

\bibitem[{{Schlawin} {et~al.}(2020){Schlawin}, {Leisenring}, {Misselt},
  {Greene}, {McElwain}, {Beatty}, \& {Rieke}}]{2020Schlawin}
{Schlawin}, E., {Leisenring}, J., {Misselt}, K., {et~al.} 2020, \aj, 160, 231,
  \dodoi{10.3847/1538-3881/abb811}

\bibitem[{{Schruba} {et~al.}(2019){Schruba}, {Kruijssen}, \&
  {Leroy}}]{schruba19}
{Schruba}, A., {Kruijssen}, J.~M.~D., \& {Leroy}, A.~K. 2019, \apj, 883, 2,
  \dodoi{10.3847/1538-4357/ab3a43}

\bibitem[{{Schruba} {et~al.}(2010){Schruba}, {Leroy}, {Walter}, {Sandstrom}, \&
  {Rosolowsky}}]{schruba10}
{Schruba}, A., {Leroy}, A.~K., {Walter}, F., {Sandstrom}, K., \& {Rosolowsky},
  E. 2010, \apj, 722, 1699, \dodoi{10.1088/0004-637X/722/2/1699}

\bibitem[{{Semenov} {et~al.}(2017){Semenov}, {Kravtsov}, \&
  {Gnedin}}]{semenov17}
{Semenov}, V.~A., {Kravtsov}, A.~V., \& {Gnedin}, N.~Y. 2017, \apj, 845, 133,
  \dodoi{10.3847/1538-4357/aa8096}

\bibitem[{{Shaya} {et~al.}(2017){Shaya}, {Tully}, {Hoffman}, \&
  {Pomar{\`e}de}}]{shaya17}
{Shaya}, E.~J., {Tully}, R.~B., {Hoffman}, Y., \& {Pomar{\`e}de}, D. 2017,
  \apj, 850, 207, \dodoi{10.3847/1538-4357/aa9525}

\bibitem[{{Smith} {et~al.}(2007){Smith}, {Draine}, {Dale}, {Moustakas},
  {Kennicutt}, {Helou}, {Armus}, {Roussel}, {Sheth}, {Bendo}, {Buckalew},
  {Calzetti}, {Engelbracht}, {Gordon}, {Hollenbach}, {Li}, {Malhotra},
  {Murphy}, \& {Walter}}]{smith07}
{Smith}, J.~D.~T., {Draine}, B.~T., {Dale}, D.~A., {et~al.} 2007, \apj, 656,
  770, \dodoi{10.1086/510549}

\bibitem[{{Smithsonian Astrophysical
  Observatory}(2000)}]{smithsonian_astrophysical_observatory_saoimage_2000}
{Smithsonian Astrophysical Observatory}. 2000, Astrophysics Source Code
  Library, ascl:0003.002

\bibitem[{{Soifer} {et~al.}(2008){Soifer}, {Helou}, \& {Werner}}]{soifer08}
{Soifer}, B.~T., {Helou}, G., \& {Werner}, M. 2008, \araa, 46, 201,
  \dodoi{10.1146/annurev.astro.46.060407.145144}

\bibitem[{{Soifer} {et~al.}(1987){Soifer}, {Neugebauer}, \& {Houck}}]{soifer87}
{Soifer}, B.~T., {Neugebauer}, G., \& {Houck}, J.~R. 1987, \araa, 25, 187,
  \dodoi{10.1146/annurev.aa.25.090187.001155}

\bibitem[{{Sun} {et~al.}(2018){Sun}, {Leroy}, {Schruba}, {Rosolowsky},
  {Hughes}, {Kruijssen}, {Meidt}, {Schinnerer}, {Blanc}, {Bigiel}, {Bolatto},
  {Chevance}, {Groves}, {Herrera}, {Hygate}, {Pety}, {Querejeta}, {Usero}, \&
  {Utomo}}]{sun18}
{Sun}, J., {Leroy}, A.~K., {Schruba}, A., {et~al.} 2018, \apj, 860, 172,
  \dodoi{10.3847/1538-4357/aac326}

\bibitem[{{Sun} {et~al.}(2020){Sun}, {Leroy}, {Schinnerer}, {Hughes},
  {Rosolowsky}, {Querejeta}, {Schruba}, {Liu}, {Saito}, {Herrera}, {Faesi},
  {Usero}, {Pety}, {Kruijssen}, {Ostriker}, {Bigiel}, {Blanc}, {Bolatto},
  {Boquien}, {Chevance}, {Dale}, {Deger}, {Emsellem}, {Glover}, {Grasha},
  {Groves}, {Henshaw}, {Jimenez-Donaire}, {Kim}, {Klessen}, {Kreckel}, {Lee},
  {Meidt}, {Sand strom}, {Sardone}, {Utomo}, \& {Williams}}]{sun20}
{Sun}, J., {Leroy}, A.~K., {Schinnerer}, E., {et~al.} 2020, \apjl, 901, L8,
  \dodoi{10.3847/2041-8213/abb3be}

\bibitem[{{Sun} {et~al.}(2022){Sun}, {Leroy}, {Rosolowsky}, {Hughes},
  {Schinnerer}, {Schruba}, {Koch}, {Blanc}, {Chiang}, {Groves}, {Liu}, {Meidt},
  {Pan}, {Pety}, {Querejeta}, {Saito}, {Sandstrom}, {Sardone}, {Usero},
  {Utomo}, {Williams}, {Barnes}, {Benincasa}, {Bigiel}, {Bolatto}, {Boquien},
  {Chevance}, {Dale}, {Deger}, {Emsellem}, {Glover}, {Grasha}, {Henshaw},
  {Klessen}, {Kreckel}, {Kruijssen}, {Ostriker}, \& {Thilker}}]{sun22}
{Sun}, J., {Leroy}, A.~K., {Rosolowsky}, E., {et~al.} 2022, \aj, 164, 43,
  \dodoi{10.3847/1538-3881/ac74bd}

\bibitem[{{Thilker} {et~al.}(subm.)}]{THILKER_PHANGSJWST}
{Thilker}, D., {et~al.} subm., \apjl

\bibitem[{{Thilker} {et~al.}(2022){Thilker}, {Whitmore}, {Lee}, {Deger},
  {Chandar}, {Larson}, {Hannon}, {Ubeda}, {Dale}, {Glover}, {Grasha},
  {Klessen}, {Kruijssen}, {Rosolowsky}, {Schruba}, {White}, \&
  {Williams}}]{thilker22}
{Thilker}, D.~A., {Whitmore}, B.~C., {Lee}, J.~C., {et~al.} 2022, \mnras, 509,
  4094, \dodoi{10.1093/mnras/stab3183}

\bibitem[{{Tielens}(2008)}]{tielens08}
{Tielens}, A.~G.~G.~M. 2008, \araa, 46, 289,
  \dodoi{10.1146/annurev.astro.46.060407.145211}

\bibitem[{{Turner} {et~al.}(2021){Turner}, {Dale}, {Lee}, {Boquien}, {Chandar},
  {Deger}, {Larson}, {Mok}, {Thilker}, {Ubeda}, {Whitmore}, {Belfiore},
  {Bigiel}, {Blanc}, {Emsellem}, {Grasha}, {Groves}, {Klessen}, {Kreckel},
  {Kruijssen}, {Leroy}, {Rosolowsky}, {Sanchez-Blazquez}, {Schinnerer},
  {Schruba}, {Van Dyk}, \& {Williams}}]{turner21}
{Turner}, J.~A., {Dale}, D.~A., {Lee}, J.~C., {et~al.} 2021, \mnras, 502, 1366,
  \dodoi{10.1093/mnras/stab055}

\bibitem[{{Utomo} {et~al.}(2018){Utomo}, {Sun}, {Leroy}, {Kruijssen},
  {Schinnerer}, {Schruba}, {Bigiel}, {Blanc}, {Chevance}, {Emsellem},
  {Herrera}, {Hygate}, {Kreckel}, {Ostriker}, {Pety}, {Querejeta},
  {Rosolowsky}, {Sandstrom}, \& {Usero}}]{utomo18}
{Utomo}, D., {Sun}, J., {Leroy}, A.~K., {et~al.} 2018, \apjl, 861, L18,
  \dodoi{10.3847/2041-8213/aacf8f}

\bibitem[{{Ward} {et~al.}(2022){Ward}, {Kruijssen}, {Chevance}, {Kim}, \&
  {Longmore}}]{ward22}
{Ward}, J.~L., {Kruijssen}, J.~M.~D., {Chevance}, M., {Kim}, J., \& {Longmore},
  S.~N. 2022, \mnras, 516, 4025, \dodoi{10.1093/mnras/stac2467}

\bibitem[{{Watkins} {et~al.}(subm.)}]{WATKINS_PHANGSJWST}
{Watkins}, E., {et~al.} subm., \apjl

\bibitem[{{Weisz} {et~al.}(2009){Weisz}, {Skillman}, {Cannon}, {Dolphin},
  {Kennicutt}, {Lee}, \& {Walter}}]{weisz09}
{Weisz}, D.~R., {Skillman}, E.~D., {Cannon}, J.~M., {et~al.} 2009, \apj, 704,
  1538, \dodoi{10.1088/0004-637X/704/2/1538}

\bibitem[{{Werner} {et~al.}(2004){Werner}, {Roellig}, {Low}, {Rieke}, {Rieke},
  {Hoffmann}, {Young}, {Houck}, {Brandl}, {Fazio}, {Hora}, {Gehrz}, {Helou},
  {Soifer}, {Stauffer}, {Keene}, {Eisenhardt}, {Gallagher}, {Gautier}, {Irace},
  {Lawrence}, {Simmons}, {Van Cleve}, {Jura}, {Wright}, \&
  {Cruikshank}}]{werneretal04}
{Werner}, M.~W., {Roellig}, T.~L., {Low}, F.~J., {et~al.} 2004, \apjs, 154, 1,
  \dodoi{10.1086/422992}

\bibitem[{{Whitcomb} {et~al.}(2020){Whitcomb}, {Sandstrom}, {Murphy}, \&
  {Linden}}]{whitcomb20}
{Whitcomb}, C.~M., {Sandstrom}, K., {Murphy}, E.~J., \& {Linden}, S. 2020,
  \apj, 901, 47, \dodoi{10.3847/1538-4357/abaef6}

\bibitem[{{Whitmore} {et~al.}(2022){Whitmore}, {Chandar}, \&
  {Lee}}]{whitmore22}
{Whitmore}, B., {Chandar}, R., \& {Lee}, J.~C. 2022

\bibitem[{{Whitmore} {et~al.}(subm.)}]{WHITMORE_PHANGSJWST}
{Whitmore}, B., {et~al.} subm., \apjl

\bibitem[{{Whitmore} {et~al.}(2021){Whitmore}, {Lee}, {Chandar}, {Thilker},
  {Hannon}, {Wei}, {Huerta}, {Bigiel}, {Boquien}, {Chevance}, {Dale}, {Deger},
  {Grasha}, {Klessen}, {Kruijssen}, {Larson}, {Mok}, {Rosolowsky},
  {Schinnerer}, {Schruba}, {Ubeda}, {Van Dyk}, {Watkins}, \&
  {Williams}}]{whitmore21}
{Whitmore}, B.~C., {Lee}, J.~C., {Chandar}, R., {et~al.} 2021, \mnras, 506,
  5294, \dodoi{10.1093/mnras/stab2087}

\bibitem[{{Wild} \& {Hewett}(2005)}]{2005Wild}
{Wild}, V., \& {Hewett}, P.~C. 2005, \mnras, 358, 1083,
  \dodoi{10.1111/j.1365-2966.2005.08844.x}

\bibitem[{{Williams} {et~al.}(subm.)}]{WILLIAMS_PHANGSJWST}
{Williams}, T., {et~al.} subm., \apjl

\bibitem[{{Wright} {et~al.}(2010){Wright}, {Eisenhardt}, {Mainzer}, {Ressler},
  {Cutri}, {Jarrett}, {Kirkpatrick}, {Padgett}, {McMillan}, {Skrutskie},
  {Stanford}, {Cohen}, {Walker}, {Mather}, {Leisawitz}, {Gautier}, {McLean},
  {Benford}, {Lonsdale}, {Blain}, {Mendez}, {Irace}, {Duval}, {Liu}, {Royer},
  {Heinrichsen}, {Howard}, {Shannon}, {Kendall}, {Walsh}, {Larsen}, {Cardon},
  {Schick}, {Schwalm}, {Abid}, {Fabinsky}, {Naes}, \& {Tsai}}]{wise}
{Wright}, E.~L., {Eisenhardt}, P.~R.~M., {Mainzer}, A.~K., {et~al.} 2010, \aj,
  140, 1868, \dodoi{10.1088/0004-6256/140/6/1868}

\end{thebibliography}

\end{document}